\documentclass[
    10pt,
    twocolumn,
    superscriptaddress,
    nofootinbib,
    amsmath,
    amssymb,
    aps,
    prx
]{revtex4-2}

\usepackage{xcolor}
\usepackage{graphicx}
\usepackage[colorlinks=true, linkcolor=blue, citecolor=red]{hyperref}
\usepackage{amsthm}
\usepackage{array}

\usepackage[usestackEOL]{stackengine}
\usepackage{tikz}
\usetikzlibrary{quantikz}

\newcommand{\mbf}{\mathbf}
\newcommand{\mbb}{\mathbb}
\newcommand{\mc}{\mathcal}
\newcommand{\tr}[1]{\text{Tr}\left[ #1 \right]} % trace
\newcommand{\ip}[1]{\left\langle #1 \right\rangle} % expectation/inner product
\newcommand{\op}[2]{\left|#1\right\rangle\!\left\langle #2\right|} % outer product
\newcommand{\conv}[1]{\text{Conv}\left( #1\right)}

\newcommand{\Net}{\text{Net}}
\newcommand{\Bell}{\text{Bell}}
\newcommand{\PNet}{\mbf{P}_{\Net}}
\newcommand{\rhoNet}{\rho^{\Net}}
\newcommand{\psiNet}{\psi^{\Net}}
\newcommand{\NNet}{\mc{N}^{\Net}}

\newcommand{\PiNet}{\Pi^{\Net}}

\newcommand{\UNet}{U^{\Net}}
\newcommand{\LNet}{\mc{L}_{\Net}}
\newcommand{\Cost}{\text{Cost}}

\newcommand{\CHSH}{\text{CHSH}}
\newcommand{\av}{\vec{a}}
\newcommand{\bv}{\vec{b}}

\newcommand{\xv}{\vec{x}}
\newcommand{\yv}{\vec{y}}
\newcommand{\zv}{\vec{z}}

\newcommand{\thetav}{\vec{\theta}}
\newcommand{\phiv}{\vec{\phi}}
\newcommand{\gammav}{\vec{\gamma}}
\newcommand{\lambdav}{\vec{\lambda}}
\newcommand{\sigmav}{\vec{\sigma}}
\newcommand{\N}{\mc{N}}

\newcommand{\QNet}{\mc{Q}_{\Net}}

\newcommand{\Prep}{\text{Prep}}
\newcommand{\Meas}{\text{Meas}}

\theoremstyle{definition}

\newtheorem{definition}{Definition}

\newtheorem{proposition}{Proposition}

\newtheorem{theorem}{Theorem}

\newcolumntype{C}[1]{>{\centering\let\newline\\\arraybackslash\hspace{0pt}}m{#1}}

\begin{document}

\title{Variational Quantum Optimization of Nonlocality in Noisy Quantum Networks}

\author{Brian Doolittle}
\affiliation{Department of Physics, University of Illinois at Urbana-Champaign, Urbana, Illinois  61801,USA}

\author{Tom Bromley}
\affiliation{Xanadu, Toronto, Ontario, M5G 2C8, Canada}

\author{Nathan Killoran}
\affiliation{Xanadu, Toronto, Ontario, M5G 2C8, Canada}

\author{Eric Chitambar}
\affiliation{Department of Electrical and Computer Engineering, Coordinated Science Laboratory,University of Illinois at Urbana-Champaign, Urbana, Illinois 61801, USA}

\date{\today}

\begin{abstract}
    The inherent noise and complexity of quantum communication networks leads to challenges in designing quantum network protocols using classical methods.
    To address this issue, we develop a variational quantum optimization framework that simulates quantum networks on quantum hardware and optimizes the network using differential programming techniques.
    We use our hybrid framework to optimize nonlocality in noisy quantum networks.
    On the noisy IBM quantum computers, we demonstrate our framework’s ability to maximize quantum nonlocality.
    On a classical simulator with a static noise model, we investigate the noise robustness of quantum nonlocality with respect to unital and nonunital channels.
    In both cases, we find that our optimization methods can reproduce known results, while uncovering interesting phenomena.
    When unital noise is present we find numerical evidence suggesting that maximally entangled state preparations yield maximal nonlocality. 
    When nonunital noise is present we find that nonmaximally entangled states can yield maximal nonlocality.
    Thus, we show that variational quantum optimization is a practical design tool for quantum networks in the near-term.
    In the long-term, our variational quantum optimization techniques show promise of scaling beyond classical approaches and can be deployed on quantum network hardware to optimize quantum communication protocols against their inherent noise.
\end{abstract}

\maketitle

\section{Introduction}

The world is progressing towards the quantum internet \cite{Kimble2008quantum_internet,Simon2017global_quantum_network,Wehner2018quantum_internet,kozlowski2019towards}, a global network of quantum devices linked by quantum communication.
The quantum internet will revolutionize science and technology by providing advantages in distributed sensing \cite{Kmr2014clock_synchronization,khabiboulline2019_distributed_sensing, khabiboiulline2019_distributed_sensing2}, communications \cite{Sun2016teleportation,Cozzolino2019_high_dimensional_communication,Herbert2020superdense}, network security \cite{scarani2009_qkd,Broadbent2009blind_qc,Lee2018_network_di,Pirandola2020, Luo2022_di_network}, and distributed information processing \cite{buhrman2010_communication_complexity,Cuomo2020_distributed_qc}.
Unfortunately, these applications are not readily available due to, largely, the presence of noise in quantum hardware.
Nevertheless, we are at the forefront of the quantum internet.
Using existing technology, we can build and scale rudimentary quantum networks, and as new quantum technologies emerge, we can upgrade the functionality of these networks \cite{Wehner2018quantum_internet}.

As quantum communication networks scale, two daunting design challenges emerge.
First, network protocols must be designed around hardware noise, however, the difficulty of exactly characterizing quantum noise grows exponentially with the number of qubits  \cite{Chaung1997_process_tomography, harper2020efficient, Onorati2021_noise_tomography}.
Second, classical simulation and optimization methods fail to scale with the exponential growth of the Hilbert space dimension.
While tensor networks \cite{Orus2019_tn} are a promising classical approach, their efficiency requires certain symmetries and entanglement structures to hold.
In general, these assumptions might not hold because quantum repeaters \cite{Briegel1998_quantum_repeater,sangouard2011_quantum_repeater} can perform entanglement swapping protocols \cite{Bennet1993_teleportation_entanglement,Zukowski1993_entanglement_swapping,Bose1998_generalized_entanglement_swapping} to create complex, multi-qubit entanglement across distant network nodes.
Thus, how can we design protocols for noisy, complex quantum networks when classical approaches fail? 

Quantum problems often have quantum solutions.
Variational quantum optimization (VQO) \cite{Mitarai2018_quantum_circuit_learning,Moll2018_vqo,Cerezo2021} is a promising tool for quantum network design.
This hybrid algorithm simulates a quantum system on a quantum computer and optimizes it using a classical computer.
Hybrid, quantum-classical algorithms have demonstrated success across a wide range of simulation and optimization problems \cite{Cerezo2021,Yuan2019_vqs, Endo2020_general_vqs}.
% For example, the variational quantum eigensolver \cite{Peruzzo2014_vqe,tilly2021variational_vqe} can find the ground state of a given Hamiltonian.
In the near-term, VQO techniques are predicted to show practical advantages on noisy intermediate-scale quantum (NISQ) devices \cite{Preskill2018}.
Furthermore, VQO can be deployed on devices in quantum networks to optimize communication protocols against the noise inherent to the quantum network hardware.

To demonstrate that VQO can be a design tool for quantum networks, we use it to find the optimal state preparations and measurements for nonlocality in quantum networks.
Nonlocality refers to a phenomenon where entanglement is used to create stronger-than-classical correlations between distant, noncommunicating systems \cite{bell1964epr,brunner2014nonlocality,aspect1981,Giustina2015loop_hole_free,Hensen2015loop_hole_free, Shalm2015_loophole_free}.
These nonlocal correlations are not only a matter of theoretical interest, but can test quantum systems in a black-box manner \cite{brunner2014nonlocality, pironio2010random, gallego2010_dim_witnessing,Renou2018_self-test_network,bancal2018_self-test_network, supic2020, Supic2022_network_self-testing}, and securely perform device-independent cryptography on untrusted devices \cite{ekert1991,Barrett2005,acin2006_di,acin2007_di,Liu2018, Lee2018,vazirani2019diqkd,Luo2022_di_network}.
In recent years, machine learning has successfully solved complex problems that arise in the study of nonlocality \cite{deng2018machine,canabarro2019ml_nonlocality,bharti2019teach}.
Most notably, a hybrid, quantum-classical reinforcement learning algorithm was shown to find the optimal state preparations and measurements that maximize nonlocality \cite{bharti2019teach}.
Furthermore, hybrid optimization techniques have been used to maximize nonlocality in simple photonic systems \cite{Suprano2021,Poderini2022_black-box}.
Our VQO methods have many similarities with these previous approaches, thus, we expect VQO to show similar success.

\begin{figure}[h] 
    \centering
    \includegraphics[width=.48\textwidth]{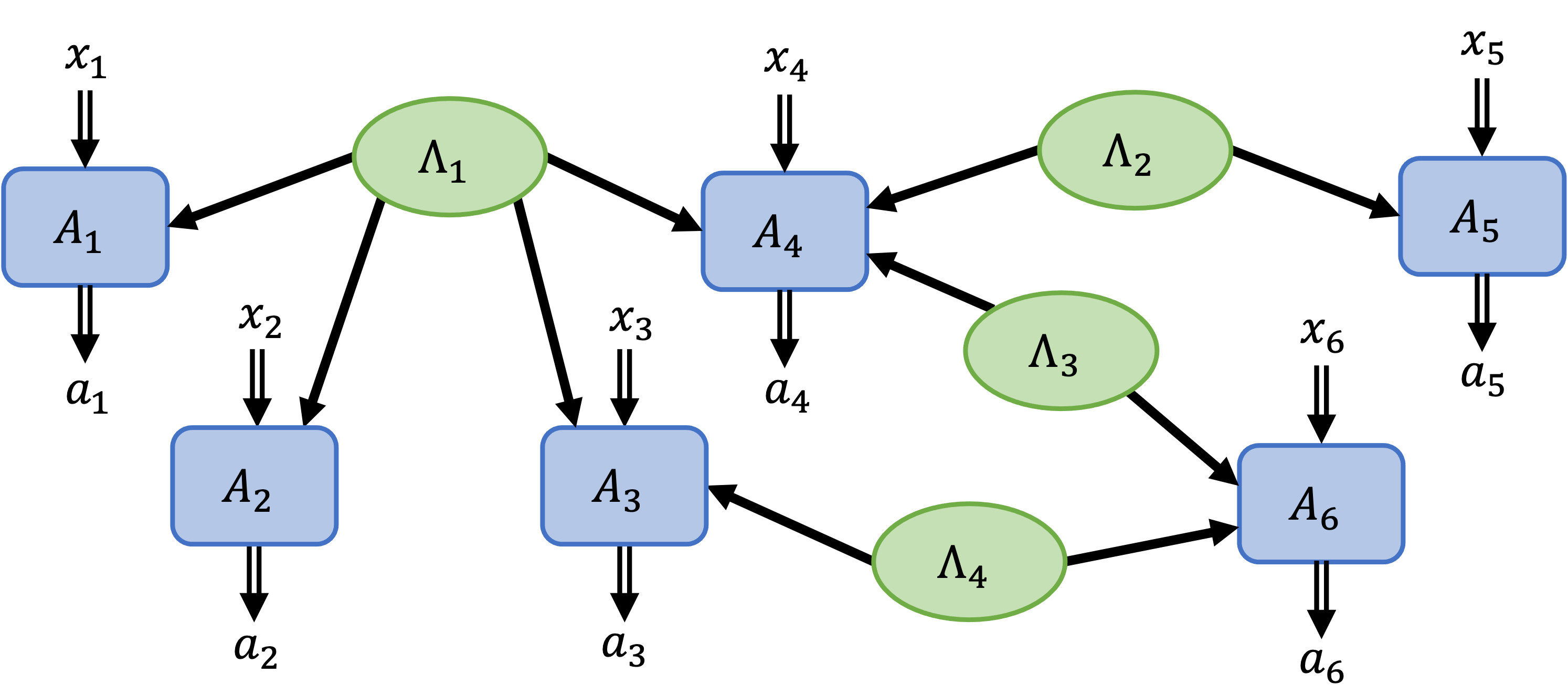}
    \caption{\linespread{1}\selectfont{\small
    \textbf{Example $n$-local network.}
    An $n$-local network can be represented by a directed acyclic graph.
    Sources (green ovals) correlate the nodes (blue rectangles).
    The double-line arrows depict the classical input $x_j$ and output $a_j$ of each node.
    The solid arrows show the links between sources and nodes.
}}
\label{fig:n-local_network_example}
\end{figure}

Nonlocal correlations can be generalized to networks having multiple entanglement sources (see Fig. \ref{fig:n-local_network_example}).
These so-called $n$-local quantum networks resemble quantum repeater chains \cite{Briegel1998_quantum_repeater,sangouard2011_quantum_repeater} and entanglement swapping scenarios \cite{Bennet1993_teleportation_entanglement,Zukowski1993_entanglement_swapping,Bose1998_generalized_entanglement_swapping} that will form the backbone of the quantum internet \cite{Kimble2008quantum_internet,Simon2017global_quantum_network,Wehner2018quantum_internet,kozlowski2019towards}.
Entanglement in $n$-local quantum networks can lead to  non-$n$-local correlations \cite{Branciard_2010_bilocal_correlations,Branciard_2012_bilocal_v_nonbilocal,Tavakoli_2014_star,Mukherjee2015chain,Rosset_2016_nonlinear_bell_inequalities,Tavakoli2016tree,Yang2021_tree_network,tavakoli2021network_nonlocality} that have been experimentally verified \cite{ carvacho2017experimental_bilocal, saunders2017_bilocal_expt, andreoli2017_bilocal_expt, sun2019experimental_bilocal, poderini2020experimental}.
Unfortunately, non-$n$-local correlations deteriorate in the presence of noise \cite{cabello2005_colored_noise,Pal2015,Zhang2020} making them difficult to apply in practice \cite{carvacho2017experimental_bilocal, saunders2017_bilocal_expt, andreoli2017_bilocal_expt, sun2019experimental_bilocal, poderini2020experimental}.
Previous works have studied the effect of white noise on non-$n$-local correlations \cite{Tavakoli_2014_star,Mukherjee2015chain,Rosset_2016_nonlinear_bell_inequalities,Tavakoli2016tree,Yang2021_tree_network}, however, general noise models have only been studied in the simplest cases \cite{Pal2015,Zhang2020}.
To help bridge this gap, we use VQO to maximize non-$n$-locality in noisy quantum networks and investigate the noise robustness of non-$n$-local correlations across a wide range of noise models. 

Hence we show that VQO can provide practical value to the design and development of noisy quantum networks.
In Section \ref{section:n-local_quantum_networks}, we give an overview of $n$-local networks.
In Section \ref{section:vqo_noisy_quantum_networks}, we describe our VQO framework for noisy quantum networks.
In Section \ref{section:quantum_non-n-locality}, we discuss quantum non-$n$-locality and how it can be maximized using VQO.
In Section \ref{section:vqo_quantum_hardware}, we demonstrate VQO of non-$n$-locality on noisy quantum hardware.
In Section \ref{sec:vqo-classical-simulator}, we apply VQO on a classical simulator to investigate the noise robustness of non-$n$-locality.
In summary, we justify our VQO framework for quantum networks by reproducing known results, obtaining new results, and showing its promise of practical advantages when run on quantum hardware.

Our work is accessible, transparent, and reproducible on a laptop computer.
Our VQO software for quantum networks is released as a Python package called qNetVO \cite{qNetVO}, which is built upon PennyLane \cite{pennylane2018}.
All numerics and data are available on GitHub in a supplementary codebase \cite{supp_codebase}.

\section{\textit{n}-Local Networks}\label{section:n-local_quantum_networks}

In this section, we outline the theoretical model of $n$-local quantum networks.
We first introduce $n$-local networks in the classical setting and then, extend this model to the noisy quantum setting.
The experienced reader may proceed to Section \ref{section:vqo_noisy_quantum_networks} where our variational quantum optimization framework is discussed in detail.

\subsection{\textit{n}-Local Classical Networks}

In the classical setting, an $n$-local network consists of $n$ sources $\Lambda_1$, $\dots$, $\Lambda_n$ that distribute randomness to $m$ nonsignaling devices or nodes $A_1, \dots, A_m$ (see Fig. \ref{fig:n-local_network_example}).
The $i^{th}$ source outputs classical value $\lambda_i$ drawn randomly from the distribution $\Omega^{\Lambda_i}$ with probability $P(\lambda_i)$ and sends $\lambda_i$ to all linked devices.
All sources in the network are assumed to be independent such that \begin{equation}\label{eq:source_independence}
    P(\lambdav) = \prod_{i=1}^nP(\lambda_i),
\end{equation}
where $\lambdav=(\lambda_i)_{i=1}^n$ contains the random values output from each independent source.
The sources $\{\Lambda_i\}_{i=1}^n$ each distribute their random value to the nodes $\{A_j\}_{j=1}^m$ through a collection of links $\{L_k\}_{k=1}^l$.
The node $A_j$ may receive multiple random values, thus, we denote with $\lambdav^{A_j}\subseteq\lambdav$ the set of random values received by node $A_j$.

The $j^{th}$ node has classical input and output alphabets $\mc{X}_j:=\{1,\dots,|\mc{X}_j|\}$ and $\mc{A}_j:=\{1,\dots,|\mc{A}_j|\}$ respectively where the input and output alphabets for the entire network are denoted $\mc{X}:=\mc{X}_1\times\dots\times \mc{X}_m$ and $\mc{A}:=\mc{A}_1\times\dots\times \mc{A}_m$ respectively.
Hence the network processes the classical input $\xv\in\mc{X}$ to produce the classical output $\av\in\mc{A}$ where $\xv=(x_j\in\mc{X}_j)_{j=1}^m$ and $\av=(a_j\in\mc{A}_j)_{j=1}^m$.
Since, the nodes are nonsignaling, their joint probability distribution must satisfy
\begin{equation}\label{eq:nonsignaling_constraint}
    P\left(\av\big|\xv,\lambdav\right) = \prod_{j=1}^m P\left(a_j\big|x_j,\lambdav^{A_j}\right),
\end{equation}
for all $\xv\in\mc{X}$, $\av\in\mc{A}$, and  $\lambdav\in\{\Omega^{\Lambda_i}\}_{i=1}^n$.

We characterize $n$-local networks using only their input-output statistics.
Hence we consider a scenario where many identical and independent experiments are performed.
In each experiment, a classical input $\xv\in\mc{X}$ is drawn from a uniform random distribution.
The network processes the input $\xv$ to produce the output $\av\in\mc{A}$.
After many repetitions, an approximate conditional probability distribution $\{P(\av|\xv)\}_{\av\in\mc{A},\xv\in\mc{X}}$ is constructed.
These conditional probabilities fully characterize the network and are represented as a column stochastic matrix referred to as the \textit{network behavior}
\begin{equation}\label{eq:network_behavior}
    \mbf{P}=\sum_{\av\in\mc{A}}\sum_{\xv\in\mc{X}}P(\av|\xv)|\av\rangle\langle\xv|,
\end{equation}
where $\{\ket{\xv}\}_{\xv\in\mc{X}}$ and $\{|\av\rangle\}_{\av\in\mc{A}}$ form classical orthonormal bases over the input and output sets respectively.
The transition probabilities $P(\av|\xv)$ decompose as \cite{Rosset_2016_nonlinear_bell_inequalities},
\begin{align}\label{eq:n-local-condition}
    P(\av|\xv) = \sum_{\lambda_1\in\Omega^{\Lambda_1}}\dots\sum_{\lambda_n\in\Omega^{\Lambda_n}}&\prod_{i=1}^nP(\lambda_i)\times \notag \\
    &\times\prod_{j=1}^m P\left(a_j|x_j,\lambdav^{A_j}\right).
\end{align}

\definition\label{def:n-local_classical_set}
$\LNet$ is the set of network behaviors whose probabilities $P(\av|\xv)$ decompose as Eq. \eqref{eq:n-local-condition} for all inputs $\xv\in\mc{X}$ and outputs $\av\in\mc{A}$ for a given $n$-local network.

\subsection{\textit{n}-Local Quantum Networks}

An $n$-local network is extended to the quantum setting by replacing the shared randomness at each source with quantum entanglement.
The entanglement is distributed amongst the nodes, which subsequently measure their local quantum state to produce a classical output. 

This work models an $n$-local quantum network as an $N$-qubit system where each qubit is indexed by an integer $q_i\in[N]$.
An $M$-qubit subsystem is referenced using the sequence $(q_i)_{i=1}^M\subset[N]$.
Sources $\Lambda_i$, nodes $A_j$, and links $L_k$ are uniquely described by their local qubits, $\Lambda_i, A_j, L_k \subset [N]$.
Qubits cannot be shared between multiple sources, links, or nodes, hence, $\Lambda_i\cap \Lambda_{i'}= \emptyset$ $\forall$ $i\neq i'$ and similarly for $A_j$ and $L_k$.

In the quantum setting, the $n$ sources collectively prepare the state $\ket{\psi^{\Net}} = \bigotimes_{i=1}^n \ket{\psi^{\Lambda_i}}$ where $\ket{\psi^{\Net}}\in \mc{H}^{\Net}_{\Prep}$.
We define $\mc{H}^{\Net}_{\text{Prep}} = \bigotimes_{i=1}^n\mc{H}^{\Lambda_i}_{\Prep}$ as the joint Hilbert space where source independence is ensured by the separability across states.
For mixed states, we denote the density operator  as $\rho^{\Lambda_i}\in D(\mc{H}_{\Prep}^{\Lambda_i})$. 

Quantum channels model the link between source and measurement devices and noise in the network.
A quantum channel is represented by the completely-positive trace-preserving (CPTP) map \cite{Nielsen2009} $\N^{L_k} : D(\mc{H}^{L_k}_{\text{Prep}}) \to D(\mc{H}^{L_k}_{\text{Meas}})$,
where the $L_k$ denotes the qubits on which the channel act.
For convenience, the input and output Hilbert spaces have equal dimension.
A quantum channel is expressed in either the operator-sum  representation \cite{Nielsen2009, Kraus1983}
\begin{equation}\label{eq:kraus_op_rep}
    \N(\rho) = \sum_i K_i \rho K_i^{\dagger}, \; \text{where}\;\sum_i K_i^{\dagger}K_i = \mbb{I},
\end{equation}
where $\{K_i\}_i$ are Kraus operators \cite{Kraus1983},
or the the system-environment representation \cite{Nielsen2009, stinespring1955}
\begin{equation}\label{eq:system-environment_noise}
    \mc{N}(\rho) = \text{Tr}_E[U_{\mc{N}}(\rho\otimes\op{0\dots 0}{0\dots 0}^E)U_{\mc{N}}^\dagger],
\end{equation}
where $E\subseteq[N']$ is an ancillary set of qubits that represent the environment, $\text{Tr}_E[\cdot]$ denotes the partial trace over the environment, and $U_{\mc{N}}$ is a unitary applied to both system and environment.
Independent quantum channels combine to describe the network noise as $\NNet=\bigotimes_{k=1}^l \mc{N}^{L_k}$ where in the noiseless case, $\NNet(\rhoNet)=\text{id}(\rhoNet)=\rhoNet$.
Note that noise can also be applied to the source or measurement devices.

% The nodes receive a .
% If an $M$-qubit source $\Lambda_i = (q_k)_{k=1}^M\subseteq[N]$ prepares the state $\ket{\psi^{\Lambda_i}}$, then we assume that each qubit state $\ket{\psi^{q_k}}$ is sent to a separate node.
% Each node then holds the qubits $A_j = (q_{k'})_{k'=1}^{M'}\subseteq [N]$ which it has received from different entanglement sources.
% Hence device $A_j$ can share entanglement with a number of different nodes.

The measurement at node $A_j$ is modeled using a projection-valued measure (PVM) $\{\Pi^{A_j}_{a_j|x_j}\}_{a_j\in\mc{A}_j}$ that forms a set of orthogonal projectors satisfying $\sum_{a_j\in\mc{A}_j}\Pi^{A_j}_{a_j|x_j} = \mbb{I}^{A_j}$.
The node measures its local qubits $\rho^{A_j}\in D(\mc{H}_{\Meas}^{A_j})$ that were received from its linked sources.
In aggregate, the network applies the projector $\Pi^{\Net}_{\av|\xv} = \bigotimes_{j=1}^m \Pi^{A_j}_{a_j|x_j}$, where the PVM applied at each node is conditioned upon the classical input $x_j\in\mc{X}_k$.
% If the projector is rank-one it can be written as  $\Pi^{A_j}_{a_j|x_j} = \op{\pi^{A_j}_{a_j|x_j}}{\pi^{A_j}_{a_j|x_j}}$.
Upon measurement, the classical output $\av$ is obtained with probability,
\begin{equation}\label{eq:quantum-conditional-probabilities-born-rule}
    P(\av|\xv) = \tr{\PiNet_{\av|\xv}\NNet(\rhoNet)},
\end{equation}
where any permutations needed to map $\mc{H}^{\Net}_{\Prep}$ to $\mc{H}^{\Net}_{\Meas}$ are included implicitly.

The Hermitian observable at the $j^{th}$ node is expressed $O^{A_j}_{x_j} = \sum_{a_j\in\{\pm 1\}} a_j \Pi^{A_j}_{a_j|x_j}$ where $\Pi^{A_j}_{+|x_j}$ and $\Pi^{A_j}_{-|x_j}$ constitute a PVM.
For a fixed state $\rho^{A_j}$, the expectation of the observable is
\begin{equation}
    \ip{O^{A_j}_{x_j}}_{\rho^{A_j}} = \tr{O^{A_j}_{x_j}\rho^{A_j}},
\end{equation}
where the expectation of multiple observables,
\begin{equation}\label{eq:m-partite_correlator}
    \ip{O^{A_1}_{x_1}\dots O^{A_m}_{x_m}}_{\rhoNet} = \prod_{j=1}^m \ip{O^{A_j}_{x_j}}_{\rho^{A_j}},
\end{equation}
is understood as the expected parity of $\av$.

\definition \label{def:n-local_quantum_set} $\QNet$ is the collection of network behaviors $\PNet$ whose probabilities $P(\av|\xv)$ decompose as Eq. \eqref{eq:quantum-conditional-probabilities-born-rule} for some choice of network state preparation $\rhoNet$, noise model $\NNet$, and PVM measurements $\{\{\PiNet_{\av|\xv}\}_{\av\in\mc{A}}\}_{\xv\in\mc{X}}$.

\section{Variational Quantum Optimization of Noisy Quantum Networks}\label{section:vqo_noisy_quantum_networks}

In this section, we develop a variational quantum optimization (VQO) framework for maximizing non-$n$-locality in noisy quantum networks.
At a high-level, our VQO framework simulates the quantum network as quantum circuit and optimizes the circuit parameters using differential programming and gradient descent.
We implement our methods in a Python package called qNetVO: the Quantum Network Variational Optimizer \cite{qNetVO}.
The qNetVO software is built upon PennyLane, a free and open-source quantum differential programming framework \cite{pennylane2018}.
PennyLane enables our VQO framework to be easily run on a wide range of quantum devices and classical simulators.

\subsection{Simulating Noisy Quantum Networks}\label{section:simulating_quantum_networks}

\begin{figure}[t] 
    \centering
    \includegraphics[width=.48\textwidth]{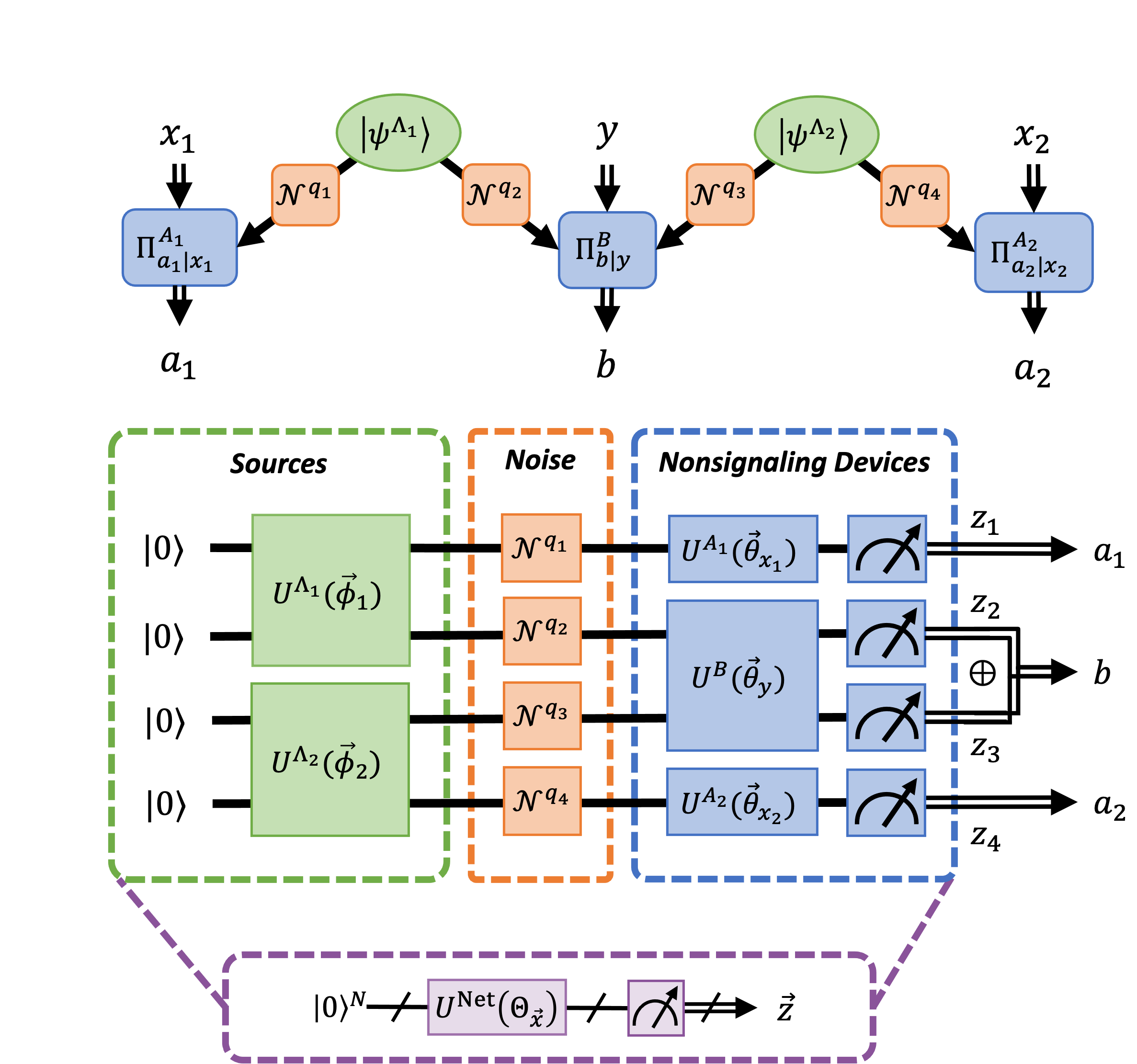}
    \caption{\linespread{1}\selectfont{\small
    \textbf{Noisy network ansatz circuit.}
    (Top) A quantum network with two sources (green), communication noise (orange), and PVM measurements (blue).
    (Bottom) The three-layer quantum circuit ansatz that simulates the network.
    In the source layer, the quantum state $|\psi^{\Lambda_i}\rangle$ is prepared.
    In the noise layer, a static noise model $\mc{N}^{q_k}$ is applied.
    In the measurement layer, the projector $\Pi^{A_j}_{z_j|x_j}$ is a measurement in the computational basis.
    The ansatz circuit is combined into a single unitary $\UNet(\Theta_{\xv})$ and the outcome probabilities are calculated using Eq. \eqref{eq:ansatz_outcome_probabilities}. 
    When a measurement includes multiple qubits, an XOR operation is used to map the raw output to a single bit, \textit{e.g.}, $b= z_2\oplus z_3$.
}}
\label{fig:noisy_bilocal_network_ansatz}
\end{figure}

To simulate a noisy $n$-local quantum network on a quantum computer, the state preparations, measurements, and noise are encoded into the unitary evolution of a quantum circuit.
We define the network-specific ansatz as a unitary operator $\UNet(\Theta_{\xv})$ that is parameterized by a collection of real values $\Theta_{\xv}$ referred to as the circuit settings.
The ansatz circuit parameters are organized as $\Theta_{\xv}=\{\phiv, \thetav_{\xv}\}$ where
\begin{equation}
    \phiv = \left(\phiv_i\right)_{i=1}^n \quad\text{and}\quad  \thetav_{\xv} = \left(\thetav_{x_j}\right)_{j=1}^m,
\end{equation}
parameterize the state preparations and measurements respectively.

The parameterized unitary $\UNet(\Theta_{\xv})$ acts upon the $N$-qubit zero state $|0\rangle^N$ and is measured in the computational basis $\{\ket{\zv}\}_{\zv\in\mc{Z}}$ where $\mc{Z}:=\{0,1\}^N$ is the set of all $N$-bit strings.
When a noiseless quantum computer executes the ansatz circuit, the bitstring $\zv$ is output with probability
\begin{equation}\label{eq:ansatz_outcome_probabilities}
    P(\zv|\xv) = \left\vert \langle\zv|\UNet(\Theta_{\xv})\ket{0}^N \right\vert^2.
\end{equation}
Using Eqs. \eqref{eq:network_behavior} and \eqref{eq:ansatz_outcome_probabilities}, we express the parameterized quantum circuit behavior $\mbf{P}_{\text{QC}}(\Theta)$ as the column stochastic matrix
\begin{equation}\label{eq:quantum_circuit_behavior}
    \mbf{P}_{\text{QC}}(\Theta) = \sum_{\zv\in\mc{Z}}\sum_{\xv\in\mc{X}} \left\vert \bra{\zv}\UNet(\Theta_{\xv})\ket{0}^N \right\vert^2 |\zv\rangle\langle\xv|.
\end{equation}
The network settings $\Theta$ contain the settings for all inputs $\xv\in\mc{X}$ and are organized as
\begin{equation}\label{eq:network_settings}
    \Theta = \left\{\phiv\;, \quad\left\{\{\thetav_{x_j}\}_{x_j\in\mc{X}_j}\right\}_{j=1}^m \right\},
\end{equation}
where the network settings $\Theta$ are distinct from the circuit settings $\Theta_{\xv}$ because $\Theta_{\xv}$ contains only the settings pertaining to input $\xv\in\mc{X}$.

If a quantum network's output does not conform to an $N$-bit string, then a classical postprocessing map $\mbf{L} : \mc{Z}\to\mc{A}$ is needed to obtain the simulated network behavior $\mbf{P}$ from the quantum circuit behavior $\mbf{P}_{\text{QC}}$.
That is,
\begin{equation}
    \mbf{P}(\Theta) = \mbf{L}\;\mbf{P}_{\text{QC}}(\Theta),
\end{equation}
where $\mbf{P}_{\text{QC}}(\Theta)$ is defined in Eq. \eqref{eq:quantum_circuit_behavior} and the postprocessing map is represented as a column stochastic matrix,
\begin{equation}
    \mbf{L} = \sum_{\av\in\mc{A}}\sum_{\zv\in\mc{Z}}P(\av|\zv)|\av\rangle\langle\zv|.
\end{equation}
On a quantum computer, the network behavior $\mbf{P}(\Theta)$ is obtained by repeatedly executing the ansatz circuit  $\UNet(\Theta_{\xv})$ across all inputs $\xv\in\mc{X}$ to estimate the probabilities $P(\zv|\xv)$ for all inputs.

\subsubsection{Simulating Noiseless \textit{n}-Local Quantum Networks}

A noiseless $n$-local quantum network ansatz decomposes into preparation and measurement layers,
\begin{equation}
    \UNet(\Theta_{\xv}) = \UNet_{\text{Meas}}(\thetav_{\xv})\UNet_{\text{Prep}}(\phiv),
\end{equation}
where each layer is modularized as
\begin{align}
    &\UNet_{\text{Prep}}(\phiv) = \bigotimes_{i=1}^n U^{\Lambda_i}(\phiv_i) \quad \text{and} \label{eq:prep_ansatz_decomposition} \\
    &\UNet_{\text{Meas}}(\thetav_{\xv}) = \bigotimes_{j=1}^m U^{A_j}(\thetav_{x_j}). \label{eq:meas_ansatz_decomposition}
\end{align}
The network state preparation and measurements are expressed as
\begin{align}
    &\ket{\psiNet}= \UNet_{\text{Prep}}(\phiv)\ket{0}^N, \label{eq:parameterized_state_preparation}\\
    \Pi^{\Net}_{\zv|\xv}=& \left(\UNet_{\text{Meas}}(\thetav_{\xv})\right)^\dagger\op{\zv}{\zv}\UNet_{\text{Meas}}(\thetav_{\xv}). \label{eq:parameterized_measurement}
\end{align}
Combining Eqs. \eqref{eq:parameterized_state_preparation} and \eqref{eq:parameterized_measurement} with Eq. \eqref{eq:ansatz_outcome_probabilities} yields,
\begin{equation}\label{eq:noiseless_ansatz_probabilities}
    P(\zv|\yv) = \ip{\psiNet\Big|\Pi^{\Net}_{\zv|\xv}\Big|\psiNet}.
\end{equation}

\subsubsection{Simulating Noisy \textit{n}-Local Quantum Networks}\label{section:simulating_noisy_n-local_quantum_networks}

On quantum hardware, noise is modeled using a unitary operator $U_{\N}$ to implement the system-environment representation of a quantum channel expressed in Eq. \eqref{eq:system-environment_noise}.
Ancilla qubits are required to implement nonunitary dynamics. In the worst case, the Hilbert space dimension of the ancilla space is $d_E = d_S^2$ where $d_S$ is the dimension of the system the quantum channel operates upon \cite{stinespring1955}.

If the quantum circuit is run on a classical simulator such as the PennyLane \texttt{default.mixed} mixed state simulator \cite{pennylane2018}, then ancillary qubits are not needed and the operator-sum representation expressed in Eq. \eqref{eq:kraus_op_rep} can be used instead.
In practice, using a classical simulator to model noisy dynamics is a good idea because the inherent noise of the quantum hardware is not included.
Furthermore, the Kraus operators give precise control over the action of quantum channel description that might not easily be implemented by the unitary gates available on a quantum computer.

Detector errors can be modeled using a classical postprocessing map $\mbf{E}$.
That is, a noiseless network behavior $\PNet$ can have noise added as
\begin{equation}
    \PNet' = \mbf{E}^{\Net}\PNet,
\end{equation}
where $\mbf{E}$ is a column stochastic matrix.
This model of detector errors describes unflagged errors that occur without the experimenter's knowledge.
Hence the errors cannot be corrected or removed by postselection.

\subsection{Optimizing Quantum Networks}

The goal of our VQO framework is to find the optimal settings $\Theta^{\star}$ that yield a network behavior $\PNet(\Theta^{\star})$ that is optimal for a particular task, \textit{e.g.}, violating a Bell inequality.
We define a problem-specific cost function $\Cost(\PNet(\Theta))$ that quantifies the network's performance at this task.
The optimization objective is then expressed as a minimization of the cost function,
\begin{equation}\label{eq:arg-min-cost}
    \Theta^{\star} = \arg\min_{\Theta}\Cost(\PNet(\Theta)).
\end{equation}
The cost function can quantify a wide range of network properties such as entropic quantities, the distance to a desired network behavior, or the winning probability of a multipartite game.

The optimization problem in Eq. \eqref{eq:arg-min-cost} is solved using gradient descent \cite{ruder2016gradient_descent,Bottou2018gradient_descent}
\begin{equation}\label{eq:gradient-descent}
    \Theta' = \Theta - \eta\nabla_{\Theta}\Cost(\PNet(\Theta)),
\end{equation}
where $\eta\in\mbb{R}$ is the step size and $\nabla_{\Theta}\Cost(\Theta)$ is the gradient of the cost function evaluated at $\Theta$.
This iterative optimization procedure incrementally traverses the path of steepest descent to find a local minimum of the cost function where in each step, the network settings are updated as $\Theta\to\Theta'$.
The gradient $\nabla_{\Theta}\Cost(\PNet(\Theta))$ is evaluated numerically using automatic differentation \cite{griewank1989autodiff,Baydin2018autodiff}.
On a classical simulator, the gradient of a quantum circuit can be evaluated  using a chain rule method known as \textit{back propagation} \cite{Baydin2018autodiff}.
On quantum hardware, the gradient of a quantum circuit is evaluated using the parameter-shift rule \cite{Schuld2019_parameter_shift}.
In practice, the parameter-shift rule first runs the network simulation over a collection of ``shifted" settings $\{\Theta^{g}\}_g = \hat{\Theta}_{\text{PS}}$ and then, the classical optimizer uses the resulting circuit behaviors $\{\mbf{P}_{\text{QC}}(\Theta^g)\}_{\Theta^g\in\hat{\Theta}_{\text{PS}}}$ to construct the gradient.

\subsection{Variational Quantum Optimization Algorithm}

\begin{figure}[t] 
    \centering
    \includegraphics[width=.48\textwidth]{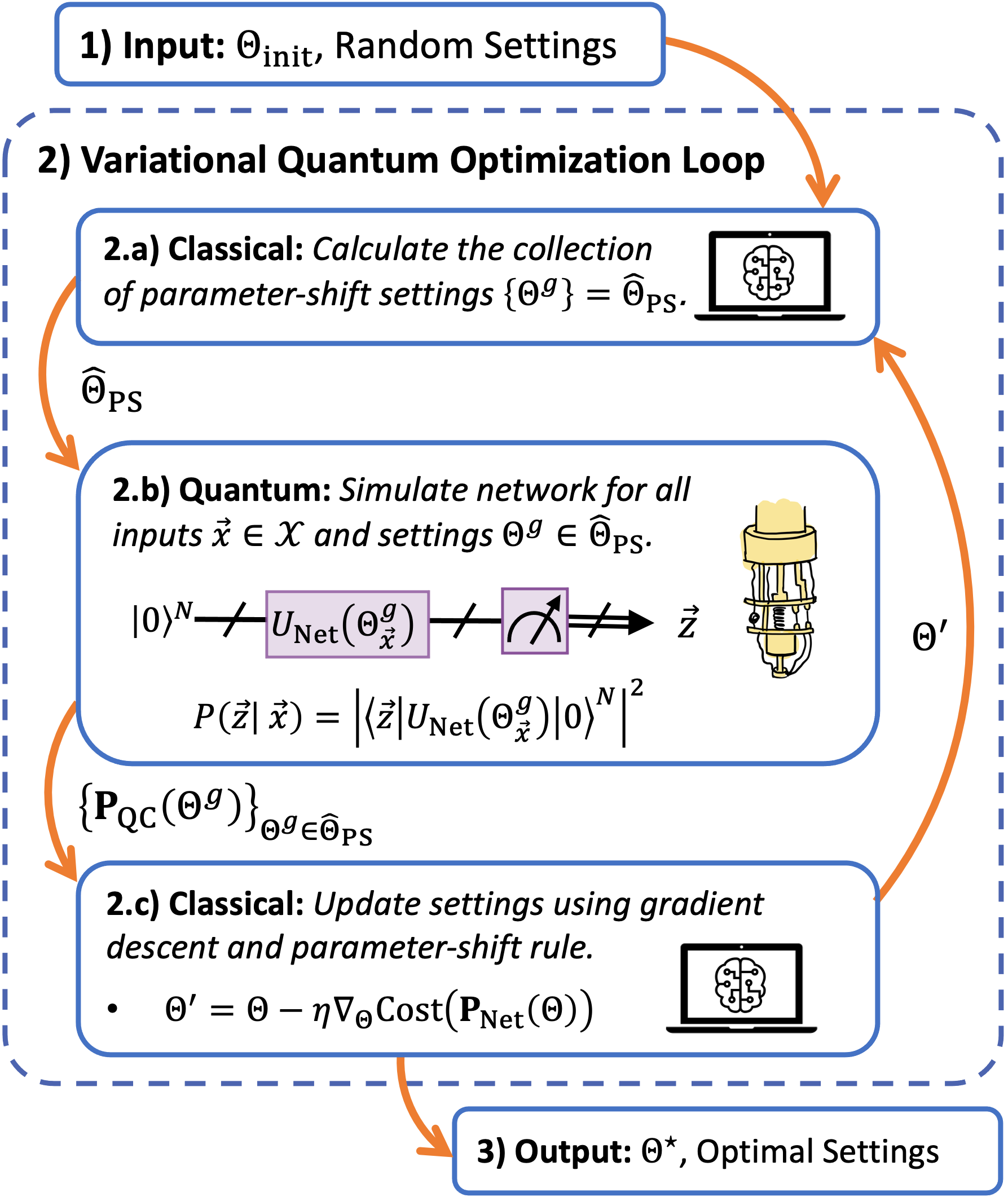}
    \caption{\linespread{1}\selectfont{\small
    \textbf{Variational Quantum Optimization.}
    In this hybrid algorithm, the optimal settings $\Theta^\star$ that minimize the $\Cost(\PNet(\Theta))$ are computed.
    Steps 2.a) and 2.c) are performed on a classical computer while step 2.b) is performed on a quantum computer. 
}}
\label{fig:vqo_diagram}
\end{figure}

Having established the key components of our VQO framework, we now explicitly describe our hybrid optimization algorithm implemented by the qNetVO software \cite{qNetVO}.
The algorithm requires a parameterized network ansatz $\UNet(\Theta_{\xv})$ and cost function $\Cost(\PNet(\Theta))$.
As hyperparameters, the algorithm accepts the step size $\eta$ and the number of gradient descent iterations \texttt{num\_steps}.
The following steps depicted in Fig. \ref{fig:vqo_diagram}.

\begin{enumerate}
    \item The network settings $\Theta_{\text{init}}$ are randomly initialized.
    \item The hybrid VQO loop repeats \texttt{num\_steps} times. In each step the following actions are performed:
    \begin{enumerate}
        \item The classical optimizer constructs the collection of settings $\{\Theta^g\}_g=\hat{\Theta}_{\text{PS}}$ needed for the parameter-shift rule.
        \item The quantum computer evaluates $\mbf{P}_{\text{QC}}(\Theta^g)$ for all settings in $\Theta^g \in \hat{\Theta}_{\text{PS}}$.
        \item The classical optimizer evaluates the gradient $\nabla_{\Theta} \Cost(\PNet(\Theta))$ using the parameter-shift rule and updates the network settings using Eq.~\eqref{eq:gradient-descent} to perform gradient descent.
    \end{enumerate}
    \item The optimization algorithm exits after performing \texttt{num\_steps} of gradient descent iterations. As output, the algorithm provides the collection of settings evaluated in each step, and the cost $\Cost(\PNet(\Theta))$ achieved by those settings.
\end{enumerate}

The VQO algorithm is not guaranteed to find the global optimum, however, the probability of finding the global optimum is increased by repeating the optimization with randomly initialized settings $\Theta_{\text{init}}$.
Whether the global optimum can be found is highly dependent on the cost function and network ansatz.
Furthermore, the algorithm provides a upper bound on the true minimum of the cost function. 
That is, it does not find the exact settings $\Theta^\star$ that minimize the cost, but settings that are close to optimal, \textit{i.e.}, $\Theta' = \Theta^\star \pm \epsilon$ for some small value $\epsilon$.
The precision of the optimization can be adjusted using the step size $\eta$.
In practice, qNetVO \cite{qNetVO} provides convenient descriptions of ansatz, network settings, and cost functions, while PennyLane \cite{pennylane2018} provides the machinery for automatic differentiation and the execution of circuits on classical and quantum hardware.

\section{Quantum Non-\textit{n}-Locality}\label{section:quantum_non-n-locality}

In this section, we introduce the concept of quantum non-$n$-locality.
We first discuss Bell inequalities and how they can be used to witness quantum non-$n$-locality. 
Next, we introduce the considered $n$-local networks, their Bell inequalities, and their maximal quantum violations in the noiseless case.
Finally, we describe how our variational quantum optimization methods can be used to maximize quantum non-$n$-locality.

\subsection{Witnessing Quantum  Non-\textit{n}-Locality}

In the noiseless setting, quantum entanglement allows for stronger-than-classical correlations to form across distant network nodes.
That is, the set of classical network behaviors $\LNet$  from Definition \ref{def:n-local_classical_set} and the set of quantum network behaviors $\QNet$ from Definition \ref{def:n-local_quantum_set} relate as $\LNet\subseteq\QNet$ when both network have the same topology \cite{brunner2014nonlocality, tavakoli2021network_nonlocality}.
Quantum non-$n$-locality is then defined as follows.

\begin{definition}
    A quantum network behavior $\mbf{P}\in\QNet$ is \textit{non-$n$-local} if and only if $\mbf{P}\notin\LNet$ where both classical and quantum $n$-local networks have the same topology. 
\end{definition}

To decide whether a network behavior $\PNet$ is non-$n$-local, we must characterize the geometry of the set of classical network behaviors $\LNet$.
In general, $\LNet$ is closed, connected, and contains a finite set of vertices $\mc{V}$ corresponding to deterministic behaviors \cite{tavakoli2021network_nonlocality}
\begin{equation}
    \mc{V}=\left\{\mbf{P}\in\LNet\;|\;P(\av|\xv)\in\{0,1\}\;\forall\;\av\in\mc{A},\;\xv\in\mc{X} \right\}.
\end{equation}
Deterministic network behaviors do not require shared randomness, hence, the set of vertices $\mc{V}$ is determined solely by the network nodes $\{A_j\}_{j=1}^m$.
As a result, if we denote $\LNet^\star$ as the local ($n=1$) set having a single source and network nodes $\{A_j\}_{j=1}^m$, then $\LNet^\star = \conv{\LNet}$ \cite{brunner2014nonlocality, tavakoli2021network_nonlocality}.
In general, $\LNet\subseteq\LNet^\star$ where equality only occurs in trivial cases, \textit{e.g.}, the network contains a single measurement device. 
It follows that $\LNet$ is nonconvex for all $\LNet\subset\LNet^\star$ \cite{tavakoli2021network_nonlocality}.

The set of $n$-local network behaviors $\LNet$ is bounded by inequalities referred to as Bell inequalities \cite{brunner2014nonlocality,tavakoli2021network_nonlocality}
\begin{equation}\label{eq:bell_inequality}
    S_{\Bell}(\mbf{P})\leq \beta,
\end{equation}
where $S_{\Bell}(\cdot)$ is a function referred to as the \textit{Bell score} and $\beta$ is the classical upper bound.
All behaviors $\mbf{P}\in\LNet$ must satisfy the Bell inequalities that bound $\LNet$.
For the local ($n=1$) case, the Bell inequalities bounding $\LNet^\star$ are linear half-space inequalities \cite{brunner2014nonlocality}.
In general, the Bell score $S_{\Bell}(\mbf{P})$ is a nonlinear function of the probabilities $P(\av|\xv)$ \cite{Fritz_2012,Branciard_2010_bilocal_correlations,Branciard_2012_bilocal_v_nonbilocal,Tavakoli_2014_star,Mukherjee2015chain,chaves_2016_polynomial,Tavakoli2016tree,Rosset_2016_nonlinear_bell_inequalities, Rosset_2016_nonlinear_bell_inequalities, tavakoli2021network_nonlocality, Yang2021_tree_network}.

The Bell inequalities bounding $\LNet$ can be used to witness quantum non-$n$-locality.
That is, if $S_{\text{Bell}}(\mbf{P})\not\leq \beta$, then $\mbf{P}\notin\LNet$.
In this case, the Bell inequality is violated and the behavior $\mbf{P}$ is witnessed to be \textit{non-n-local}.
Therefore, the study of quantum non-$n$-locality reduces to finding $n$-local quantum behaviors $\mbf{P}\in\QNet$ that violate a particular Bell inequality.
Fortunately, many Bell inequalities have been derived that tightly bound important $n$-local networks including star \cite{Tavakoli_2014_star}, chain \cite{Mukherjee2015chain}, and tree \cite{Rosset_2016_nonlinear_bell_inequalities, Tavakoli2016tree,Yang2021_tree_network} topologies.
Thus, non-$n$-locality can be studied broadly without deriving new $n$-local bounds.

\begin{figure*}[t] 
    \centering
    \includegraphics[width=0.8\textwidth]{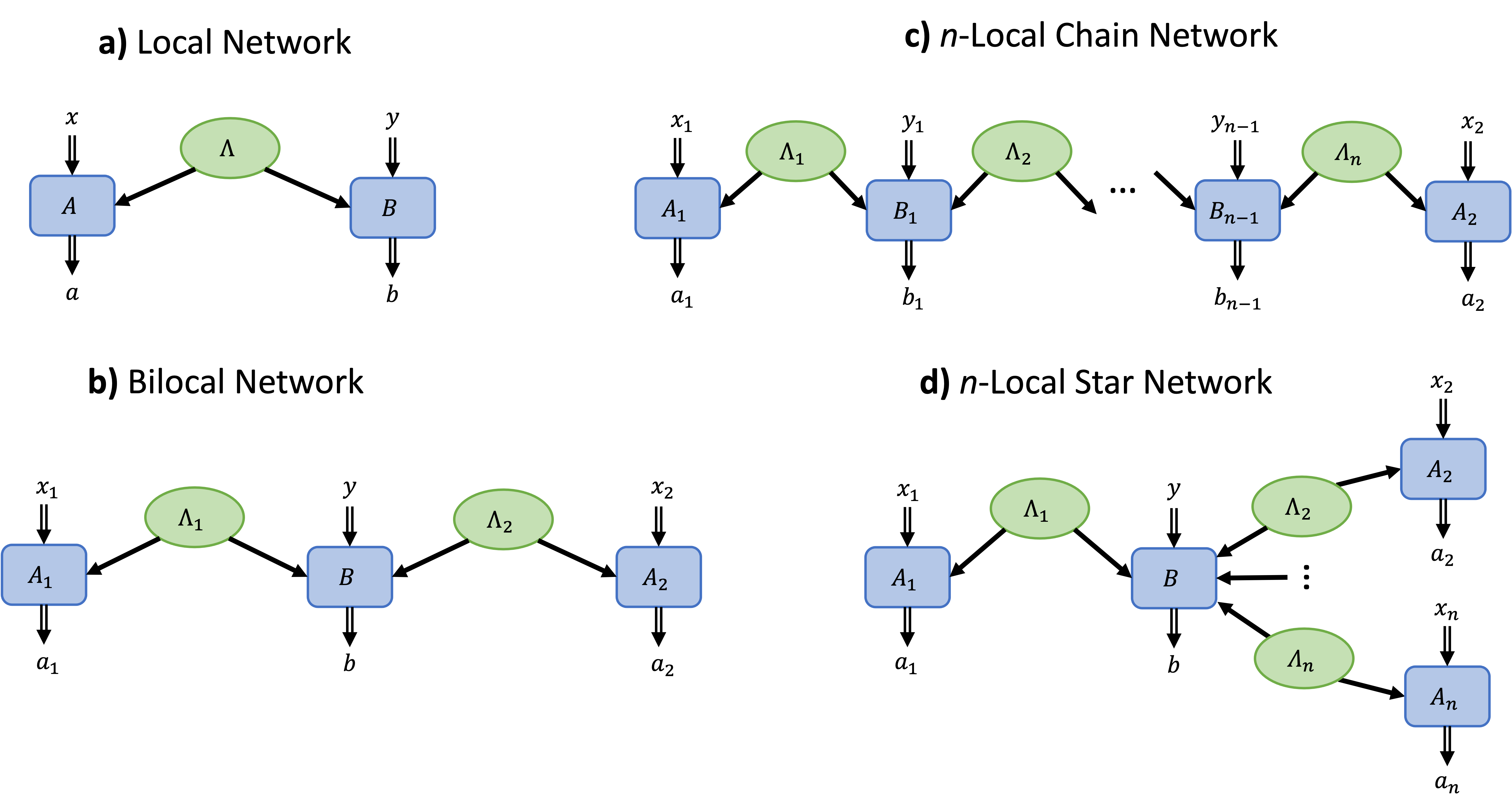}
    \caption{\linespread{1}\selectfont{\small
        \textbf{$n$-Local Networks.}
        This work focuses on four important $n$-local networks, a) local network, b) bilocal network, c) $n$-local chain, and d) $n$-local star.
        Sources are depicted as green ovals and nodes are blue rectangles.
        The labels $A_j$ and $B_j$ distinguish between exterior and interior nodes respectively.
    }}
    \label{fig:n-local_networks}
\end{figure*}
\subsection{Bell Inequalities for \textit{n}-Local Networks}\label{section:bell_inequalities_for_n-local_networks}

This work considers general $n$-local star \cite{Tavakoli_2014_star} and chain \cite{Mukherjee2015chain} networks where each network node has binary inputs $x_j,y_j\in\{0,1\}$ and outputs $a_j, b_j\in\{\pm1\}$ (see Fig.~\ref{fig:n-local_networks}).
Star networks are important because they can implement generalized entanglement swapping protocols \cite{Bennet1993_teleportation_entanglement,Zukowski1993_entanglement_swapping,Bose1998_generalized_entanglement_swapping, Tavakoli_2014_star} where the exterior nodes in the star can become arbitrarily entangled based upon the measurement applied in the central node.
Similarly, chain networks are important because they model quantum repeater chains that enable long-distance quantum communication via relay nodes that perform entanglement swapping \cite{Briegel1998_quantum_repeater,sangouard2011_quantum_repeater}.
In both cases, non-$n$-locality is important for device-independent certification of the entanglement sources and quantum measurements required to implement entanglement swapping protocols and network security protocols \cite{vazirani2019diqkd,Liu2018,Lee2018_network_di, Luo2022_di_network}.

For the $n$-local networks depicted in Fig.~\ref{fig:n-local_networks}, we now introduce the Bell inequalities and their quantum violations.
In all cases, the optimal noiseless strategies are achieved using maximally entangled state preparations $\ket{\psi^{\Lambda_i}} = \ket{\Phi^+} = (\ket{00}+\ket{11})/\sqrt{2}$ and linear combinations of the Pauli observables $\sigma_x$ and $\sigma_z$.

The $n$-local star network depicted in Fig. \ref{fig:n-local_networks}.d consists of $n$, two-qubit sources $\{\Lambda_i=(q_i, q_{n+i})\}_{i=1}^n$ and $m=n+1$ measurement nodes arranged in a star formation with $n$ nodes $\{A_j=(q_j)\}_{j=1}^n$ serving as points of the star connected to a single central node $B=(q_{n+j})_{j=1}^n$.
The $n$-local star network Bell inequality is defined as \cite{Tavakoli_2014_star}
\begin{equation}\label{eq:n-local_star_inequality}
    S_{n\text{-Star}} := |I_{n,0}|^{1/n} + |I_{n,1}|^{1/n} \leq 1,
\end{equation}
where
\begin{equation}\label{eq:I_correlator_combo}
    I_{n,y} = \frac{1}{2^n}\sum_{x_1,\dots,x_n}(-1)^{y(\sum_j x_j)}\langle O^{A_1}_{x_1}\dots O^{A_n}_{x_n} O^B_{y} \rangle.
\end{equation}
In Eq.~\eqref{eq:I_correlator_combo}, $O^{A_j}_{x_j}$ and $O^B_{y}$ are dichotomic observables and the $m$-partite correlator is evaluated using Eq.~\eqref{eq:m-partite_correlator}.
The maximal quantum violation of the star inequality is  $S_{n-\text{Star}}^\star = \sqrt{2} \not\leq 1$, where the exterior nodes each apply the observable $O^{A_j}_{x_j} = (\sigma_z + (-1)^{x_j}\sigma_x)/\sqrt{2}$ and the interior node applies the observable $O^B_y = (1-y)\bigotimes_{i=1}^n \sigma_z + y\bigotimes_{i=1}^n \sigma_x$ \cite{Tavakoli_2014_star}.

In the $n=1$ case, the star network reduces to two measurement devices depicted in Fig.~\ref{fig:n-local_networks}.a.
This simple two-qubit network is the fundamental example of nonlocality \cite{brunner2014nonlocality} and is bounded by the CHSH inequality \cite{chsh-inequality1969}
\begin{equation}\label{eq:chsh-inequality}
    S_{\text{CHSH}} := \left \vert\sum_{x,y=0}^1 (-1)^{x y}\ip{O_x^A O^B_y}\right \vert \leq 2.
\end{equation}
Rearranging the CHSH inequality in Eq.~\eqref{eq:chsh-inequality} we find the $n=1$ case for the star inequality in Eq.~\eqref{eq:n-local_star_inequality}
\begin{align}\label{eq:1-star-inequality}
    \frac{1}{2}\Big|\ip{O^A_0O^B_0}&+\ip{O^A_1O^B_0}\Big| + \notag \\
    &+ \frac{1}{2}\Big|\ip{O^A_0O^B_1} - \ip{O^A_1O^B_1}\Big| \leq 1.
\end{align}
The maximal violation of Eq.~\eqref{eq:chsh-inequality} is $S^\star_{\text{CHSH}} = 2\sqrt{2} \not\leq 2$ and is achieved by the source preparation $\ket{\psi^{\Lambda}}=\ket{\Phi^+}$ measured using the observables $O^A_x = (1-x)\sigma_z + x\sigma_x$ and $O^B_y=(\sigma_z + (-1)^y \sigma_x)/\sqrt{2}$ \cite{Cirelson1980}.

In the $n=2$ case, the star network reduces to the bilocal network depicted in Fig.~\ref{fig:n-local_networks}.b.
This network contains two sources $\Lambda_1=(q_1,q_2)$ and $\Lambda_2=(q_3,q_4)$.
The two exterior nodes $A_1=(q_1)$ and $A_2=(q_4)$ hold the first and last qubit while the central holds the qubits $B=(q_2,q_3)$.
The bilocal network is bounded by the Bell inequality \cite{Branciard_2010_bilocal_correlations,Branciard_2012_bilocal_v_nonbilocal}
\begin{equation}\label{eq:bilocal_inequality}
    S_{\text{Biloc}} := \sqrt{|I_{2,y=0}|} + \sqrt{|I_{2,y=1}|} \leq 1,
\end{equation}
which is exactly the $n=2$ case of $n$-local star Bell inequality in Eq. \eqref{eq:n-local_star_inequality}.
The maximal violation is $S_{\text{Biloc}}^{\star} = \sqrt{2} \not\leq 1$ and is achieved using the observables $O^{A_j}_{x_j} = (\sigma_z + (-1)^{x_j}\sigma_x)/\sqrt{2}$ and $O^{B}_y = (1-y)\sigma_z\otimes\sigma_z + y\sigma_x\otimes\sigma_x$ \cite{Branciard_2012_bilocal_v_nonbilocal}.

The $n$-local chain network is an important extension of the bilocal network.
In this network, the $n$ sources $\{\Lambda_i = (q_{2i-1},q_{2i})\}_{i=1}^{n}$ connect the $m=n+1$ nonsignaling measurement devices in a chain depicted in Fig.~\ref{fig:n-local_networks}.c.
The exterior nodes $A_1 = (q_1)$ and $A_2 = (q_{2n})$ hold the first and last qubits while the interior nodes each hold two qubits $\{B_j = (q_{2j-2},q_{2j-1}) \}_{j=2}^n$.
The set of $n$-local chain behaviors is bounded by the inequality \cite{Mukherjee2015chain},
\begin{equation} \label{eq:n-local-chain-bell-inequality}
    S_{n\text{-Chain}}:=\sqrt{|I_{2,y=0}|} + \sqrt{|I_{2,y=1}|} \leq 1,
\end{equation}
where $S_{n\text{-Chain}}$ is very similar to the bilocal inequality described by Eq. \eqref{eq:bilocal_inequality}.
The distinction arises in the central observable $O^{B}_y$, which for the $n$-local chain, decomposes as
\begin{equation}
    O^B_y = \bigotimes_{j=2}^n O^{B_j}_{y_j=y},
\end{equation}
where $O^{B_j}_{y_j=y}$ is the two-qubit observable applied at each interior node in the chain.
In Eq.~\eqref{eq:n-local-chain-bell-inequality} the collection of interior measurement devices $\{B_j\}_{j=2}^n$ is treated as a single observable $O^B_y$ conditioned on a single input $y\in\{0,1\}$.
The optimal quantum violation of Eq.~\eqref{eq:n-local-chain-bell-inequality} is $S_{n\text{-Chain}}^{\star} = \sqrt{2} \not\leq 1$ using Bell state preparations and the observables $O^{B_j}_{y_j} = (1-y_j)\sigma_z\otimes\sigma_z + y_j\sigma_x\otimes\sigma_x$ and $O^{A_j}_{x_j} = (\sigma_z + (-1)^{x_j}\sigma_x)/\sqrt{2}$ \cite{Mukherjee2015chain}.

\subsection{VQO of Quantum Non-\textit{n}-Locality} \label{section:max_nonlocality_quantum_networks}

To summarize, non-$n$-locality is witnessed by the violation of a Bell inequality $S_{\text{Bell}}(\cdot)\leq \beta$ that bounds the $n$-local set $\LNet$.
Thus, our objective is to find
\begin{equation}\label{eq:non-n-locality_maximization}
    S_{\text{Bell}}^{\star} = \max_{\mbf{P}\in \QNet} S_{\text{Bell}}(\mbf{P}),
\end{equation}
where $S_{\text{Bell}}^{\star}$ is the maximal Bell score that can be achieved by $n$-local quantum network and the maximization is performed over all network behaviors in the $n$-local quantum set $\QNet$ (see Definition \ref{def:n-local_quantum_set}).
This maximization requires us to find the optimal state preparations and measurements for maximizing non-$n$-locality given a static network noise model $\NNet$.

The maximization of non-$n$-locality expressed in Eq.~\eqref{eq:non-n-locality_maximization} is well suited for our VQO framework.
The $n$-local network ansatz is constructed similarly to Fig.~\ref{fig:noisy_bilocal_network_ansatz} where the state preparations and measurements are modeld by  Eq.~\eqref{eq:prep_ansatz_decomposition} and Eq.~\eqref{eq:meas_ansatz_decomposition} respectively.
Furthermore, the network noise $\NNet$ can be modeled as described in Section~\ref{section:simulating_noisy_n-local_quantum_networks}.
The cost function is simply expressed as
\begin{equation}
    \Cost(\mbf{P}(\Theta)) = - S_{\text{Bell}}(\mbf{P}(\Theta)),
\end{equation}
where the minus sign transforms the minimization of the cost into the maximization of the Bell score.
It is then a simple matter of using software such as qNetVO \cite{qNetVO} to perform the algorithm shown in Fig.~\ref{fig:vqo_diagram}.

\section{VQO of Non-\textit{n}-Locality on Quantum Hardware}\label{section:vqo_quantum_hardware}

\begin{figure}[b] 
    \centering
    \includegraphics[width=.4\textwidth]{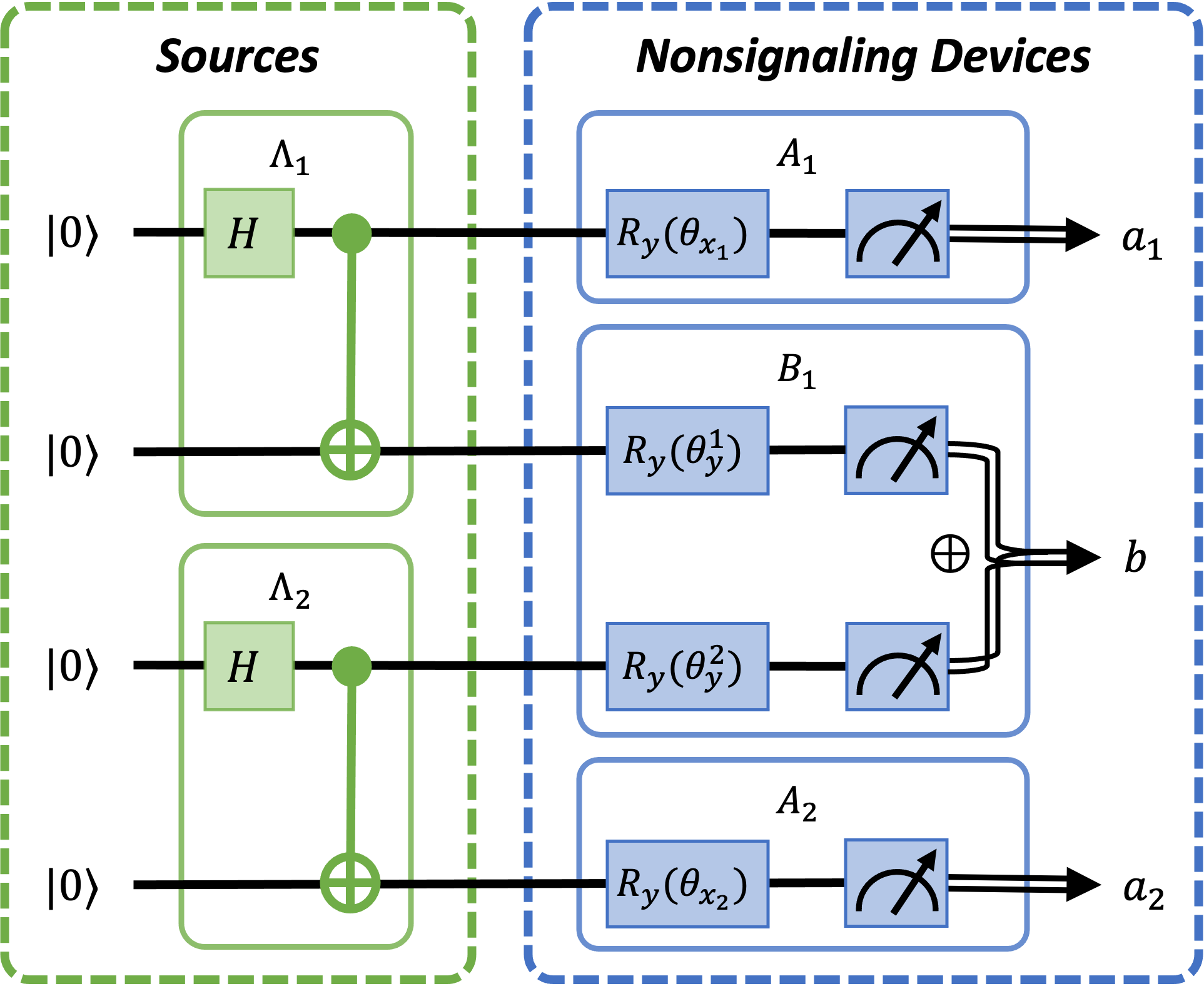}
    \caption{\linespread{1}\selectfont{\small
        \textbf{VQO hardware ansatz for bilocal network.}
        When applying VQO on hardware, we consider a simple ansatz.
        Each source (green) prepares the Bell state $\ket{\Phi^+}$ using a Hadamard and CNOT gate.
        Each nonsignaling device (blue) applies a local rotation about the $y$-axis to each qubit before measurement in the computational basis.
        When a measurement device contains more than one local qubit, the XOR is taken to convert the bit string into a binary output. 
    }
}
\label{fig:vqo-hardware-ansatz}
\end{figure}

In this section, we show that our VQO framework can be run on quantum hardware and discuss how our optimization methods show promise in demonstrating practical advantages.
We first demonstrate that our VQO techniques can successfully maximize non-$n$-locality on IBM quantum hardware.
Then, we discuss how our methods can be scaled to demonstrate practical advantages on quantum computers.
Finally, we discuss the advantages that might be gained by deploying our optimization methods directly on quantum network hardware.

\subsection{VQO of Non-\textit{n}-Locality on Noisy IBM Quantum Computers}

\begin{figure*}[t] 
    \centering
    \includegraphics[width=.8\textwidth]{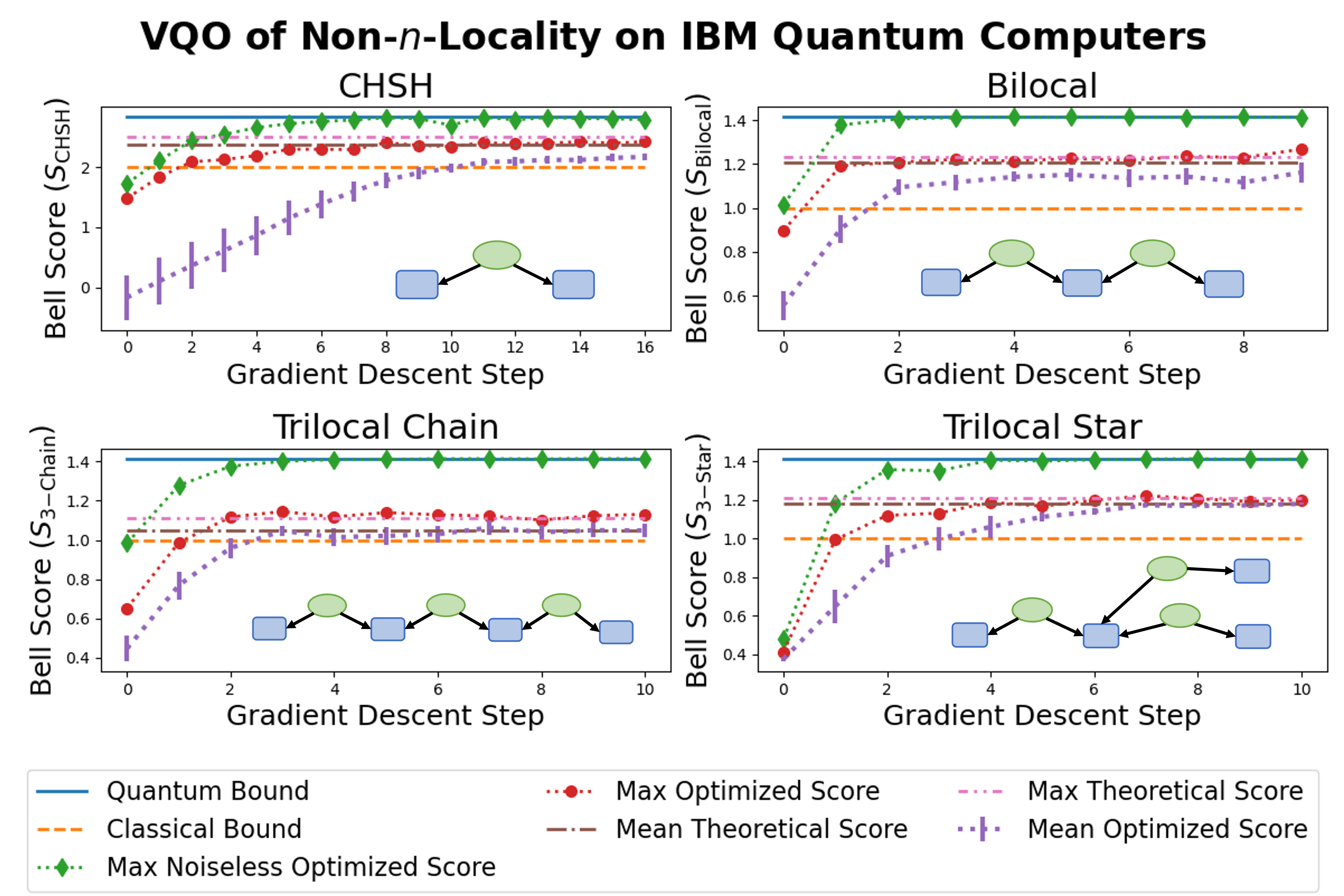}
    \caption{\linespread{1}\selectfont{\small
        \textbf{VQO of Non-$n$-Locality on IBM Quantum Computers.}
        Non-$n$-locality is optimized in four different $n$-local networks.
        The $x$-axis shows the step of the gradient descent optimization while the $y$-axis shows the Bell score.
        The noiseless quantum bound is shown by the solid blue line and the classical bound is shown by the dashed orange line.
        For each Bell inequality, we aggregate data across several different optimization runs.
        At the end of each optimization, the theoretically optimal score is evaluated on noisy hardware to serve as a baseline.
        The mean theoretical score is shown by the dash-dotted brown line and
        the max theoretical score is shown by the dash-dot-dotted pink line.
        In each step, the max score across all optimization is shown by the dotted red line with circle markers while the mean score across all optimizations is shown by the dotted purple line with error bars showing the standard error.
        The dotted green line with diamond markers shows the noiseless optimal score obtained by running the settings for the maximal score in each step on a noiseless classical simulator.
    }
}
\label{fig:vqo-hardware-plots}
\end{figure*}

Using IBM quantum computers, we use the variational quantum optimization algorithm depicted in Fig.~\ref{fig:vqo_diagram} to maximize the non-$n$-locality in quantum networks using Eq.~\eqref{eq:non-n-locality_maximization}.
We use VQO to maximize the CHSH score $S_{\text{CHSH}}$, the bilocal score $S_{\text{Bilocal}}$, the trilocal chain score $S_{3\text{-Chain}}$, and the trilocal star score $S_{3\text{-Star}}$.
Each optimization uses the same preparation and measurement ansatzes depicted in Fig. \ref{fig:vqo-hardware-ansatz} where the Bell state $\ket{\Phi^+}= \frac{1}{\sqrt{2}}(\ket{00}+\ket{11})$ is prepared at each source and local qubit measurements are free to rotate about the $y$-axis.
This ansatz reflects experimental setups of $n$-local networks \cite{carvacho2017experimental_bilocal, andreoli2017_bilocal_expt, saunders2017_bilocal_expt,sun2019experimental_bilocal,poderini2020experimental} where the free parameters correspond a rotation applied to each measurement apparatus

Our variational quantum optimization results are shown in Fig. \ref{fig:vqo-hardware-plots}.
In all optimizations, each circuit evaluation was run using 6000 shots and randomized initial settings.
The CHSH case was optimized using the 5-qubit \texttt{ibmq\_belem} device while the bilocal, trilocal chain, and trilocal star networks were optimized using the 7-qubit \texttt{ibmq\_casablanca} and \texttt{ibmq\_jakarta} devices.
The CHSH plot shown in the upper-left of Fig.~\ref{fig:vqo-hardware-plots} aggregates data from 11 separate optimizations using a step size of $\eta=0.12$.
The bilocal plot shown in the upper-right of Fig.~\ref{fig:vqo-hardware-plots} aggregates data from 5 optimizations and using a step size ranging from $\eta=1.4$ to $\eta=1.5$.
The trilocal chain plot shown in the bottom-left of Fig.~\ref{fig:vqo-hardware-plots} aggregates data from 6 optimization using a step size ranging from $\eta=1.6$ to $\eta=2$.
The trilocal star plot shown in the bottom-right of Fig.~\ref{fig:vqo-hardware-plots} aggregates data from 5 optimizations using a step sizes ranging from $\eta=1.6$ to $\eta=1.8$.

As a baseline measure of noise on the IBM quantum computers, we calculate the theoretical score using the optimal settings in the noiseless case described in Section \ref{section:bell_inequalities_for_n-local_networks}.
The noise on quantum hardware deteriorates the non-$n$-local correlations causing a separation from the noiseless quantum bound.
In particular, the bottom-left plot shows the theoretical quantum violation of the trilocal chain inequality to be close the classical bound, especially when compared with the trilocal star network (bottom-right).
This difference is due to the fact that the $S_{n\text{-Star}}$ is more robust to noise than $S_{n\text{-Chain}}$ \cite{Tavakoli_2014_star,Mukherjee2015chain}.
Finally, the noise on the quantum computers is not constant, it can fluctuate throughout the day \cite{dasgupta2021stability}, hence, the optimal settings in the noiseless case do not always produce the same value.

All plots in Fig.~\ref{fig:vqo-hardware-plots} show the mean optimized score exceeding the classical bound, thus, VQO is able to find non-$n$-local settings in all cases.
In most cases, the error bars shrink as the optimization step increases indicating convergence to an optimum.
In the bilocal network optimization, the significant error bars on the final step are likely a result of the step size being too large. 
In the trilocal chain and star optimizations, the mean optimization score converges to the mean theoretical score showing that the optimization consistently finds the theoretical maximum.
In the bilocal and CHSH optimizations, the mean optimization score does not reach the mean theoretical score.
This is the result of some optimizations finding the local optimum of the classical bound. 

In all plots, the max optimized score converges to a value consistent with the max theoretical score.
In all cases except for the CHSH optimization, an optimized score is found that exceed the max theoretical score.
While there is a statistical chance that that the optimized score may be larger than the max theoretical score, it is also possible that our VQO framework may be finding optimal settings tailored for the quantum hardware.
That is, there may be biases in the gate operations and measurement bases that VQO may be optimizing against.
On the other hand, the theoretical settings are na{\"i}ve to any such hardware biases. 

The final feature shown in the data of Fig. \ref{fig:vqo-hardware-plots} is that the setting optimized on the IBM hardware correspond to the optimal settings in the noiseless case.
That is, when we take the settings for the maximal score in each optimization step and run them on a noiseless classical simulator, they maximally violate the Bell inequality in question.
While this remarkable feature is interesting, its value is questionable.
While we may be able to obtain the optimal settings on a large NISQ device, the quantum computer will still output a nonoptimal answer that cannot be checked on a classical computer due to computing constraints.
However, there may exist some applications where the optimal settings provide value on their own.
At any rate, the prevalence and usefulness of this feature should be a matter of further study.

% We briefly remark on the sources of error in our quantum computing results.
% Most notably, the performance of quantum hardware fluctuates considerably \cite{dasgupta2021stability}.
% While the noise fluctuation brings reproducibility of our results into question, we argue that we obtained comparable results from three different quantum processors across several weeks.
% Thus, we believe our results to be reproducible.
% In extreme cases, both theoretically optimal scores and optimized scores were unable to surpass the classical bound.
% We deemed such results as failed experiments and did not incorporate the data into the results in Fig. \ref{fig:vqo-hardware-plots}.

\subsection{Scaling VQO for Practical Advantages on Quantum Computers}\label{section:scaling_vqo}

It is challenging to scale our optimizations beyond what was considered in the previous section.
As more circuit parameters are introduced, the number of circuit executions required by the parameter-shift rule grows polynomially \cite{Schuld2019_parameter_shift,pennylane2018}.
Furthermore, some cost functions require additional correlators to be evaluated as the network scales. 
For example, the $n$-local star Bell inequality in Eq. \eqref{eq:n-local_star_inequality} contains $2^{n+1}$ correlator terms that each require a quantum circuit to be run or differentiated.
Unfortunately, additional circuit executions are have a significant overhead due to the latency of remote execution, queue wait times, and serial circuit execution on a quantum computer.
We mitigate this overhead by reserving quantum hardware and batching circuit executions, however, more can be done to scale the optimizations.

First, we need wide parallelization across quantum computers.
In the worst case, the cost function will require $|\mc{X}| = \prod_{j=1}^m |\mc{X}_j|$ circuits to construct the network behavior $\PNet(\Theta)$ for all network inputs.
Furthermore, the parameter-shift rule scales the number of circuit executions polynomially.
Fortunately, these circuits are independent and can be run in parallel.
Thus, if we wish to scale VQO techniques, it will be exceedingly important parallelize circuit executions across many NISQ devices.

Second, to overcome the latency of remote execution, we need classical and quantum hardware to be run in close proximity.
The quantum computing industry is taking steps in this direction, for example, the Qiskit Runtime environment.

Third, larger NISQ devices will allow larger networks to be simulated.
In this regime, we may find practical simulation advantages.
However, NISQ devices have yet to show a simulation advantage \cite{Gerritsma2010_simulation,Lanyon2011_simulation,Barends2015,Martinez2016_schwinger_simulation,rost2020simulation,rost2021demonstrating}.
Nevertheless, NISQ devices are predicted to provide simulation advantages in the near-term \cite{Childs2018, Preskill2018}
As an aside, we note that larger networks can be simulated using smaller quantum devices where an exponential increase in the number of circuit evaluations is accrued \cite{Peng2020_simulating_large_circuits, Barratt2021_parallel_nisq_simulation}.
Such methods are only feasible if wide parallelization across NISQ devices can offset the exponential increase in circuit evaluations.

\subsection{Adapting VQO to Quantum Network Hardware}

In principle, the VQO framework depicted in Fig.~\ref{fig:vqo_diagram} can be run on a quantum network rather than a quantum computer.
The key requirement is that the network devices have free parameters that can be tuned continuously, and are therefore, differentiable.
For example, rotating the polarization of a photon in an optics setup is similar to the ansatzes shown in Fig.~\ref{fig:vqo-hardware-ansatz} where qubits are rotated about the $y$-axis.
In fact, hybrid optimization techniques on photonic systems have previously demonstrated the ability to maximize the violation of the CHSH inequality \cite{Suprano2021,Poderini2022_black-box}.
Hence, our VQO framework could be extended to similar photonic implementations of $n$-local networks \cite{carvacho2017experimental_bilocal, andreoli2017_bilocal_expt, saunders2017_bilocal_expt,sun2019experimental_bilocal,poderini2020experimental}.

A key advantage of extending our VQO framework to quantum network hardware is that network protocols can be optimized against the noise inherent to the quantum network hardware.
That is, the noise and biases on quantum computers may not accurately reflect the noise and biases on quantum network hardware.
Hence, communications protocols optimized on quantum network hardware will be more robust because they are tailored specially for the hardware they run on.
Therefore, the cost of noise tomography \cite{Chaung1997_process_tomography,harper2020efficient,Onorati2021_noise_tomography} can be avoided altogether.

Deploying VQO on quantum networks may provide a means of automating device integration and maintenance in quantum networks.
Using VQO, quantum network devices may be able to automatically align photon polarization bases, maintain the communication capacity of channels, or set up device-independent protocols.
Such self-organization amongst network devices may significantly reduce the manual work and expertise needed to build, scale, and maintain quantum networks.

\section{Using VQO to Investigate the Noise Robustness of Quantum Non-\textit{n}-Locality}\label{sec:vqo-classical-simulator}

In this section, we demonstrate that our VQO framework can provide practical value when run on a classical simulator. 
To do this, we use VQO to investigate the noise robustness of quantum non-$n$-locality.
We verify that VQO can reproduce theoretical noise robustness results while also uncovering interesting new phenomena.

We begin by discussing the effect of noise on non-$n$-local correlations.
Then, we discuss the maximal violations for arbitrary mixed states.
Next, we overview our use of VQO to investigate the noise robustness of quantum non-$n$-locality.
Finally, we observe using VQO that maximally entangled state preparations are optimal for unital noise models, but not necessarily for nonunital noise models.

\subsection{The Noise Robustness of Quantum Non-\textit{n}-Locality}

The \textit{noise robustness} of quantum non-$n$-locality quantifies the amount of noise that an $n$-local quantum network can tolerate before its behaviors become $n$-local.
Noise causes the deformation of $\QNet$, the set of quantum network behaviors.
That is, if $\QNet^{\text{id}}$ denotes the set of noiseless quantum network behaviors, then the set of quantum behaviors for a noisy network $\QNet$ satisfies $\QNet\subset\QNet^{\text{id}}$.
Furthermore, if a sufficient amount of noise is present, then $\QNet\subseteq\LNet$ and the non-$n$-locality of the network is broken.
We extend the concept non-$n$-locality breaking from references \cite{Pal2015,Zhang2020}, which introduce the general concept of nonlocality breaking noise.

To quantify noise, a quantum channel $\N_{\gamma}$ or classical postprocessing map $\mbf{E}_{\gamma}$ is parameterized by $\gamma\in[0,1]$.
We let $\gamma=0$ denote the noiseless case while all $\gamma\in[0,1]$ correspond to CPTP maps for quantum channels and column stochastic maps for classical postprocessing models.
If the noise model is parameterized by multiple values, then we denote the parameters as $\gammav=(\gamma_k)_{k=1}^M$ where $M$ specifies the number of parameters in the model.

The robustness of quantum non-$n$-locality can then be quantified by the critical noise parameter $\gamma_0$ at which non-$n$-locality is broken.
In general, it is difficult to determine if $\QNet\subseteq\LNet$ because all Bell inequalities must be known.
However, it is reasonable to determine the noise parameter $\gamma_0$ that breaks the non-$n$-locality with respect to a particular Bell inequality $S_{\text{Bell}}$, which is the approach taken in references \cite{Pal2015, Zhang2020}.

\subsection{``Maximal" \textit{n}-Local Violations Using Noisy Quantum States}\label{section:maximal_qubit_violation}

In Section \ref{section:bell_inequalities_for_n-local_networks}, the optimal quantum strategies for non-$n$-locality are described in the noiseless case.
In the presence of noise, these strategies are not guaranteed to be optimal.
Fortunately, maximal violations have been derived for arbitrary mixed state preparations when separable qubit PVM measurements are applied at each measurement device \cite{Horodecki1995,gisin2017_bilocal_criterion,andreoli2017maximal_star_violation,kundu2020_nlocal_max_qubit_violations}.

These bounds for maximal violation are based upon the necessary and sufficient conditions for violation of the CHSH inequality \cite{Horodecki1995}.
That is, we first construct the real-valued $3\times 3$ matrix $T_{\rho^{\Lambda}}$ with elements
\begin{equation}\label{eq:correlation_matrix}
    t_{i,j} = \tr{\rho^{\Lambda} \sigma^{q_1}_i \otimes \sigma^{q_2}_j},
\end{equation}
where $\sigma^{q_1}_i$ and $\sigma^{q_2}_j$ are Pauli operators acting on qubits $q_1$ and $q_2$ respectively.
We then construct the matrix
\begin{equation}\label{eq:U-matrix}
    R_{\rho^{\Lambda}} = T_{\rho^{\Lambda}}^T T_{\rho^{\Lambda}}
\end{equation}
and define its two largest eigenvalues as $\mu_1(R_{\rho^{\Lambda}})$ and $\mu_2(R_{\rho^{\Lambda}})$ respectively.
If $\mu_1(R_{\rho^{\Lambda}}) + \mu_2(R_{\rho^{\Lambda}})>1$ then the CHSH inequality is maximally violated by the score
\begin{equation}\label{eq:chsh_violation_condition}
    S^\star_{\CHSH} = 2\sqrt{\mu_1(R_{\rho^{\Lambda}}) + \mu_2(R_{\rho^{\Lambda}})}.
\end{equation} 
The maximal $n$-local star score is \cite{gisin2017_bilocal_criterion,andreoli2017maximal_star_violation,kundu2020_nlocal_max_qubit_violations}
\begin{equation}\label{eq:maximal_star_violation}
    S^\star_{n\text{-Star}} = \sqrt{\sum_{j=1}^2\left(\prod_{i=1}^n\mu_j(R_{\rho^{\Lambda_i}})\right)^{1/n}},
\end{equation}
and the maximal $n$-local chain score is \cite{kundu2020_nlocal_max_qubit_violations},
\begin{equation}\label{eq:maximal_chain_violation}
    S_{n\text{-Chain}}^{\star} = \sqrt{\sqrt{\prod_{i=1}^n\mu_1(R_{\rho^{\Lambda_i}})} + \sqrt{\prod_{i=1}^n\mu_2(R_{\rho^{\Lambda_i}})}}.
\end{equation}
The maximal violations for the star network relate to the maximal CHSH score $S_{\text{CHSH}}^{\Lambda_i\star}$ for source $\Lambda_i$ as \cite{gisin2017_bilocal_criterion,andreoli2017maximal_star_violation,kundu2020_nlocal_max_qubit_violations}
\begin{equation}
    S^\star_{n\text{-Star}} \leq \left(\frac{1}{2^n}\prod_{i=1}^n S^{\Lambda_i \star}_{\text{CHSH}}\right)^{1/n},
\end{equation}
where equality is achieved when the sources are identical such that $\rho^{\Lambda_i}=\rho^{\Lambda_{i'}}$ for all $i,i'\in[n]$ \cite{gisin2017_bilocal_criterion}.
These equations do not account for entangled measurements or POVMs, which may improve the Bell score.
However, in the bilocal case of Eq. \eqref{eq:maximal_star_violation} the correlations of Bell state measurements are proven to be strict subset of local PVM measurements \cite{andreoli2017maximal_star_violation}.

Unfortunately, the ``maximal" violations expressed in Eqs. \eqref{eq:maximal_star_violation} and \eqref{eq:maximal_chain_violation} do not reflect the optimal strategy in some cases.
Our variational quantum optimization results in Sections \ref{section:vqo_unital_channels} and \ref{section:vqo_nonunital_channels} uncover key edge cases in which Eqs. \eqref{eq:maximal_star_violation} and \eqref{eq:maximal_chain_violation} do not hold.
Namely, classical strategies applied on a subset of sources can sometimes achieve greater Bell scores.

First, consider the quantum $n$-local star network with one classical source $\rho^{\Lambda_1} = \op{00}{00}$ and $n-1$ quantum sources $\rho^{\Lambda_i} = \op{\Phi^+}{\Phi^+}$ for all $i\in[2,n]$.
Since $\rho^{\Lambda_1}$ is a classical state, $\mu_1(R_{\rho^{\Lambda_1}})=1$ and $\mu_2(R_{\rho^{\Lambda_1}})=0$.
Using Eq. \eqref{eq:maximal_star_violation}, we find that $S^\star_{n\text{-Star}} = 1$ and is therefore $n$-local.
However, if the classical state $\rho^{\Lambda_1}=\op{00}{00}$ is measured using the local observables $O^{A_1}_{x_1} = (1-x_1)\sigma_z + x_1 \sigma_x$ and $O^{B_1}_y = \sigma_z$, then the two-body correlator is $\ip{O^{A_1}_{x_1}O^{B_1}_y}_{\rho^{\Lambda_1}} = (1-x_1)$.
This strategy is equivalent to the case where the measurement on the central node $B_1$ outputs 1 with certainty while the measurement at node $A_1$ outputs 1 if $x_1=0$, and otherwise, outputs $\pm 1$ with equal probability.
The remainder of the network nodes implement the measurements for the optimal quantum star network strategy.
Then, using Eq. \eqref{eq:I_correlator_combo} and the fact that all correlator terms with $x_1=1$ vanish due to the random output from $A_1$, we find
\begin{align}
    I_{n,y} = \frac{1}{2}\frac{1}{2^{(n-1)/2}},
\end{align}
from which it follows,
\begin{equation}
    S_{n\text{-Star}} = 2\left(\frac{1}{2}\frac{1}{2^{(n-1)/2}}\right)^{1/n} = \left(\sqrt{2}\right)^{(n-1)/n} \geq 1.
\end{equation}
Furthermore, we can extend this argument into a more general case where the star network has $k$ classical sources that prepare the state $\op{00}{00}$, then, the described strategy can achieve a Bell score of
\begin{equation}\label{eq:star_classical_sources}
    S_{n\text{-Star}} = (\sqrt{2})^{(n-k)/n} \geq 1.
\end{equation}
Thus, we have described a partially classical strategy that outperforms the ``maximal" score predicted by Eq. \eqref{eq:maximal_star_violation}.

As a second example, we consider the $n$-local chain network with the interior devices preparing the classical state $\rho^{\Lambda_i}=\op{00}{00}$ for all $i\in[2,n-1]$.
In this case, Eq.~\eqref{eq:maximal_chain_violation} predicts that $S_{n\text{-Chain}}=1$, however, we show that this is not the case.
Let the observables on qubits $q_3,\dots,q_{n-2}$ be $\sigma_z$ for all $y$.
Then $\ip{O^{q_3}_y \dots O^{q_{n-2}}_y} = 1$ for all $y\in\{0,1\}$ because the sources prepare classical states on these qubits.
However, if sources $\rho^{\Lambda_1}$ and $\rho^{\Lambda_n}$ are quantum, then they can achieve $S^\star_{\text{Biloc}}$ using the $n=2$ case using Eq. \eqref{eq:maximal_star_violation}.
Hence, the $n$-local chain inequality in Eq. \eqref{eq:n-local-chain-bell-inequality} can be maximally violated even when all interior sources are classical.

Thus, we show that edge cases exist for which the maximal violations predicted by Eqs.~\eqref{eq:maximal_chain_violation} and \eqref{eq:maximal_star_violation} break down.
Nevertheless, these equations still present a useful baseline to which we compare our VQO results.
Furthermore, the partially classical strategies utilized in the edge cases will be useful in understanding the noise robustness results obtained from VQO.

\subsection{VQO of the Noise Robustness of Non-\textit{n}-Locality}

\begin{figure}[h] 
    \centering
    \includegraphics[width=.48\textwidth]{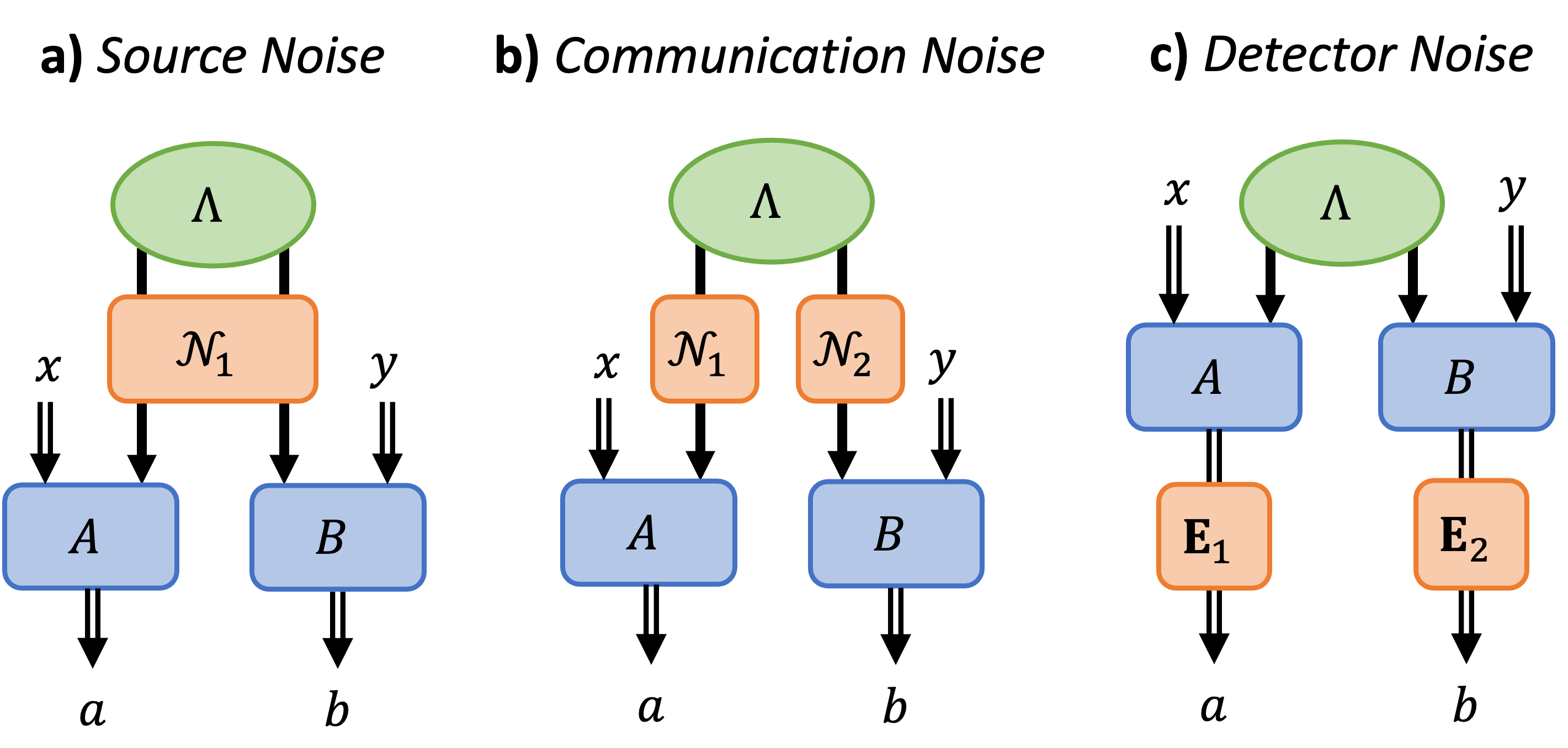}
    \caption{\linespread{1}\selectfont{\small
    \textbf{Noise Models in $n$-Local Networks.}
    a) Source Noise: A quantum channel is applied to all qubits at a given source.
    b) Communication Noise: A quantum channel is applied independently to each qubit.
    c) Detector Noise: Classical post-processing is applied to the classical data output from the $n$-local network.
}}
\label{fig:n-local_noise_models}
\end{figure}

Our objectives are to verify that our VQO framework can reproduce known noise robustness results and also, uncover interesting new results.
In doing so, we demonstrate the practical value of our VQO techniques when applied on a classical simulator.
Our investigative approach to noise robustness is distinct from previous works \cite{Pal2015, Zhang2020} that evaluate the precise noise parameters at which nonlocality is broken.
Instead, we use VQO to find maximal violations of a Bell inequality given a static noise model $\NNet_{\gammav}$.
By scanning through the noise parameters, we create a picture of how the non-$n$-locality deteriorates as the amount of noise increases.
Hence we are able to easily compare the relative noise robustness across different $n$-local networks.

We consider noise applied during the preparation, communication, and measurement stages of an $n$-local network (see Fig. \ref{fig:n-local_noise_models}).
Source noise occurs during the state preparation at each source and is modeled as $\NNet{\gammav} = \bigotimes_{i=1}^n \N^{\Lambda_i}_{\gamma_i}$.
Communication noise occurs during transmission of quantum states and is modeled as $\NNet_{\gammav} = \bigotimes_{k=1}^l \N^{L_k}_{\gamma_k}$.
Detector noise occurs during measurement and is modeled as $\mbf{E}^{\text{Net}}_{\gammav} = \bigotimes_{j=1}^m \mbf{E}^{A_j}_{\gamma_j}$.
Alternatively, detector noise can be modeled as an adjoint channel applied to the measurement $\N^{\text{Net}\dagger}_{\gamma}(\PiNet_{\av|\xv})$.

To simplify our investigation, we characterize the network noise using one noise parameter.
First, we consider networks where noise is applied to only one source, link, or measurement device.
This setting models a faulty component in an ideal network.
Second, we consider the case where noise is applied uniformly to all sources, links, or measurement devices.
This setting is more realistic and assumes that all devices are equally noisy.
Our approach can easily be extended to nonuniform noise models as it introduces no additional computational overhead in comparison to the uniform noise model.

To evaluate the noise robustness we begin with static noise model $\NNet_{\gamma}$.
To create a high-level overview of the noise robustness, we scan through the noise parameter $\gamma\in[0,1]$ using an interval of 0.05.
For each $\gamma$, we use VQO to find the optimal state preparations and measurements that maximize non-$n$-locality with respect to the Bell inequality $S_{\text{Bell}}$.
We repeat this procedure for all considered $n$-local networks depicted in Fig.~\ref{fig:n-local_networks} and compare their relative noise robustness.
Furthermore, we compare the optimized results with theoretical bounds on the max violation.
For some channels, we derive max violation directly and show that our optimizations reproduce the expected results.
In other cases, we use Eqs.~\eqref{eq:chsh_violation_condition}, \eqref{eq:maximal_star_violation}, and \eqref{eq:maximal_chain_violation} to derive the maximal violation for a fixed state such as the Bell state $\ket{\Phi^+} = (\ket{00} + \ket{11})/\sqrt{2}$.

We organize our investigation into two broad classes of noise, unital and nonunital.
In each case, we consider noise applied to sources, qubit communication, and detectors.
For each noise model we compare the results across the Bell inequalities expressed in Eqs. \eqref{eq:1-star-inequality}, \eqref{eq:bilocal_inequality}, \eqref{eq:n-local_star_inequality}, and \eqref{eq:n-local-chain-bell-inequality}.
We note that in the case of the CHSH inequality, we use the Bell score $\frac{1}{2}S_{\text{CHSH}}$ so that it can be compared directly with the Bell scores of the other $n$-local networks.
Finally, to investigate how entangled state preparations or measurements relate to the noise robustness of non-$n$-locality, we explore a wide range of state preparation and measurement ansatzes.
For details about the ansatzes, please refer to Appendix \ref{appendix:ansatz_table}

\subsection{Unital Channels}\label{section:vqo_unital_channels}

A quantum channel $\mc{U}$ is unital if it satisfies $\mc{U}(\mbb{I}) = \mbb{I}$.
Unital noise cannot improve the purity of a state, hence, $\tr{\mc{U}(\rho)^2}\leq\tr{\rho^2}$.
To motivate our investigation of unital noise, we refer to Theorem \ref{thm:two-sided_unital_chsh} proven in Appendix \ref{appendix:unital_channels}.
This theorem states that the noisy state $\mc{U}_1\otimes\mc{U}_2(\op{\Phi^+}{\Phi^+})$ maximally violates the CHSH inequality for any two unital qubit channels $\mc{U}_1$ and $\mc{U}_2$.
Hence, unital qubit noise does not affect the optimal state preparation for violating the CHSH inequality.
Thus, we ask two questions.
Does the optimality of maximally entangled states extend to the $n$-local networks with unital qubit noise models?
Does the optimality of maximally entangled states extend to $n$-local networks with unital multi-qubit noise models?

To help answer these questions, we use our VQO framework to investigate the noise robustness of various unital channel including qubit depolarizing noise, source depolarizing noise, detector white noise, and qubit dephasing noise.
In all cases, we find that maximally entangled states and local qubit measurements are sufficient to achieve maximal violations.
That is, we find no example of arbitrary state preparations or measurements leading to a larger Bell score.
While our results provide evidence that maximally entangled state preparation may be optimal in the case of general unital noise, we cannot make any significant claim outside of the considered networks and noise models.

Furthermore, we find that VQO is able to reproduce known results and find interesting new quantum strategies. 
In all cases, we show that our VQO results align closely with known or derived maximal violations.
Furthermore, in the case of the dephasing channel, we find that the ``maximal" $n$-local violations of Eqs. \eqref{eq:maximal_star_violation} and \eqref{eq:maximal_chain_violation} do not hold because there exist partially classical strategies that are more robust to dephasing noise.

% \begin{conjecture} \label{conj:qubit_unital_noise_n-local}
%     Consider an $n$-local network with unital noise applied to each qubit $\NNet = \bigotimes_{k=1}^N \mc{U}^{q_k}$.
%     The maximal violation each of the $n$-local Bell inequalities of Eqs. \eqref{eq:bilocal_inequality}, \eqref{eq:n-local_star_inequality}, and \eqref{eq:n-local-chain-bell-inequality} is achieved by the state preparation $\ket{\psi^{\Lambda_i}} = \ket{\Phi^+}$.
% \end{conjecture}

% \begin{figure}[t] 
%     \centering
%     \includegraphics[width=.48\textwidth]{figures/unital_noise_robustness_plots.png}
%     \caption{\linespread{1}\selectfont{\small
%         \textbf{Unital noise robustness of non-$n$-locality.}
%         We plot the dependence of the Bell score $S_{\text{Bell}}$ on the noise parameter $\gamma$ for the CHSH, bilocal, chain, and star networks.
%         The horizontal lines show the quantum bound (solid blue) and the classical bound (dashed orange).
%         The markers show the Bell score maximized using our VQO framework across the range $[0,1]$ with an interval of $0.05$.
%         The solid curves show the theoretically optimal score.
%         The left column shows noise applied to a single qubit, source, or detector while the right column shows when noise is applied uniformly to each qubit, source, or detector.
%     }
% }
% \label{fig:unital_noise_robustness_plots}
% \end{figure}

\subsubsection{Qubit Depolarizing Noise Robustness}

\begin{figure}[h] 
    \centering
    \includegraphics[width=.48\textwidth]{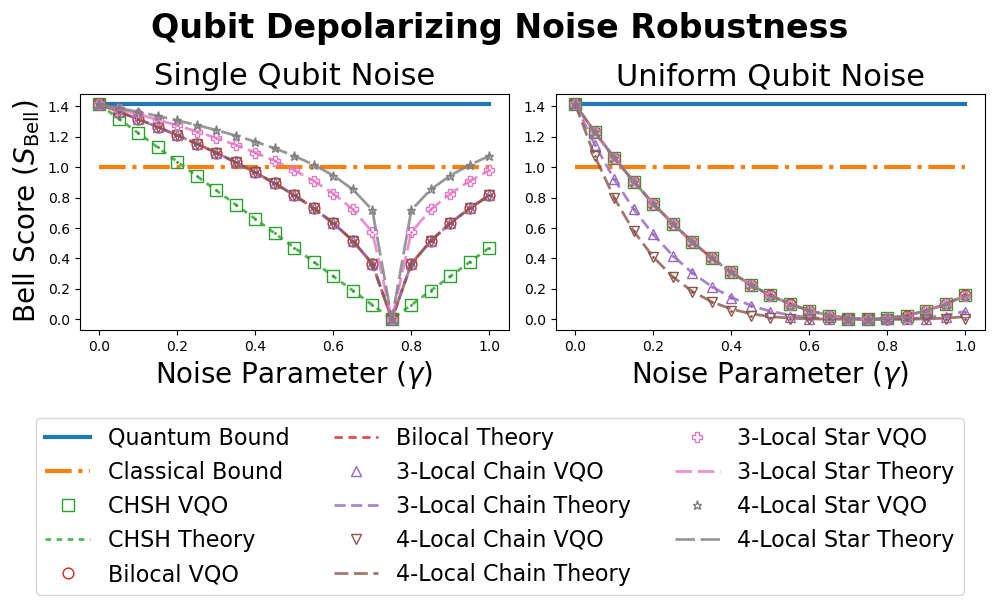}
    \caption{\linespread{1}\selectfont{\small
        \textbf{Qubit depolarizing noise robustness of non-$n$-locality.}
        (Left) Qubit depolarizing noise is applied to qubit $q_1$. (Right) Qubit depolarizing noise is applied uniformly to all qubits. 
        The markers show the maximal Bell score achieved using VQO.
        The dashed lines show the theoretical maximal score found using Eqs. \eqref{eq:chsh_violation_condition}, \eqref{eq:maximal_star_violation}, and \eqref{eq:maximal_chain_violation} and the Bell state preparation $\ket{\Phi^+}$ at each source. 
    }
}
\label{fig:qubit_depolarizing_noise_robustness}
\end{figure}

Qubit depolarizing noise mixes the maximally mixed state with the input qubit state as
\begin{equation}
    \mc{D}_{v}(\rho) = v\rho + \frac{(1-v)}{2}\mbb{I}_2\tr{\rho},
\end{equation}
where $v$ is a parameter commonly referred to as the \textit{visibility}.
The visibility relates to the noise parameter as
\begin{equation}\label{eq:gamma-visibility-relation-depolarizing}
    \gamma = \frac{3}{4}(1-v).
\end{equation}
The Kraus operators for the qubit depolarizing channel are expressed as Pauli operators
\begin{align}
    &K_0 = \sqrt{1-\gamma}\;\mbb{I}_2, \quad &K_1 = \sqrt{\frac{\gamma}{3}}\;\sigma_x, \notag \\
    &K_2 = \sqrt{\frac{\gamma}{3}}\;\sigma_y, \quad &K_3 = \sqrt{\frac{\gamma}{3}}\;\sigma_z. \label{eq:qubit_depolarizing_kraus_ops}
\end{align}
Using these Kraus operators, we simulate qubit depolarizing noise on the PennyLane \texttt{default.mixed} classical simulator \cite{pennylane2018} (see Fig. \ref{fig:qubit_depolarizing_noise_robustness}).
We find a close correspondence between the Bell scores maximized using VQO and the theoretical maximum derived using Eqs. \eqref{eq:chsh_violation_condition}, \eqref{eq:maximal_star_violation}, and \eqref{eq:maximal_chain_violation} and the Bell state preparation $\ket{\Phi^+}$ at each source.

Using our VQO framework, we find the maximal violation is achieved using the Bell state $\ket{\Phi^+}$.
However, we also consider arbitrary preparation and measurement ansatzes (see Appendix \ref{appendix:ansatz_table}).
In the CHSH and bilocal networks, we find that arbitrary state preparations and measurements do not yield a larger Bell score.
For star and chain networks,  we are unable to explore the arbitrary state preparation and measurement ansatzes because the PennyLane \texttt{default.mixed} simulator does not permit efficient optimization. 

\subsubsection{Source Depolarizing Noise Robustness}

\begin{figure}[t] 
    \centering
    \includegraphics[width=.48\textwidth]{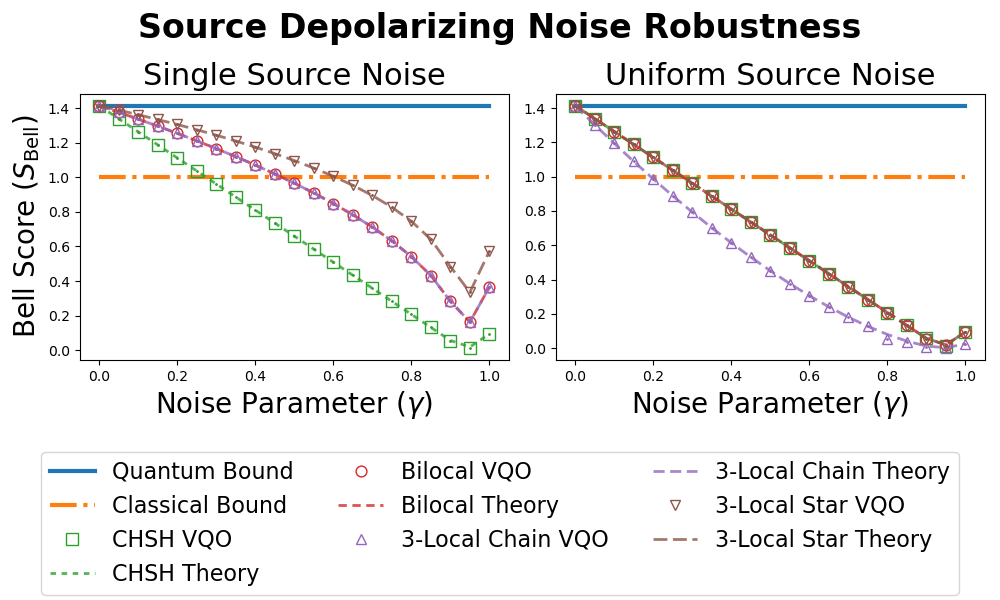}
    \caption{\linespread{1}\selectfont{\small
        \textbf{Source depolarizing noise robustness of non-$n$-locality.}
        (Left) Depolarizing noise is applied to source $\Lambda_1$. (Right) Depolarizing noise is applied uniformly to all sources. 
        The markers show the maximal Bell score achieved using VQO.
        The dashed lines show the theoretical maximal score from Eqs. \eqref{eq:source_depolarizing_theoretical_star} and \eqref{eq:source_depolarizing_theoretical_chain}. 
    }
}
\label{fig:source_depolarizing_noise_robustness}
\end{figure}

Source depolarizing noise mixes the two-qubit maximally mixed state with the input state,
\begin{equation}
    \mc{D}_{v}(\rho) = v\rho + \frac{(1-v)}{4}\mbb{I}_4 \tr{\rho},
\end{equation}
where the visibility $v$ relates to the noise parameter as
\begin{equation}\label{eq:gamma-visibility-relation-source_depolarizing}
    \gamma = \frac{15}{16}(1 - v).  
\end{equation}
The Kraus operators for a two-qubit depolarizing channel are expressed as
\begin{align}
    &K_{0,0} = \sqrt{1-\gamma}\mbb{I}, \\
    \Big\{&K_{i,j} = \sqrt{\frac{\gamma}{15}}\sigma_i\otimes\sigma_j \Big\}_{i,j=0}^3,\label{eq:two-qubit_depolarizing_kraus_ops}
\end{align}
where all $i$ and $j$ are considered except $i=j=0$.

Maximal quantum violations for $n$-local networks with source depolarizing noise have been derived in terms of the visibility $v$.
Namely, the max violation of the star network is $S^\star_{n\text{-Star}} = \sqrt{2}\left(\prod_{i=1}^n v_i\right)^{1/n}$ \cite{Branciard_2012_bilocal_v_nonbilocal,Tavakoli_2014_star} and the max violation of the chain network is $S^\star_{n\text{-Chain}}=\sqrt{2}\sqrt{\prod_{i=1}^n v_i}$ \cite{Mukherjee2015chain}.
Using Eq. \eqref{eq:gamma-visibility-relation-source_depolarizing}, we redefine these maximal violations in terms of the visibility to find
\begin{align}
    S^\star_{n\text{-Star}} &= \sqrt{2}\left(\prod_{i=1}^n \Big|1-\frac{16}{15} \gamma_i\Big| \right)^{1/n}, \label{eq:source_depolarizing_theoretical_star}\\
    S^\star_{n\text{-Chain}} &= \sqrt{2}\sqrt{\prod_{i=1}^n \Big|1-\frac{16}{15} \gamma_i\Big|}, \label{eq:source_depolarizing_theoretical_chain}
\end{align}
where $S^\star_{\text{Bilocal}}$ and $S^\star_{\text{CHSH}}$ are treated as the respective $n=2$ and $n=1$ cases of $S^\star_{n\text{-Star}}$.

The results in Fig. \ref{fig:source_depolarizing_noise_robustness} verify that our VQO framework can reproduce the theoretical results in Eqs.~\eqref{eq:source_depolarizing_theoretical_star} and \eqref{eq:source_depolarizing_theoretical_chain}.
To simulate these networks, we use the PennyLane \texttt{default.mixed} simulator \cite{pennylane2018} and the Kraus operators expressed in Eq.~\eqref{eq:two-qubit_depolarizing_kraus_ops}.
We find that the state preparation $\ket{\Phi^+}$ and local qubit measurements optimized over rotations about the $y$-axis are sufficient for maximal violations in all cases.
Unfortunately, we are unable to efficiently scale our optimizations beyond the 3-Local chain and 3-Local star networks because there is a significant computational overhead in applying the two-qubit Kraus operators expressed in Eq.~\eqref{eq:two-qubit_depolarizing_kraus_ops}.

We provide further evidence supporting the optimality of maximally entangled state preparations and local qubit measurements through the consideration of general ansatzes.
In both the CHSH and bilocal cases, we consider arbitrary state preparation and measurement ansatzes (see Appendix \ref{appendix:ansatz_table}).
We find no example of states or measurements that exceed the theoretical bound in Eqs. \eqref{eq:source_depolarizing_theoretical_star} and \eqref{eq:source_depolarizing_theoretical_chain}.

\subsubsection{White Noise Detector Errors}

\begin{figure}[b] 
    \centering
    \includegraphics[width=.48\textwidth]{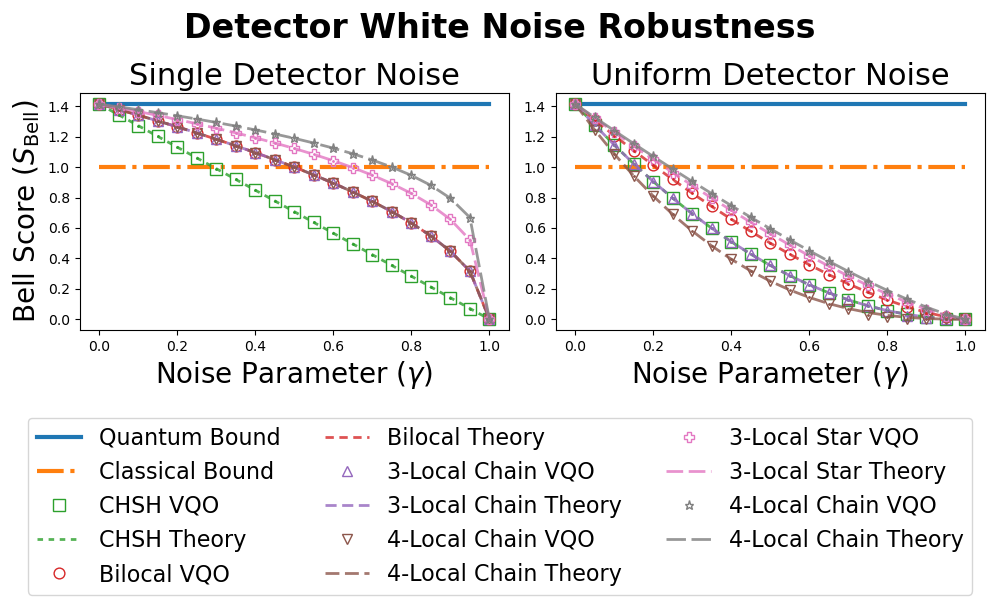}
    \caption{\linespread{1}\selectfont{\small
        \textbf{Detector white noise robustness of non-$n$-locality.}
        (Left) Detector white noise is applied to measurement device $A_1$. (Right) Detector white noise is applied uniformly to all measurement devices. 
        The markers show the maximal Bell score achieved using VQO.
        The dashed lines show the theoretical maximal score from Eqs. \eqref{eq:detector_white_noise_theoretical_star} and \eqref{eq:detector_white_noise_theoretical_chain}. 
    }
}
\label{fig:detector_white_noise_robustness}
\end{figure}

For a detector with binary outputs, we define a white noise error as the classical postprocessing map
\begin{equation}\label{eq:white_noise_detector_error}
    \mbf{W}^j_{\gamma} = (1-\gamma)\mbb{I} + \frac{\gamma}{2}\begin{pmatrix}
        1 & 1 \\ 1 & 1
    \end{pmatrix},
\end{equation}
where with probability $\gamma$, the detector outputs a binary value drawn from a uniform random distribution.
The white noise detector error is described by the $M$-qubit POVM with elements
\begin{equation}\label{eq:detector_white_noise_povm}
    \Pi'_{\pm|x} = (1-\gamma)\Pi_{\pm|x} + \frac{\gamma}{2}\mbb{I}_M,
\end{equation}
where $\Pi_{\pm|x}$ are projectors onto even and odd parity subspaces.
In Proposition \ref{prop:detector_white_noise_unitality} of Appendix \ref{appendix:unitality of detector white noise}, we prove that the white noise detector error is unital and  equivalent to the $M$-qubit depolarizing channel acting upon the detector's local state $\rho^{A_j}$.
Furthermore, in Proposition \ref{prop:theoretical_white_noise_detector_error} of Appendix \ref{appendix:unitality of detector white noise}, we prove that the maximal Bell score for detector errors on all measurement devices is
\begin{align}
        &S^\star_{n\text{-Star}} = \sqrt{2} \left(\prod_{j=1}^{n+1} (1-\gamma_j)\right)^{1/n}, \label{eq:detector_white_noise_theoretical_star}\\
        &S^\star_{n\text{-Chain}} = \sqrt{2}\sqrt{\prod_{j=1}^{n+1}(1-\gamma_j)}. \label{eq:detector_white_noise_theoretical_chain}
    \end{align}

In Fig. \ref{fig:detector_white_noise_robustness}, we verify that our VQO framework can reproduce the theoretical violations of Proposition \ref{prop:theoretical_white_noise_detector_error}.
To simulate the network we use the PennyLane \texttt{default.qubit} pure state simulator \cite{pennylane2018} and the postprocessing white noise models of Eq. \eqref{eq:white_noise_detector_error}.
For all networks, we find that the state preparation $\ket{\Phi^+}$ and local qubit measurements optimized over rotations about the $y$-axis are sufficient to achieve the theoretical maximum.

Since the noisy network simulation does not require ancillary qubits or mixed state simulation, it performs with greater computational efficiency.
Therefore, we consider arbitrary state preparations in the CHSH, bilocal, 3-local chain, 4-local chain, and 3-local star.
We find no improvement over the theoretical bound in Proposition \ref{prop:theoretical_white_noise_detector_error} when more general simulation ansatzes are considered.

\subsubsection{Qubit Dephasing Noise}

\begin{figure}[h] 
    \centering
    \includegraphics[width=.48\textwidth]{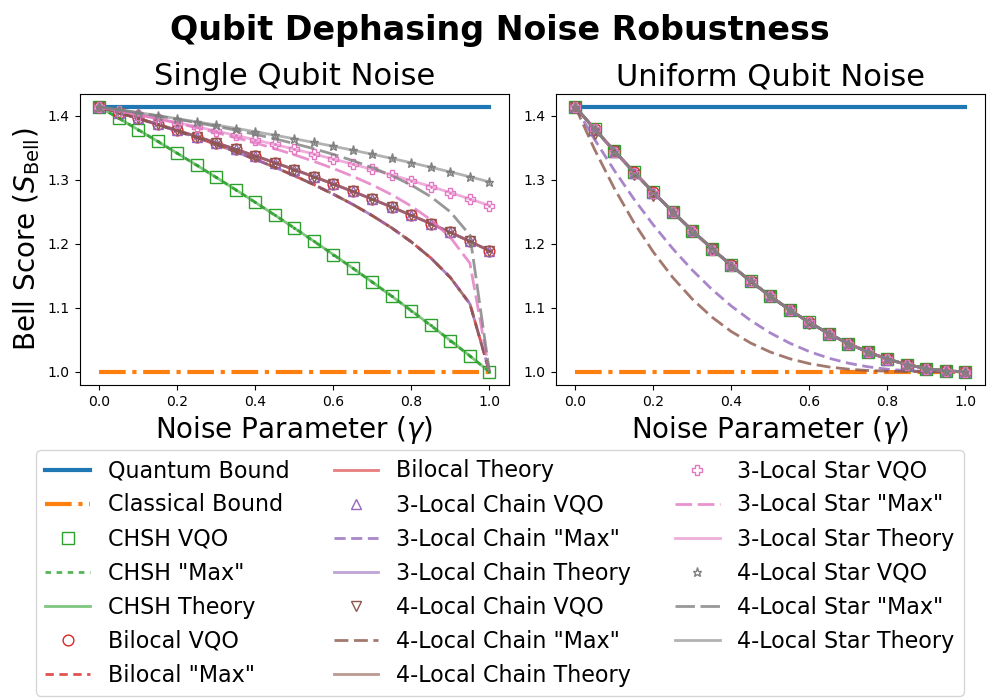}
    \caption{\linespread{1}\selectfont{\small
        \textbf{Qubit dephasing noise robustness of non-$n$-locality.}
        (Left) Dephasing noise is applied to qubit $q_1$. (Right) Dephasing noise is applied uniformly to all qubits. 
        The markers show the maximal Bell score achieved using VQO.
        The dashed lines show the ``maximal" score predicted by Eqs. \eqref{eq:chsh_violation_condition}, \eqref{eq:maximal_star_violation}, and \eqref{eq:maximal_chain_violation}.
        The solid lines match the markers using the theoretical max scores described by Eqs. \eqref{eq:theoretical_uniform_dephasing} and \eqref{eq:theoretical_single_qubit_dephasing}.
    }
}
\label{fig:qubit_dephasing_noise_robustness}
\end{figure}

\begin{figure}[b] 
    \centering
    \includegraphics[width=.48\textwidth]{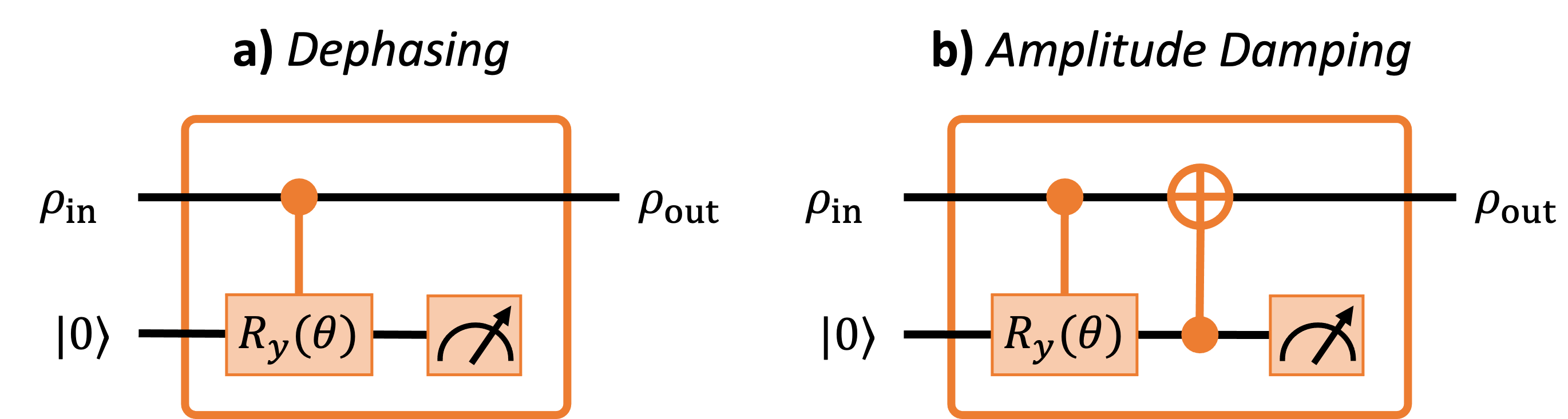}
    \caption{\linespread{1}\selectfont{\small
        \textbf{Qubit dephasing and amplitude damping circuits} \cite{Nielsen2009}.
        a) The qubit dephasing channel $\mc{P}_\gamma$ is implemented using one ancillary qubit and a controlled rotation about the $y$-axis.
        b) The qubit amplitude damping channel $\mc{A}_{\gamma}$ is implemented using one anillary qubit and a controlled rotation about the $y$-axis followed by a CNOT gate.
        In both circuits, the rotation parameter $\theta$ relates to the noise parameter $\gamma$ as $\theta = 2\sin^{-1}(\sqrt{\gamma})$.
    }
}
\label{fig:amplitude_phase_damping_circuits}
\end{figure}

Qubit dephasing is a unital channel describing the decoherence process as
\begin{align}
    \mc{P}_\gamma(\rho) &=  \frac{1}{2}(1+\sqrt{1-\gamma})\rho + \frac{1}{2}(1-\sqrt{1-\gamma})\sigma_z\rho \sigma_z,
    % &= \begin{pmatrix}
    %     \rho_{00} & \sqrt{1-\gamma}\rho_{01} \\ \sqrt{1-\gamma}\rho_{10} & \rho_{11}
    % \end{pmatrix}.
\end{align}
where the offdiagonal terms of $\rho$ go to zero as the noise parameter $\gamma$ increases.
The Kraus operators for the dephasing channel are
\begin{equation}\label{eq:dephasing_kraus_ops}
    K_0 = \begin{pmatrix}
        1 & 0 \\ 0 & \sqrt{1-\gamma}
    \end{pmatrix} \; \text{and} \; \quad K_1 = \begin{pmatrix}
        0 & 0 \\ 0 & \sqrt{\gamma}
    \end{pmatrix},
\end{equation}
however, the circuit expressed in Fig. \ref{fig:amplitude_phase_damping_circuits}.a) leads to the most efficient computation.

As proven in Proposition \ref{prop:two-sided_dephasing} of Appendix \ref{appendix:unital_channels}, the maximal violation of the CHSH inequality given two-sided dephasing noise $\mc{P}_{\gamma_1}\otimes\mc{P}_{\gamma_2}$ is
\begin{equation}
    S^\star_{\text{CHSH}} = 2\sqrt{1 + (1 - \gamma_1)(1- \gamma_2)}.
\end{equation}
For uniform state preparations, the maximal violation for the $n$-local star inequality is \cite{gisin2017_bilocal_criterion}
\begin{equation}
    S^\star_{n\text{-Star}} = \left(\frac{1}{2^n}\prod_{i=1}^n S^{\Lambda_i \star}_{\text{CHSH}}\right)^{1/n},
\end{equation}
which, in the case of uniform dephasing noise, reduces to
\begin{equation} \label{eq:theoretical_uniform_dephasing}
    S^\star_{n\text{-Star}} = \sqrt{1 + (1-\gamma)^2}.
\end{equation}
For the $n$-local chain, the partially classical strategy can implemented using the interior sources because classical states are not affected by dephasing noise.
Thus, $S^\star_{n\text{-Chain}}$ is simply the bilocal case, which is encompassed by Eq. \eqref{eq:theoretical_uniform_dephasing}.
In the single-qubit dephasing case, the maximal star violation is then
\begin{equation} \label{eq:theoretical_single_qubit_dephasing}
    S^\star_{n\text{-Star}} = \left(\sqrt{1 + (1-\gamma)}2^{(n-1)/2}\right)^{1/n},
\end{equation}
where the $n$-local chain simply corresponds to the bilocal case.

Using the dephasing circuit implementation in Fig.~\ref{fig:amplitude_phase_damping_circuits} and the PennyLane \texttt{default.qubit} simulator \cite{pennylane2018}, we demonstrate that our VQO framework can find the optimal state preparations and measurements for non-$n$-locality.
In all cases, the maximum is achieved by preparing the $\ket{\Phi^+}$ state at each source and optimizing over local qubit measurements.
Since each noisy qubit requires an ancilla, our simulations require twice the number of qubits needed to simulate the noiseless network.
We are able to optimize over arbitrary state preparations and measurements in the CHSH, bilocal, 3-local chain, and 4-local chain.
For the star networks, we optimize over arbitrary states preparations, but measurements are optimized over local qubit rotations due to inefficiency in computational performance.
In general, we find that arbitrary state preparation and measurement ansatzes do not improve the Bell score beyond maximally entangled states and separable measurements.

As shown by the dashed lines in Fig. \ref{fig:qubit_dephasing_noise_robustness} the predictions by Eqs. \eqref{eq:maximal_star_violation} and \eqref{eq:maximal_chain_violation} fall off of the VQO data as $\gamma$ increases, hence, these ``maximal" violation do not hold for dephasing noise.
Our VQO can find the theoretical maximum described by Eqs. \eqref{eq:theoretical_uniform_dephasing} and \eqref{eq:theoretical_single_qubit_dephasing}.
These violations can be achieved because the dephasing channel preserves classical states.
That is, $\mc{P}_{\gamma=1}\otimes\text{id}(\op{\Phi^+}{\Phi^+}) = \mc{P}_{\gamma=1}\otimes\mc{P}_{\gamma=1}(\op{\Phi^+}{\Phi^+}) = (\op{00}{00} + \op{11}{11})/2$, a classical mixed state describing uniform shared randomness.

\subsection{Nonunital Noise}\label{section:vqo_nonunital_channels}

In this section, we investigate the noise robustness of quantum non-$n$-locality with respect to nonunital noise models.
Nonunital noise describes quantum channels satisfying $\mc{N}(\mbb{I}) \neq \mbb{I}$.
For unital channels, we found evidence that the Bell state preparation $\ket{\Phi^+}$ was optimal for achieving the maximal violations of the considered $n$-local Bell inequalities.

On the contrary, we find that maximally entangled state preparations are not optimal for nonunital noise models.
Using our VQO framework, we find that nonmaximally entangled state preparations can achieve greater Bell scores than maximally entangled states when qubit amplitude damping noise and biased detector errors are present. 
Furthermore, in the case of colored source noise, we find that the there is a preference of which maximally entangled state is optimal.
That is, the state $\ket{\Psi^+}= (\ket{01} + \ket{10})/\sqrt{2}$ can achieve higher Bell scores than the state $\ket{\Phi^+}= (\ket{00}+\ket{11})/\sqrt{2}$.

In our investigation, we first verify that our VQO results agree with the maximal violations for maximally entangled states predicted by Eqs. \eqref{eq:maximal_star_violation} and \eqref{eq:maximal_chain_violation}, or with partially classical strategies that outperform these predictions.
Then, we show that our VQO can find stronger violations when arbitrary state preparations are permitted.
Hence we show that maximally entangled states are not always optimal in the presence of nonunital noise.

\subsubsection{Qubit Amplitude Damping Noise}

\begin{figure}[t] 
    \centering
    \includegraphics[width=.48\textwidth]{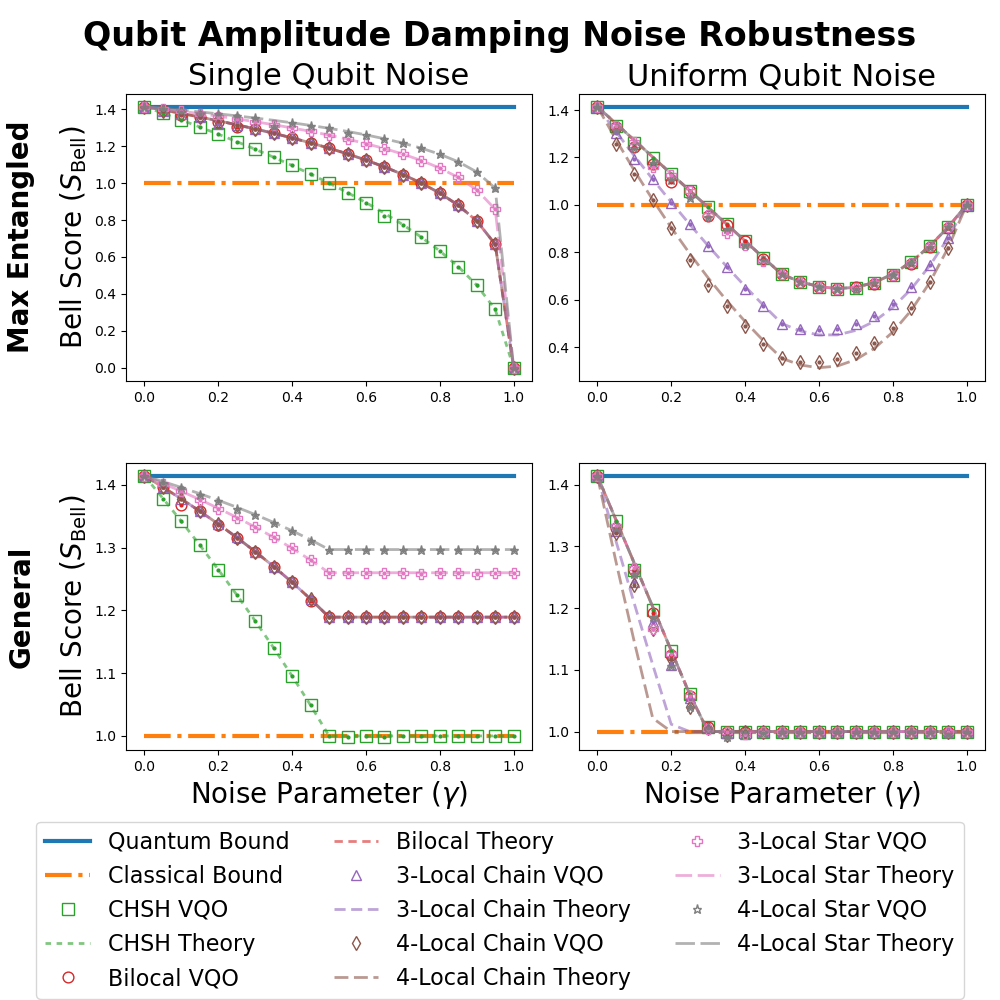}
    \caption{\linespread{1}\selectfont{\small
        \textbf{Qubit amplitude damping noise robustness}
        Amplitude damping noise is applied to a single qubit (left column) and uniformly to all qubits (right column).
        In all plots, the markers show the maximal Bell score achieved using VQO (top row) over maximally entangled state preparations and local qubit measurements, (bottom row) over arbitrary state preparations and measurements.
        The dashed lines show the theoretically maximal score from Eqs. \eqref{eq:chsh_violation_condition}, \eqref{eq:maximal_star_violation}, and \eqref{eq:maximal_chain_violation} when the Bell state preparation is used.
        In the bottom row, the dashed line saturates at the bound for classical sources described by Eq.~\eqref{eq:star_classical_sources}.
    }
}
\label{fig:qubit_amplitude_damping_noise_robustness}
\end{figure}

The qubit amplitude damping channel $\mc{A}_\gamma$ describes the process of energy dissipation where a higher energy qubit state transitions into a lower energy state.
The Kraus operators are defined as
\begin{equation}\label{eq:amplitude_damping_kraus_ops}
    K_0 = \begin{pmatrix}
        1 & 0 \\ 0 & \sqrt{1-\gamma}
    \end{pmatrix}, \quad K_1 = \begin{pmatrix}
        0 & \sqrt{\gamma} \\ 0 & 0
    \end{pmatrix}
\end{equation}
where the effect on a qubit density matrix is
\begin{equation}
    \mc{A}_{\gamma}(\rho) = \begin{pmatrix}
        \gamma + (1-\gamma)\rho_{00} & \sqrt{1-\gamma}\rho_{01} \\
        \sqrt{1-\gamma}\rho_{10} & (1-\gamma)\rho_{11}
    \end{pmatrix}.
\end{equation}
We find that it is most efficient to simulate the amplitude channel on the PennyLane \texttt{default.qubit} classical simulator \cite{pennylane2018} using an ancillary qubit for each channel as described in Fig.~\ref{fig:amplitude_phase_damping_circuits}.b.
Furthermore, the amplitude damping channel preserves the classical state $\mc{A}_{\gamma}(\op{0}{0}) = \op{0}{0}$, hence the optimal partially classical strategies described in Section \ref{section:maximal_qubit_violation} can be used with amplitude damping channel.

As shown in the top row of Fig. \ref{fig:qubit_amplitude_damping_noise_robustness}, we perform our VQO over maximally entangled state preparation and local qubit measurement ansatzes.
We find a close correspondence between the maximal violations predicted by Eqs.~\eqref{eq:chsh_violation_condition}, \eqref{eq:maximal_star_violation}, and \eqref{eq:maximal_chain_violation} when Bell state preparations are used.
As discussed in Proposition \ref{prop:chsh-breaking_amplitude_damping} in Appendix \ref{appendix:amplitude_damping_channel}, once the nonlocality of the CHSH inequality is broken, there exists maximally entangled states that achieve larger CHSH scores than the Bell state.
This feature can be observed in the top-right plot of Fig.~\ref{fig:qubit_amplitude_damping_noise_robustness} where in the region $\gamma\in[0.55,0.75]$, our VQO results for the 3-local and 4-local chains score slightly higher than the prediction made using Bell state preparations and Eq.~\eqref{eq:maximal_chain_violation}.

In the bottom row of Fig.~\ref{fig:qubit_amplitude_damping_noise_robustness}, we perform VQO over arbitrary state preparation and measurement ansatzes.
We compare our VQO results to the Bell score calculated by taking the larger of the predictions by Eq. \eqref{eq:star_classical_sources} and \eqref{eq:maximal_star_violation} where Bell state preparations are used in each case.
In the single-qubit case, our results closely match the theoretical prediction where the classical strategy becomes maximal at $\gamma=0.5$, which corresponds to exactly the point where the qubit amplitude damping channel breaks the nonlocality of the CHSH inequality \cite{Pal2015}.
In the uniform case, there is also a close correspondence between theoretical and VQO data.
However, we note that 3-local and 4-local chain scores drop off more quickly than the VQO data because the maximally entangled state cannot be used to implement the classical strategy on the interior chain network nodes.

In Appendix \ref{appendix:amplitude_damping_channel}, we prove for the CHSH inequality that nonmaximally entangled state preparations can perform better than maximally entangled states.
If the state preparation and noise is constant across all sources \cite{gisin2017_bilocal_criterion}, then the optimality of nonmaximally entangled state preparations can be extended to all considered $n$-local networks.
However, as shown in Fig.~\ref{fig:chsh_uniform_amplitude_damping_high_res} the region of improved Bell violation is small.
Within in the 0.05 interval for $\gamma$ in  Fig.~\ref{fig:qubit_amplitude_damping_noise_robustness} the improvement of the Bell score is only present for $\gamma=0.3$. 
As outlined in Appendix \ref{appendix:amplitude_damping_channel} the uniform qubit amplitude breaking noise breaks non-$n$-locality of maximally entangled states when $\gamma_0=(1-1/\sqrt{2})<0.3$.
However, nonmaximally entangled states can be seen in Fig. \ref{fig:qubit_amplitude_damping_noise_robustness} to be slightly above the classical bound at $\gamma=0.3$, hence our VQO finds the slight improvement over maximally entangled states.

\subsubsection{Source Colored Noise}

\begin{figure}[h] 
    \centering
    \includegraphics[width=.48\textwidth]{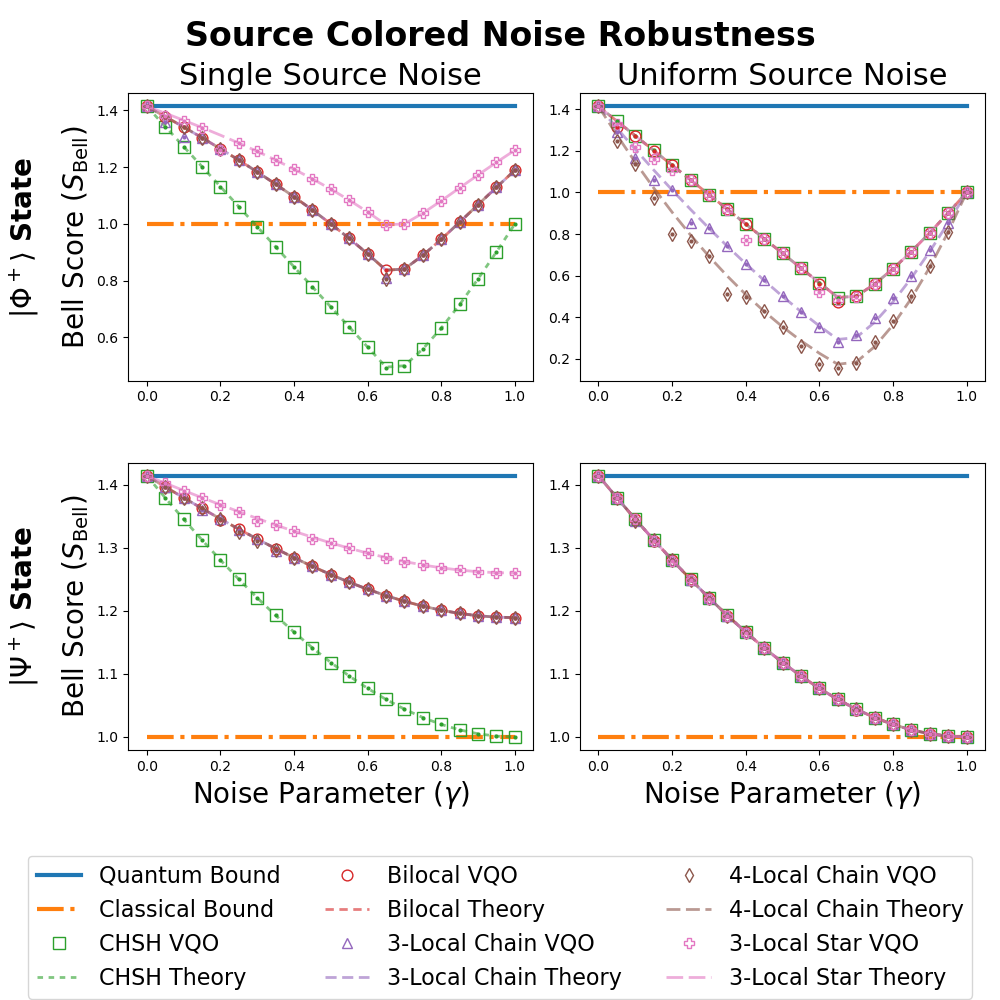}
    \caption{\linespread{1}\selectfont{\small
        Colored noise is applied to a single source (left column) and uniformly to all sources (right column).
        In all plots, the markers show the maximal Bell score achieved using VQO over $\ket{\Phi^+}$ state preparations and local qubit measurements (top row) and over $\ket{\Psi^+}$ state preparations and arbitrary measurements (bottom row).
        The dashed lines show the theoretically maximal score.
        (Top row)
        For the single source noise case, Eqs. \eqref{eq:colored_noise_single_phi_state_star} and \eqref{eq:colored_noise_single_phi_state_chain} are maximal, while for the uniform case, Eqs. \eqref{eq:chsh_violation_condition}, \eqref{eq:maximal_star_violation}, and \eqref{eq:maximal_chain_violation} are maximal. 
        (Bottom row) The maximal score is achieved by Eqs.~ \eqref{eq:colored_noise_single_phi_state_star} and \eqref{eq:colored_noise_single_phi_state_chain} for single source and uniform source noise.
    }
}
\label{fig:source_colored_noise_robustness}
\end{figure}

Colored noise is important because it affects photonic entanglement sources \cite{cabello2005_colored_noise}.
Colored noise is a two-qubit noise model representing depolarization on a preferred axis of the state,
\begin{align}
    \mc{C}_{\gamma}(\rho) &= (1 - \gamma)\rho + \frac{\gamma}{2}\begin{pmatrix}
        0 & 0 & 0 & 0 \\ 0 & 1 & 0 & 0 \\ 0 & 0 & 1 & 0 \\ 0 & 0 & 0 & 0
    \end{pmatrix} \\
    &= (1 - \gamma)\rho + \frac{\gamma}{2}(\op{\Psi^+}{\Psi^+} + \op{\Psi^-}{\Psi^-}),
\end{align}
where $|\Psi^\pm\rangle = (|01\rangle \pm |10\rangle)/\sqrt{2}$.
The Kraus operators for colored noise are
\begin{align}
    K_0 &= \sqrt{1-\gamma}\mbb{I}, \notag\\
    K_{\Psi^{\pm},\Phi^{\pm}} &= \sqrt{\gamma/2}\op{\Psi^\pm}{\Phi^{\pm}}, \label{eq:colored_noise_kraus_ops} \\
    K_{\Psi^\pm,\Psi^\pm} &= \sqrt{\gamma/2}\op{\Psi^+}{\Psi^{\pm}}, \notag
\end{align}
where $\ket{\Phi^{\pm}}$ and $\ket{\Psi^{\pm}}$ constitute the Bell basis.
When $\gamma=1$, this nonunital channel outputs $\mc{C}_{\gamma=1}(\rho) = (\op{00}{00} + \op{11}{11})/2$, which is equivalent to classical shared randomness, hence the partially classical strategies described in Section \ref{section:maximal_qubit_violation} can be implemented.

Using the PennyLane \texttt{default.mixed} simulator \cite{pennylane2018}, we simulate colored noise using the Kraus operators described by Eq. \eqref{eq:colored_noise_kraus_ops}.
Our numerical results are shown in Fig. \ref{fig:source_colored_noise_robustness} where in the top row, we consider the $\ket{\Phi^+}$ state preparation and in the bottom row, we consider the $\ket{\Psi^+}$ state preparation.
When we consider the state preparation $\rho^{\Lambda_i} = \op{\Phi^+}{\Phi^+}$, we find that in the single-source noise case, our VQO results match 
\begin{equation}\label{eq:colored_noise_single_phi_state_star}
    S^\star_{n\text{-Star}} = \left(\prod_{i=1}^n S^{\Lambda_i\star}_{\text{CHSH}}\right)^{1/n},
\end{equation}
and for the chain network
\begin{align}
    S^\star_{n\text{-Chain}} = &\Big(S^{\Lambda_i \star}_{\text{CHSH}} S^{\Lambda_i \star}_{\text{CHSH}} \prod_{i=1}^{n-1}\mu_1(R_{\rho^{\Lambda_i}})\Big)^{1/2}\label{eq:colored_noise_single_phi_state_chain}.
\end{align}
In the uniform noise case, the VQO results match Eqs. \eqref{eq:maximal_star_violation} and \eqref{eq:maximal_chain_violation}.
When we consider the state preparation $\rho^{\Lambda_i} = \op{\Psi^+}{\Psi^+}$, we find that in the single-source and uniform noise case, our VQO results match Eqs. \eqref{eq:colored_noise_single_phi_state_star} and \eqref{eq:colored_noise_single_phi_state_chain}.

We also optimize over arbitrary state preparation and measurement ansatzes.
We find no example state preparation that outperforms the $\ket{\Psi^+}$ state.
Furthermore, local qubit measurements are sufficient to obtain the maximal Bell scores.
Finally, we did not optimize the 4-local star network because it was to computationally intensive to run across all noise parameters on the mixed state simulator using the Kraus operators of Eq. \eqref{eq:colored_noise_kraus_ops}.

\subsubsection{Biased Detector Errors}

\begin{figure}[b] 
    \centering
    \includegraphics[width=.48\textwidth]{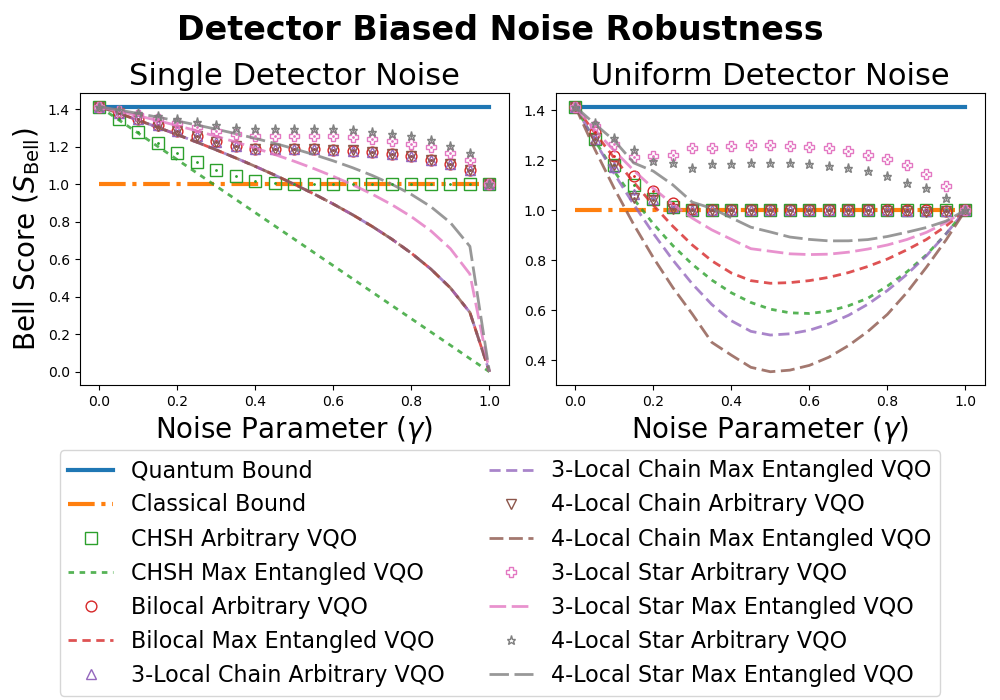}
    \caption{\linespread{1}\selectfont{\small
        \textbf{Biased detector noise robustness.}
        (Left) Biased detector noise is applied to a single detector.
        (Right) Biased detector noise is applied uniformly to all detectors.
        The markers show the maximal Bell score achieved using VQO when using arbitrary state preparations.
        The dashed lines show the maximal Bell score achieved using VQO with respect to maximally entangled states
    }
}
\label{fig:biased_detector_noise_robustness}
\end{figure}

We define a biased detector error as the classical post-processing map,
\begin{equation}
    \mbf{R}_{\gamma} = (1-\gamma)\mbb{I} + \gamma\begin{pmatrix}
        1 & 1 \\ 0 & 0
    \end{pmatrix},
\end{equation}
where the fixed classical value of $+1$ is output whenever the error occurs.
The biased detector error can be described by the POVM with elements
\begin{align}\label{eq:replacer_povm}
    \Pi'_{+|x} &= (1-\gamma)\Pi_{+|x} + \gamma \mbb{I}, \; \text{and} \\
    \Pi'_{-|x} &= (1-\gamma)\Pi_{-|x},
\end{align}
where $\Pi_{+|x}$ and $\Pi_{-|x}$ constitute an $M$-qubit PVM.
In Proposition \ref{prop:equivalence_between_replacer_and_biased_detector_errors} of Appendix \ref{appendix:nonunitality_biased_detector_noise}, we prove that the biased detector error is nonunital because it is equivalent to the $M$-qubit partial replacer channel
\begin{equation}
    \mc{R}_{\gamma,x}(\rho^{A_j}) = (1-\gamma)\rho^{A_j} + \gamma \rho'_{x}\tr{\rho^{A_j}},
\end{equation}
applied to the local qubits $\rho^{A_j}$ held by the detector.

Using the PennyLane \texttt{default.qubit} simulator \cite{pennylane2018} and the classical postprocessing map $\mbf{R}_{\gamma}$ we optimize the Bell score against biased detector noise.
In Fig. \ref{fig:biased_detector_noise_robustness}, we plot with the dashed lines our VQO over maximally entangled state preparations and local measurements.
These results are compared directly with the markers that show VQO over arbitrary state preparations and measurements.
Hence nonmaximally entangled states can improve the Bell score when biased detector errors are present.

\section{Discussion}

This work introduces a hybrid, variational quantum optimization framework for noisy quantum communication networks.
We implement our framework in the qNetVO software \cite{qNetVO}, which we use to optimize non-$n$-locality in noisy quantum networks.
We show that our VQO framework can successfully maximize non-$n$-locality on noisy IBM quantum hardware.
Furthermore, we demonstrate on a classical simulators that our VQO framework can both reproduce known results, and serve as a convenient investigative tool.
Thus, VQO techniques provide practical value to current quantum information research while also showing promise of being a key tool for quantum network design and development.

When run on a classical simulator, our VQO framework can conveniently obtain valuable insights. 
We observe that in the presence of unital noise, maximally entangled state preparations yield maximally non-$n$-local correlations.
In the presence of nonunital noise, we find that nonmaximally entangled states achieve larger violations than maximally entangled states.
This separation is formally proven in the case of the amplitude damping channel.
Our VQO techniques also find important edge cases where classical sources outperform the ``maximal" violations derived by \cite{gisin2017_bilocal_criterion, andreoli2017maximal_star_violation, kundu2020_nlocal_max_qubit_violations}.

Our variational quantum optimization techniques show promise in being a key technology in the development of quantum networks.
In principle, our methods can be deployed on quantum network hardware to optimize network protocols against the inherent hardware noise and potentially, automating the setup of protocols and maintenance of connections between quantum network nodes.
Our VQO techniques also show promise of demonstrating a practical quantum computing advantage.
In the long-term, fault-tolerant quantum computers will provide a clear simulation advantage of large, complex quantum networks \cite{feynman1982simulating,Lloyd1996_simulation}.
In the near-term, NISQ devices are predicted to show advantage in the simulation and optimization of quantum systems \cite{Childs2018, Preskill2018}.

In summary, our variational quantum optimization methods can conveniently be applied on classical hardware to solve meaningful problems while also showing promise of demonstrating practical advantages on quantum hardware.
Hence, VQO is an important tool for designing and developing quantum networks.

\subsection{Future Directions}

To demonstrate practical advantages on NISQ devices, our methods should be broadly parallelized, run with close integration between quantum and classical hardware, and scaled to the largest available devices.
Our framework is generic and interesting results may be obtained by applying VQO to new applications such as optimizing entropic quantities, multipartite games, or quantum protocols.
Additionally, it will be interesting to extend our framework to local operations and classical communication protocols such as teleportation, superdense coding, and entanglement swapping.
Finally, we found that settings optimized on the noisy IBM hardware were optimal in the noiseless case.
Investigating the extent to which this feature holds will be important to the practical application of VQO.

An advantage of our VQO framework is that it can optimize quantum network protocols against the hardware noise without needing to explicitly know the noise model.
Hence, it is important to demonstrate that the parameter-shift rule and gradient descent optimization can be extended to quantum network hardware.

Finally, we have conjectured that unital noise preserves the optimality of maximally entangled state preparations for non-$n$-locality, while nonunital noise does not.
It will be important to determine if this is the case when alternative Bell inequalities are considered, especially ones that consider more inputs or outputs on devices and different network topologies.

\subsection*{Code and Data Availability}

Our variational quantum optimization framework is released as a public python package called qNetVO \cite{qNetVO}.
All numerics and data are found in a supplementary codebase on GitHub \cite{supp_codebase}.

\subsection*{Acknowledgements}

We acknowledge the use of IBM Quantum services for this work. The views expressed are those of the authors, and do not reflect the official policy or position of IBM or the IBM Quantum team.
We acknowledge NSF Award DMR-1747426 and NSF Award 2016136 for supporting this work.

\bibliography{references}

\begin{thebibliography}{103}%
\makeatletter
\providecommand \@ifxundefined [1]{%
 \@ifx{#1\undefined}
}%
\providecommand \@ifnum [1]{%
 \ifnum #1\expandafter \@firstoftwo
 \else \expandafter \@secondoftwo
 \fi
}%
\providecommand \@ifx [1]{%
 \ifx #1\expandafter \@firstoftwo
 \else \expandafter \@secondoftwo
 \fi
}%
\providecommand \natexlab [1]{#1}%
\providecommand \enquote  [1]{``#1''}%
\providecommand \bibnamefont  [1]{#1}%
\providecommand \bibfnamefont [1]{#1}%
\providecommand \citenamefont [1]{#1}%
\providecommand \href@noop [0]{\@secondoftwo}%
\providecommand \href [0]{\begingroup \@sanitize@url \@href}%
\providecommand \@href[1]{\@@startlink{#1}\@@href}%
\providecommand \@@href[1]{\endgroup#1\@@endlink}%
\providecommand \@sanitize@url [0]{\catcode `\\12\catcode `\$12\catcode
  `\&12\catcode `\#12\catcode `\^12\catcode `\_12\catcode `\%12\relax}%
\providecommand \@@startlink[1]{}%
\providecommand \@@endlink[0]{}%
\providecommand \url  [0]{\begingroup\@sanitize@url \@url }%
\providecommand \@url [1]{\endgroup\@href {#1}{\urlprefix }}%
\providecommand \urlprefix  [0]{URL }%
\providecommand \Eprint [0]{\href }%
\providecommand \doibase [0]{https://doi.org/}%
\providecommand \selectlanguage [0]{\@gobble}%
\providecommand \bibinfo  [0]{\@secondoftwo}%
\providecommand \bibfield  [0]{\@secondoftwo}%
\providecommand \translation [1]{[#1]}%
\providecommand \BibitemOpen [0]{}%
\providecommand \bibitemStop [0]{}%
\providecommand \bibitemNoStop [0]{.\EOS\space}%
\providecommand \EOS [0]{\spacefactor3000\relax}%
\providecommand \BibitemShut  [1]{\csname bibitem#1\endcsname}%
\let\auto@bib@innerbib\@empty
%</preamble>
\bibitem [{\citenamefont {Kimble}(2008)}]{Kimble2008quantum_internet}%
  \BibitemOpen
  \bibfield  {author} {\bibinfo {author} {\bibfnamefont {H.~J.}\ \bibnamefont
  {Kimble}},\ }\href {https://doi.org/10.1038/nature07127} {\bibfield
  {journal} {\bibinfo  {journal} {Nature}\ }\textbf {\bibinfo {volume} {453}},\
  \bibinfo {pages} {1023} (\bibinfo {year} {2008})}\BibitemShut {NoStop}%
\bibitem [{\citenamefont {Simon}(2017)}]{Simon2017global_quantum_network}%
  \BibitemOpen
  \bibfield  {author} {\bibinfo {author} {\bibfnamefont {C.}~\bibnamefont
  {Simon}},\ }\href {https://doi.org/10.1038/s41566-017-0032-0} {\bibfield
  {journal} {\bibinfo  {journal} {Nature Photonics}\ }\textbf {\bibinfo
  {volume} {11}},\ \bibinfo {pages} {678} (\bibinfo {year} {2017})}\BibitemShut
  {NoStop}%
\bibitem [{\citenamefont {Wehner}\ \emph {et~al.}(2018)\citenamefont {Wehner},
  \citenamefont {Elkouss},\ and\ \citenamefont
  {Hanson}}]{Wehner2018quantum_internet}%
  \BibitemOpen
  \bibfield  {author} {\bibinfo {author} {\bibfnamefont {S.}~\bibnamefont
  {Wehner}}, \bibinfo {author} {\bibfnamefont {D.}~\bibnamefont {Elkouss}},\
  and\ \bibinfo {author} {\bibfnamefont {R.}~\bibnamefont {Hanson}},\
  }\bibfield  {journal} {\bibinfo  {journal} {Science}\ }\textbf {\bibinfo
  {volume} {362}},\ \href {https://doi.org/10.1126/science.aam9288}
  {10.1126/science.aam9288} (\bibinfo {year} {2018})\BibitemShut {NoStop}%
\bibitem [{\citenamefont {Kozlowski}\ and\ \citenamefont
  {Wehner}(2019)}]{kozlowski2019towards}%
  \BibitemOpen
  \bibfield  {author} {\bibinfo {author} {\bibfnamefont {W.}~\bibnamefont
  {Kozlowski}}\ and\ \bibinfo {author} {\bibfnamefont {S.}~\bibnamefont
  {Wehner}},\ }in\ \href@noop {} {\emph {\bibinfo {booktitle} {Proceedings of
  the Sixth Annual ACM International Conference on Nanoscale Computing and
  Communication}}}\ (\bibinfo {year} {2019})\ pp.\ \bibinfo {pages}
  {1--7}\BibitemShut {NoStop}%
\bibitem [{\citenamefont {K{\'{o}}m{\'{a}}r}\ \emph {et~al.}(2014)\citenamefont
  {K{\'{o}}m{\'{a}}r}, \citenamefont {Kessler}, \citenamefont {Bishof},
  \citenamefont {Jiang}, \citenamefont {S{\o}rensen}, \citenamefont {Ye},\ and\
  \citenamefont {Lukin}}]{Kmr2014clock_synchronization}%
  \BibitemOpen
  \bibfield  {author} {\bibinfo {author} {\bibfnamefont {P.}~\bibnamefont
  {K{\'{o}}m{\'{a}}r}}, \bibinfo {author} {\bibfnamefont {E.~M.}\ \bibnamefont
  {Kessler}}, \bibinfo {author} {\bibfnamefont {M.}~\bibnamefont {Bishof}},
  \bibinfo {author} {\bibfnamefont {L.}~\bibnamefont {Jiang}}, \bibinfo
  {author} {\bibfnamefont {A.~S.}\ \bibnamefont {S{\o}rensen}}, \bibinfo
  {author} {\bibfnamefont {J.}~\bibnamefont {Ye}},\ and\ \bibinfo {author}
  {\bibfnamefont {M.~D.}\ \bibnamefont {Lukin}},\ }\href
  {https://doi.org/10.1038/nphys3000} {\bibfield  {journal} {\bibinfo
  {journal} {Nature Physics}\ }\textbf {\bibinfo {volume} {10}},\ \bibinfo
  {pages} {582} (\bibinfo {year} {2014})}\BibitemShut {NoStop}%
\bibitem [{\citenamefont {Khabiboulline}\ \emph
  {et~al.}(2019{\natexlab{a}})\citenamefont {Khabiboulline}, \citenamefont
  {Borregaard}, \citenamefont {De~Greve},\ and\ \citenamefont
  {Lukin}}]{khabiboulline2019_distributed_sensing}%
  \BibitemOpen
  \bibfield  {author} {\bibinfo {author} {\bibfnamefont {E.~T.}\ \bibnamefont
  {Khabiboulline}}, \bibinfo {author} {\bibfnamefont {J.}~\bibnamefont
  {Borregaard}}, \bibinfo {author} {\bibfnamefont {K.}~\bibnamefont
  {De~Greve}},\ and\ \bibinfo {author} {\bibfnamefont {M.~D.}\ \bibnamefont
  {Lukin}},\ }\href {https://doi.org/10.1103/PhysRevLett.123.070504} {\bibfield
   {journal} {\bibinfo  {journal} {Phys. Rev. Lett.}\ }\textbf {\bibinfo
  {volume} {123}},\ \bibinfo {pages} {070504} (\bibinfo {year}
  {2019}{\natexlab{a}})}\BibitemShut {NoStop}%
\bibitem [{\citenamefont {Khabiboulline}\ \emph
  {et~al.}(2019{\natexlab{b}})\citenamefont {Khabiboulline}, \citenamefont
  {Borregaard}, \citenamefont {De~Greve},\ and\ \citenamefont
  {Lukin}}]{khabiboiulline2019_distributed_sensing2}%
  \BibitemOpen
  \bibfield  {author} {\bibinfo {author} {\bibfnamefont {E.~T.}\ \bibnamefont
  {Khabiboulline}}, \bibinfo {author} {\bibfnamefont {J.}~\bibnamefont
  {Borregaard}}, \bibinfo {author} {\bibfnamefont {K.}~\bibnamefont
  {De~Greve}},\ and\ \bibinfo {author} {\bibfnamefont {M.~D.}\ \bibnamefont
  {Lukin}},\ }\href {https://doi.org/10.1103/PhysRevA.100.022316} {\bibfield
  {journal} {\bibinfo  {journal} {Phys. Rev. A}\ }\textbf {\bibinfo {volume}
  {100}},\ \bibinfo {pages} {022316} (\bibinfo {year}
  {2019}{\natexlab{b}})}\BibitemShut {NoStop}%
\bibitem [{\citenamefont {Sun}\ \emph {et~al.}(2016)\citenamefont {Sun},
  \citenamefont {Mao}, \citenamefont {Chen}, \citenamefont {Zhang},
  \citenamefont {Jiang}, \citenamefont {Zhang}, \citenamefont {Zhang},
  \citenamefont {Miki}, \citenamefont {Yamashita}, \citenamefont {Terai},
  \citenamefont {Jiang}, \citenamefont {Chen}, \citenamefont {You},
  \citenamefont {Chen}, \citenamefont {Wang}, \citenamefont {Fan},
  \citenamefont {Zhang},\ and\ \citenamefont {Pan}}]{Sun2016teleportation}%
  \BibitemOpen
  \bibfield  {author} {\bibinfo {author} {\bibfnamefont {Q.-C.}\ \bibnamefont
  {Sun}}, \bibinfo {author} {\bibfnamefont {Y.-L.}\ \bibnamefont {Mao}},
  \bibinfo {author} {\bibfnamefont {S.-J.}\ \bibnamefont {Chen}}, \bibinfo
  {author} {\bibfnamefont {W.}~\bibnamefont {Zhang}}, \bibinfo {author}
  {\bibfnamefont {Y.-F.}\ \bibnamefont {Jiang}}, \bibinfo {author}
  {\bibfnamefont {Y.-B.}\ \bibnamefont {Zhang}}, \bibinfo {author}
  {\bibfnamefont {W.-J.}\ \bibnamefont {Zhang}}, \bibinfo {author}
  {\bibfnamefont {S.}~\bibnamefont {Miki}}, \bibinfo {author} {\bibfnamefont
  {T.}~\bibnamefont {Yamashita}}, \bibinfo {author} {\bibfnamefont
  {H.}~\bibnamefont {Terai}}, \bibinfo {author} {\bibfnamefont
  {X.}~\bibnamefont {Jiang}}, \bibinfo {author} {\bibfnamefont {T.-Y.}\
  \bibnamefont {Chen}}, \bibinfo {author} {\bibfnamefont {L.-X.}\ \bibnamefont
  {You}}, \bibinfo {author} {\bibfnamefont {X.-F.}\ \bibnamefont {Chen}},
  \bibinfo {author} {\bibfnamefont {Z.}~\bibnamefont {Wang}}, \bibinfo {author}
  {\bibfnamefont {J.-Y.}\ \bibnamefont {Fan}}, \bibinfo {author} {\bibfnamefont
  {Q.}~\bibnamefont {Zhang}},\ and\ \bibinfo {author} {\bibfnamefont {J.-W.}\
  \bibnamefont {Pan}},\ }\href {https://doi.org/10.1038/nphoton.2016.179}
  {\bibfield  {journal} {\bibinfo  {journal} {Nature Photonics}\ }\textbf
  {\bibinfo {volume} {10}},\ \bibinfo {pages} {671} (\bibinfo {year}
  {2016})}\BibitemShut {NoStop}%
\bibitem [{\citenamefont {Cozzolino}\ \emph {et~al.}(2019)\citenamefont
  {Cozzolino}, \citenamefont {Lio}, \citenamefont {Bacco},\ and\ \citenamefont
  {Oxenl{\o}we}}]{Cozzolino2019_high_dimensional_communication}%
  \BibitemOpen
  \bibfield  {author} {\bibinfo {author} {\bibfnamefont {D.}~\bibnamefont
  {Cozzolino}}, \bibinfo {author} {\bibfnamefont {B.~D.}\ \bibnamefont {Lio}},
  \bibinfo {author} {\bibfnamefont {D.}~\bibnamefont {Bacco}},\ and\ \bibinfo
  {author} {\bibfnamefont {L.~K.}\ \bibnamefont {Oxenl{\o}we}},\ }\href
  {https://doi.org/10.1002/qute.201900038} {\bibfield  {journal} {\bibinfo
  {journal} {Advanced Quantum Technologies}\ }\textbf {\bibinfo {volume} {2}},\
  \bibinfo {pages} {1900038} (\bibinfo {year} {2019})}\BibitemShut {NoStop}%
\bibitem [{\citenamefont {Herbert}(2020)}]{Herbert2020superdense}%
  \BibitemOpen
  \bibfield  {author} {\bibinfo {author} {\bibfnamefont {S.}~\bibnamefont
  {Herbert}},\ }\href {https://doi.org/10.1103/PhysRevA.101.062332} {\bibfield
  {journal} {\bibinfo  {journal} {Phys. Rev. A}\ }\textbf {\bibinfo {volume}
  {101}},\ \bibinfo {pages} {062332} (\bibinfo {year} {2020})}\BibitemShut
  {NoStop}%
\bibitem [{\citenamefont {Scarani}\ \emph {et~al.}(2009)\citenamefont
  {Scarani}, \citenamefont {Bechmann-Pasquinucci}, \citenamefont {Cerf},
  \citenamefont {Du\ifmmode~\check{s}\else \v{s}\fi{}ek}, \citenamefont
  {L\"utkenhaus},\ and\ \citenamefont {Peev}}]{scarani2009_qkd}%
  \BibitemOpen
  \bibfield  {author} {\bibinfo {author} {\bibfnamefont {V.}~\bibnamefont
  {Scarani}}, \bibinfo {author} {\bibfnamefont {H.}~\bibnamefont
  {Bechmann-Pasquinucci}}, \bibinfo {author} {\bibfnamefont {N.~J.}\
  \bibnamefont {Cerf}}, \bibinfo {author} {\bibfnamefont {M.}~\bibnamefont
  {Du\ifmmode~\check{s}\else \v{s}\fi{}ek}}, \bibinfo {author} {\bibfnamefont
  {N.}~\bibnamefont {L\"utkenhaus}},\ and\ \bibinfo {author} {\bibfnamefont
  {M.}~\bibnamefont {Peev}},\ }\href
  {https://doi.org/10.1103/RevModPhys.81.1301} {\bibfield  {journal} {\bibinfo
  {journal} {Rev. Mod. Phys.}\ }\textbf {\bibinfo {volume} {81}},\ \bibinfo
  {pages} {1301} (\bibinfo {year} {2009})}\BibitemShut {NoStop}%
\bibitem [{\citenamefont {Broadbent}\ \emph {et~al.}(2009)\citenamefont
  {Broadbent}, \citenamefont {Fitzsimons},\ and\ \citenamefont
  {Kashefi}}]{Broadbent2009blind_qc}%
  \BibitemOpen
  \bibfield  {author} {\bibinfo {author} {\bibfnamefont {A.}~\bibnamefont
  {Broadbent}}, \bibinfo {author} {\bibfnamefont {J.}~\bibnamefont
  {Fitzsimons}},\ and\ \bibinfo {author} {\bibfnamefont {E.}~\bibnamefont
  {Kashefi}},\ }in\ \href {https://doi.org/10.1109/focs.2009.36} {\emph
  {\bibinfo {booktitle} {2009 50th Annual {IEEE} Symposium on Foundations of
  Computer Science}}}\ (\bibinfo  {publisher} {{IEEE}},\ \bibinfo {year}
  {2009})\BibitemShut {NoStop}%
\bibitem [{\citenamefont {Lee}\ and\ \citenamefont
  {Hoban}(2018{\natexlab{a}})}]{Lee2018_network_di}%
  \BibitemOpen
  \bibfield  {author} {\bibinfo {author} {\bibfnamefont {C.~M.}\ \bibnamefont
  {Lee}}\ and\ \bibinfo {author} {\bibfnamefont {M.~J.}\ \bibnamefont
  {Hoban}},\ }\href {https://doi.org/10.1103/PhysRevLett.120.020504} {\bibfield
   {journal} {\bibinfo  {journal} {Phys. Rev. Lett.}\ }\textbf {\bibinfo
  {volume} {120}},\ \bibinfo {pages} {020504} (\bibinfo {year}
  {2018}{\natexlab{a}})}\BibitemShut {NoStop}%
\bibitem [{\citenamefont {Pirandola}\ \emph {et~al.}(2020)\citenamefont
  {Pirandola}, \citenamefont {Andersen}, \citenamefont {Banchi}, \citenamefont
  {Berta}, \citenamefont {Bunandar}, \citenamefont {Colbeck}, \citenamefont
  {Englund}, \citenamefont {Gehring}, \citenamefont {Lupo}, \citenamefont
  {Ottaviani}, \citenamefont {Pereira}, \citenamefont {Razavi}, \citenamefont
  {Shaari}, \citenamefont {Tomamichel}, \citenamefont {Usenko}, \citenamefont
  {Vallone}, \citenamefont {Villoresi},\ and\ \citenamefont
  {Wallden}}]{Pirandola2020}%
  \BibitemOpen
  \bibfield  {author} {\bibinfo {author} {\bibfnamefont {S.}~\bibnamefont
  {Pirandola}}, \bibinfo {author} {\bibfnamefont {U.~L.}\ \bibnamefont
  {Andersen}}, \bibinfo {author} {\bibfnamefont {L.}~\bibnamefont {Banchi}},
  \bibinfo {author} {\bibfnamefont {M.}~\bibnamefont {Berta}}, \bibinfo
  {author} {\bibfnamefont {D.}~\bibnamefont {Bunandar}}, \bibinfo {author}
  {\bibfnamefont {R.}~\bibnamefont {Colbeck}}, \bibinfo {author} {\bibfnamefont
  {D.}~\bibnamefont {Englund}}, \bibinfo {author} {\bibfnamefont
  {T.}~\bibnamefont {Gehring}}, \bibinfo {author} {\bibfnamefont
  {C.}~\bibnamefont {Lupo}}, \bibinfo {author} {\bibfnamefont {C.}~\bibnamefont
  {Ottaviani}}, \bibinfo {author} {\bibfnamefont {J.~L.}\ \bibnamefont
  {Pereira}}, \bibinfo {author} {\bibfnamefont {M.}~\bibnamefont {Razavi}},
  \bibinfo {author} {\bibfnamefont {J.~S.}\ \bibnamefont {Shaari}}, \bibinfo
  {author} {\bibfnamefont {M.}~\bibnamefont {Tomamichel}}, \bibinfo {author}
  {\bibfnamefont {V.~C.}\ \bibnamefont {Usenko}}, \bibinfo {author}
  {\bibfnamefont {G.}~\bibnamefont {Vallone}}, \bibinfo {author} {\bibfnamefont
  {P.}~\bibnamefont {Villoresi}},\ and\ \bibinfo {author} {\bibfnamefont
  {P.}~\bibnamefont {Wallden}},\ }\href {https://doi.org/10.1364/AOP.361502}
  {\bibfield  {journal} {\bibinfo  {journal} {Adv. Opt. Photon.}\ }\textbf
  {\bibinfo {volume} {12}},\ \bibinfo {pages} {1012} (\bibinfo {year}
  {2020})}\BibitemShut {NoStop}%
\bibitem [{\citenamefont {Luo}(2022)}]{Luo2022_di_network}%
  \BibitemOpen
  \bibfield  {author} {\bibinfo {author} {\bibfnamefont {M.-X.}\ \bibnamefont
  {Luo}},\ }\href {https://doi.org/10.1103/PhysRevResearch.4.013203} {\bibfield
   {journal} {\bibinfo  {journal} {Phys. Rev. Research}\ }\textbf {\bibinfo
  {volume} {4}},\ \bibinfo {pages} {013203} (\bibinfo {year}
  {2022})}\BibitemShut {NoStop}%
\bibitem [{\citenamefont {Buhrman}\ \emph {et~al.}(2010)\citenamefont
  {Buhrman}, \citenamefont {Cleve}, \citenamefont {Massar},\ and\ \citenamefont
  {de~Wolf}}]{buhrman2010_communication_complexity}%
  \BibitemOpen
  \bibfield  {author} {\bibinfo {author} {\bibfnamefont {H.}~\bibnamefont
  {Buhrman}}, \bibinfo {author} {\bibfnamefont {R.}~\bibnamefont {Cleve}},
  \bibinfo {author} {\bibfnamefont {S.}~\bibnamefont {Massar}},\ and\ \bibinfo
  {author} {\bibfnamefont {R.}~\bibnamefont {de~Wolf}},\ }\href
  {https://doi.org/10.1103/RevModPhys.82.665} {\bibfield  {journal} {\bibinfo
  {journal} {Rev. Mod. Phys.}\ }\textbf {\bibinfo {volume} {82}},\ \bibinfo
  {pages} {665} (\bibinfo {year} {2010})}\BibitemShut {NoStop}%
\bibitem [{\citenamefont {Cuomo}\ \emph {et~al.}(2020)\citenamefont {Cuomo},
  \citenamefont {Caleffi},\ and\ \citenamefont
  {Cacciapuoti}}]{Cuomo2020_distributed_qc}%
  \BibitemOpen
  \bibfield  {author} {\bibinfo {author} {\bibfnamefont {D.}~\bibnamefont
  {Cuomo}}, \bibinfo {author} {\bibfnamefont {M.}~\bibnamefont {Caleffi}},\
  and\ \bibinfo {author} {\bibfnamefont {A.~S.}\ \bibnamefont {Cacciapuoti}},\
  }\href {https://doi.org/10.1049/iet-qtc.2020.0002} {\bibfield  {journal}
  {\bibinfo  {journal} {{IET} Quantum Communication}\ }\textbf {\bibinfo
  {volume} {1}},\ \bibinfo {pages} {3} (\bibinfo {year} {2020})}\BibitemShut
  {NoStop}%
\bibitem [{\citenamefont {Chuang}\ and\ \citenamefont
  {Nielsen}(1997)}]{Chaung1997_process_tomography}%
  \BibitemOpen
  \bibfield  {author} {\bibinfo {author} {\bibfnamefont {I.~L.}\ \bibnamefont
  {Chuang}}\ and\ \bibinfo {author} {\bibfnamefont {M.~A.}\ \bibnamefont
  {Nielsen}},\ }\href {https://doi.org/10.1080/09500349708231894} {\bibfield
  {journal} {\bibinfo  {journal} {Journal of Modern Optics}\ }\textbf {\bibinfo
  {volume} {44}},\ \bibinfo {pages} {2455} (\bibinfo {year}
  {1997})}\BibitemShut {NoStop}%
\bibitem [{\citenamefont {Harper}\ \emph {et~al.}(2020)\citenamefont {Harper},
  \citenamefont {Flammia},\ and\ \citenamefont
  {Wallman}}]{harper2020efficient}%
  \BibitemOpen
  \bibfield  {author} {\bibinfo {author} {\bibfnamefont {R.}~\bibnamefont
  {Harper}}, \bibinfo {author} {\bibfnamefont {S.~T.}\ \bibnamefont
  {Flammia}},\ and\ \bibinfo {author} {\bibfnamefont {J.~J.}\ \bibnamefont
  {Wallman}},\ }\href
  {https://doi.org/https://doi.org/10.1038/s41567-020-0992-8} {\bibfield
  {journal} {\bibinfo  {journal} {Nature Physics}\ }\textbf {\bibinfo {volume}
  {16}},\ \bibinfo {pages} {1184} (\bibinfo {year} {2020})}\BibitemShut
  {NoStop}%
\bibitem [{\citenamefont {Onorati}\ \emph {et~al.}(2021)\citenamefont
  {Onorati}, \citenamefont {Kohler},\ and\ \citenamefont
  {Cubitt}}]{Onorati2021_noise_tomography}%
  \BibitemOpen
  \bibfield  {author} {\bibinfo {author} {\bibfnamefont {E.}~\bibnamefont
  {Onorati}}, \bibinfo {author} {\bibfnamefont {T.}~\bibnamefont {Kohler}},\
  and\ \bibinfo {author} {\bibfnamefont {T.}~\bibnamefont {Cubitt}},\
  }\href@noop {} {\bibfield  {journal} {\bibinfo  {journal} {arXiv preprint
  arXiv:2103.17243}\ } (\bibinfo {year} {2021})}\BibitemShut {NoStop}%
\bibitem [{\citenamefont {Or{\'{u}}s}(2019)}]{Orus2019_tn}%
  \BibitemOpen
  \bibfield  {author} {\bibinfo {author} {\bibfnamefont {R.}~\bibnamefont
  {Or{\'{u}}s}},\ }\href {https://doi.org/10.1038/s42254-019-0086-7} {\bibfield
   {journal} {\bibinfo  {journal} {Nature Reviews Physics}\ }\textbf {\bibinfo
  {volume} {1}},\ \bibinfo {pages} {538} (\bibinfo {year} {2019})}\BibitemShut
  {NoStop}%
\bibitem [{\citenamefont {Briegel}\ \emph {et~al.}(1998)\citenamefont
  {Briegel}, \citenamefont {D\"ur}, \citenamefont {Cirac},\ and\ \citenamefont
  {Zoller}}]{Briegel1998_quantum_repeater}%
  \BibitemOpen
  \bibfield  {author} {\bibinfo {author} {\bibfnamefont {H.-J.}\ \bibnamefont
  {Briegel}}, \bibinfo {author} {\bibfnamefont {W.}~\bibnamefont {D\"ur}},
  \bibinfo {author} {\bibfnamefont {J.~I.}\ \bibnamefont {Cirac}},\ and\
  \bibinfo {author} {\bibfnamefont {P.}~\bibnamefont {Zoller}},\ }\href
  {https://doi.org/10.1103/PhysRevLett.81.5932} {\bibfield  {journal} {\bibinfo
   {journal} {Phys. Rev. Lett.}\ }\textbf {\bibinfo {volume} {81}},\ \bibinfo
  {pages} {5932} (\bibinfo {year} {1998})}\BibitemShut {NoStop}%
\bibitem [{\citenamefont {Sangouard}\ \emph {et~al.}(2011)\citenamefont
  {Sangouard}, \citenamefont {Simon}, \citenamefont {de~Riedmatten},\ and\
  \citenamefont {Gisin}}]{sangouard2011_quantum_repeater}%
  \BibitemOpen
  \bibfield  {author} {\bibinfo {author} {\bibfnamefont {N.}~\bibnamefont
  {Sangouard}}, \bibinfo {author} {\bibfnamefont {C.}~\bibnamefont {Simon}},
  \bibinfo {author} {\bibfnamefont {H.}~\bibnamefont {de~Riedmatten}},\ and\
  \bibinfo {author} {\bibfnamefont {N.}~\bibnamefont {Gisin}},\ }\href
  {https://doi.org/10.1103/RevModPhys.83.33} {\bibfield  {journal} {\bibinfo
  {journal} {Rev. Mod. Phys.}\ }\textbf {\bibinfo {volume} {83}},\ \bibinfo
  {pages} {33} (\bibinfo {year} {2011})}\BibitemShut {NoStop}%
\bibitem [{\citenamefont {Bennett}\ \emph {et~al.}(1993)\citenamefont
  {Bennett}, \citenamefont {Brassard}, \citenamefont {Cr\'epeau}, \citenamefont
  {Jozsa}, \citenamefont {Peres},\ and\ \citenamefont
  {Wootters}}]{Bennet1993_teleportation_entanglement}%
  \BibitemOpen
  \bibfield  {author} {\bibinfo {author} {\bibfnamefont {C.~H.}\ \bibnamefont
  {Bennett}}, \bibinfo {author} {\bibfnamefont {G.}~\bibnamefont {Brassard}},
  \bibinfo {author} {\bibfnamefont {C.}~\bibnamefont {Cr\'epeau}}, \bibinfo
  {author} {\bibfnamefont {R.}~\bibnamefont {Jozsa}}, \bibinfo {author}
  {\bibfnamefont {A.}~\bibnamefont {Peres}},\ and\ \bibinfo {author}
  {\bibfnamefont {W.~K.}\ \bibnamefont {Wootters}},\ }\href
  {https://doi.org/10.1103/PhysRevLett.70.1895} {\bibfield  {journal} {\bibinfo
   {journal} {Phys. Rev. Lett.}\ }\textbf {\bibinfo {volume} {70}},\ \bibinfo
  {pages} {1895} (\bibinfo {year} {1993})}\BibitemShut {NoStop}%
\bibitem [{\citenamefont {\ifmmode~\dot{Z}\else \.{Z}\fi{}ukowski}\ \emph
  {et~al.}(1993)\citenamefont {\ifmmode~\dot{Z}\else \.{Z}\fi{}ukowski},
  \citenamefont {Zeilinger}, \citenamefont {Horne},\ and\ \citenamefont
  {Ekert}}]{Zukowski1993_entanglement_swapping}%
  \BibitemOpen
  \bibfield  {author} {\bibinfo {author} {\bibfnamefont {M.}~\bibnamefont
  {\ifmmode~\dot{Z}\else \.{Z}\fi{}ukowski}}, \bibinfo {author} {\bibfnamefont
  {A.}~\bibnamefont {Zeilinger}}, \bibinfo {author} {\bibfnamefont {M.~A.}\
  \bibnamefont {Horne}},\ and\ \bibinfo {author} {\bibfnamefont {A.~K.}\
  \bibnamefont {Ekert}},\ }\href {https://doi.org/10.1103/PhysRevLett.71.4287}
  {\bibfield  {journal} {\bibinfo  {journal} {Phys. Rev. Lett.}\ }\textbf
  {\bibinfo {volume} {71}},\ \bibinfo {pages} {4287} (\bibinfo {year}
  {1993})}\BibitemShut {NoStop}%
\bibitem [{\citenamefont {Bose}\ \emph {et~al.}(1998)\citenamefont {Bose},
  \citenamefont {Vedral},\ and\ \citenamefont
  {Knight}}]{Bose1998_generalized_entanglement_swapping}%
  \BibitemOpen
  \bibfield  {author} {\bibinfo {author} {\bibfnamefont {S.}~\bibnamefont
  {Bose}}, \bibinfo {author} {\bibfnamefont {V.}~\bibnamefont {Vedral}},\ and\
  \bibinfo {author} {\bibfnamefont {P.~L.}\ \bibnamefont {Knight}},\ }\href
  {https://doi.org/10.1103/PhysRevA.57.822} {\bibfield  {journal} {\bibinfo
  {journal} {Phys. Rev. A}\ }\textbf {\bibinfo {volume} {57}},\ \bibinfo
  {pages} {822} (\bibinfo {year} {1998})}\BibitemShut {NoStop}%
\bibitem [{\citenamefont {Mitarai}\ \emph {et~al.}(2018)\citenamefont
  {Mitarai}, \citenamefont {Negoro}, \citenamefont {Kitagawa},\ and\
  \citenamefont {Fujii}}]{Mitarai2018_quantum_circuit_learning}%
  \BibitemOpen
  \bibfield  {author} {\bibinfo {author} {\bibfnamefont {K.}~\bibnamefont
  {Mitarai}}, \bibinfo {author} {\bibfnamefont {M.}~\bibnamefont {Negoro}},
  \bibinfo {author} {\bibfnamefont {M.}~\bibnamefont {Kitagawa}},\ and\
  \bibinfo {author} {\bibfnamefont {K.}~\bibnamefont {Fujii}},\ }\href
  {https://doi.org/10.1103/PhysRevA.98.032309} {\bibfield  {journal} {\bibinfo
  {journal} {Phys. Rev. A}\ }\textbf {\bibinfo {volume} {98}},\ \bibinfo
  {pages} {032309} (\bibinfo {year} {2018})}\BibitemShut {NoStop}%
\bibitem [{\citenamefont {Moll}\ \emph {et~al.}(2018)\citenamefont {Moll},
  \citenamefont {Barkoutsos}, \citenamefont {Bishop}, \citenamefont {Chow},
  \citenamefont {Cross}, \citenamefont {Egger}, \citenamefont {Filipp},
  \citenamefont {Fuhrer}, \citenamefont {Gambetta}, \citenamefont {Ganzhorn},
  \citenamefont {Kandala}, \citenamefont {Mezzacapo}, \citenamefont
  {M\"{u}ller}, \citenamefont {Riess}, \citenamefont {Salis}, \citenamefont
  {Smolin}, \citenamefont {Tavernelli},\ and\ \citenamefont
  {Temme}}]{Moll2018_vqo}%
  \BibitemOpen
  \bibfield  {author} {\bibinfo {author} {\bibfnamefont {N.}~\bibnamefont
  {Moll}}, \bibinfo {author} {\bibfnamefont {P.}~\bibnamefont {Barkoutsos}},
  \bibinfo {author} {\bibfnamefont {L.~S.}\ \bibnamefont {Bishop}}, \bibinfo
  {author} {\bibfnamefont {J.~M.}\ \bibnamefont {Chow}}, \bibinfo {author}
  {\bibfnamefont {A.}~\bibnamefont {Cross}}, \bibinfo {author} {\bibfnamefont
  {D.~J.}\ \bibnamefont {Egger}}, \bibinfo {author} {\bibfnamefont
  {S.}~\bibnamefont {Filipp}}, \bibinfo {author} {\bibfnamefont
  {A.}~\bibnamefont {Fuhrer}}, \bibinfo {author} {\bibfnamefont {J.~M.}\
  \bibnamefont {Gambetta}}, \bibinfo {author} {\bibfnamefont {M.}~\bibnamefont
  {Ganzhorn}}, \bibinfo {author} {\bibfnamefont {A.}~\bibnamefont {Kandala}},
  \bibinfo {author} {\bibfnamefont {A.}~\bibnamefont {Mezzacapo}}, \bibinfo
  {author} {\bibfnamefont {P.}~\bibnamefont {M\"{u}ller}}, \bibinfo {author}
  {\bibfnamefont {W.}~\bibnamefont {Riess}}, \bibinfo {author} {\bibfnamefont
  {G.}~\bibnamefont {Salis}}, \bibinfo {author} {\bibfnamefont
  {J.}~\bibnamefont {Smolin}}, \bibinfo {author} {\bibfnamefont
  {I.}~\bibnamefont {Tavernelli}},\ and\ \bibinfo {author} {\bibfnamefont
  {K.}~\bibnamefont {Temme}},\ }\href
  {https://doi.org/10.1088/2058-9565/aab822} {\bibfield  {journal} {\bibinfo
  {journal} {Quantum Science and Technology}\ }\textbf {\bibinfo {volume}
  {3}},\ \bibinfo {pages} {030503} (\bibinfo {year} {2018})}\BibitemShut
  {NoStop}%
\bibitem [{\citenamefont {Cerezo}\ \emph {et~al.}(2021)\citenamefont {Cerezo},
  \citenamefont {Arrasmith}, \citenamefont {Babbush}, \citenamefont {Benjamin},
  \citenamefont {Endo}, \citenamefont {Fujii}, \citenamefont {McClean},
  \citenamefont {Mitarai}, \citenamefont {Yuan}, \citenamefont {Cincio},\ and\
  \citenamefont {Coles}}]{Cerezo2021}%
  \BibitemOpen
  \bibfield  {author} {\bibinfo {author} {\bibfnamefont {M.}~\bibnamefont
  {Cerezo}}, \bibinfo {author} {\bibfnamefont {A.}~\bibnamefont {Arrasmith}},
  \bibinfo {author} {\bibfnamefont {R.}~\bibnamefont {Babbush}}, \bibinfo
  {author} {\bibfnamefont {S.~C.}\ \bibnamefont {Benjamin}}, \bibinfo {author}
  {\bibfnamefont {S.}~\bibnamefont {Endo}}, \bibinfo {author} {\bibfnamefont
  {K.}~\bibnamefont {Fujii}}, \bibinfo {author} {\bibfnamefont {J.~R.}\
  \bibnamefont {McClean}}, \bibinfo {author} {\bibfnamefont {K.}~\bibnamefont
  {Mitarai}}, \bibinfo {author} {\bibfnamefont {X.}~\bibnamefont {Yuan}},
  \bibinfo {author} {\bibfnamefont {L.}~\bibnamefont {Cincio}},\ and\ \bibinfo
  {author} {\bibfnamefont {P.~J.}\ \bibnamefont {Coles}},\ }\href
  {https://doi.org/10.1038/s42254-021-00348-9} {\bibfield  {journal} {\bibinfo
  {journal} {Nature Reviews Physics}\ }\textbf {\bibinfo {volume} {3}},\
  \bibinfo {pages} {625} (\bibinfo {year} {2021})}\BibitemShut {NoStop}%
\bibitem [{\citenamefont {Yuan}\ \emph {et~al.}(2019)\citenamefont {Yuan},
  \citenamefont {Endo}, \citenamefont {Zhao}, \citenamefont {Li},\ and\
  \citenamefont {Benjamin}}]{Yuan2019_vqs}%
  \BibitemOpen
  \bibfield  {author} {\bibinfo {author} {\bibfnamefont {X.}~\bibnamefont
  {Yuan}}, \bibinfo {author} {\bibfnamefont {S.}~\bibnamefont {Endo}}, \bibinfo
  {author} {\bibfnamefont {Q.}~\bibnamefont {Zhao}}, \bibinfo {author}
  {\bibfnamefont {Y.}~\bibnamefont {Li}},\ and\ \bibinfo {author}
  {\bibfnamefont {S.~C.}\ \bibnamefont {Benjamin}},\ }\href
  {https://doi.org/10.22331/q-2019-10-07-191} {\bibfield  {journal} {\bibinfo
  {journal} {Quantum}\ }\textbf {\bibinfo {volume} {3}},\ \bibinfo {pages}
  {191} (\bibinfo {year} {2019})}\BibitemShut {NoStop}%
\bibitem [{\citenamefont {Endo}\ \emph {et~al.}(2020)\citenamefont {Endo},
  \citenamefont {Sun}, \citenamefont {Li}, \citenamefont {Benjamin},\ and\
  \citenamefont {Yuan}}]{Endo2020_general_vqs}%
  \BibitemOpen
  \bibfield  {author} {\bibinfo {author} {\bibfnamefont {S.}~\bibnamefont
  {Endo}}, \bibinfo {author} {\bibfnamefont {J.}~\bibnamefont {Sun}}, \bibinfo
  {author} {\bibfnamefont {Y.}~\bibnamefont {Li}}, \bibinfo {author}
  {\bibfnamefont {S.~C.}\ \bibnamefont {Benjamin}},\ and\ \bibinfo {author}
  {\bibfnamefont {X.}~\bibnamefont {Yuan}},\ }\href
  {https://doi.org/10.1103/PhysRevLett.125.010501} {\bibfield  {journal}
  {\bibinfo  {journal} {Phys. Rev. Lett.}\ }\textbf {\bibinfo {volume} {125}},\
  \bibinfo {pages} {010501} (\bibinfo {year} {2020})}\BibitemShut {NoStop}%
\bibitem [{\citenamefont {Preskill}(2018)}]{Preskill2018}%
  \BibitemOpen
  \bibfield  {author} {\bibinfo {author} {\bibfnamefont {J.}~\bibnamefont
  {Preskill}},\ }\href {https://doi.org/10.22331/q-2018-08-06-79} {\bibfield
  {journal} {\bibinfo  {journal} {{Quantum}}\ }\textbf {\bibinfo {volume}
  {2}},\ \bibinfo {pages} {79} (\bibinfo {year} {2018})}\BibitemShut {NoStop}%
\bibitem [{\citenamefont {Bell}(1964)}]{bell1964epr}%
  \BibitemOpen
  \bibfield  {author} {\bibinfo {author} {\bibfnamefont {J.~S.}\ \bibnamefont
  {Bell}},\ }\href@noop {} {\bibfield  {journal} {\bibinfo  {journal} {Physics
  Physique Fizika}\ }\textbf {\bibinfo {volume} {1}},\ \bibinfo {pages} {195}
  (\bibinfo {year} {1964})}\BibitemShut {NoStop}%
\bibitem [{\citenamefont {Brunner}\ \emph {et~al.}(2014)\citenamefont
  {Brunner}, \citenamefont {Cavalcanti}, \citenamefont {Pironio}, \citenamefont
  {Scarani},\ and\ \citenamefont {Wehner}}]{brunner2014nonlocality}%
  \BibitemOpen
  \bibfield  {author} {\bibinfo {author} {\bibfnamefont {N.}~\bibnamefont
  {Brunner}}, \bibinfo {author} {\bibfnamefont {D.}~\bibnamefont {Cavalcanti}},
  \bibinfo {author} {\bibfnamefont {S.}~\bibnamefont {Pironio}}, \bibinfo
  {author} {\bibfnamefont {V.}~\bibnamefont {Scarani}},\ and\ \bibinfo {author}
  {\bibfnamefont {S.}~\bibnamefont {Wehner}},\ }\href
  {https://doi.org/10.1103/RevModPhys.86.419} {\bibfield  {journal} {\bibinfo
  {journal} {Rev. Mod. Phys.}\ }\textbf {\bibinfo {volume} {86}},\ \bibinfo
  {pages} {419} (\bibinfo {year} {2014})}\BibitemShut {NoStop}%
\bibitem [{\citenamefont {Aspect}\ \emph {et~al.}(1981)\citenamefont {Aspect},
  \citenamefont {Grangier},\ and\ \citenamefont {Roger}}]{aspect1981}%
  \BibitemOpen
  \bibfield  {author} {\bibinfo {author} {\bibfnamefont {A.}~\bibnamefont
  {Aspect}}, \bibinfo {author} {\bibfnamefont {P.}~\bibnamefont {Grangier}},\
  and\ \bibinfo {author} {\bibfnamefont {G.}~\bibnamefont {Roger}},\ }\href
  {https://doi.org/10.1103/PhysRevLett.47.460} {\bibfield  {journal} {\bibinfo
  {journal} {Phys. Rev. Lett.}\ }\textbf {\bibinfo {volume} {47}},\ \bibinfo
  {pages} {460} (\bibinfo {year} {1981})}\BibitemShut {NoStop}%
\bibitem [{\citenamefont {Giustina}\ \emph {et~al.}(2015)\citenamefont
  {Giustina}, \citenamefont {Versteegh}, \citenamefont {Wengerowsky},
  \citenamefont {Handsteiner}, \citenamefont {Hochrainer}, \citenamefont
  {Phelan}, \citenamefont {Steinlechner}, \citenamefont {Kofler}, \citenamefont
  {Larsson}, \citenamefont {Abell\'an}, \citenamefont {Amaya}, \citenamefont
  {Pruneri}, \citenamefont {Mitchell}, \citenamefont {Beyer}, \citenamefont
  {Gerrits}, \citenamefont {Lita}, \citenamefont {Shalm}, \citenamefont {Nam},
  \citenamefont {Scheidl}, \citenamefont {Ursin}, \citenamefont {Wittmann},\
  and\ \citenamefont {Zeilinger}}]{Giustina2015loop_hole_free}%
  \BibitemOpen
  \bibfield  {author} {\bibinfo {author} {\bibfnamefont {M.}~\bibnamefont
  {Giustina}}, \bibinfo {author} {\bibfnamefont {M.~A.~M.}\ \bibnamefont
  {Versteegh}}, \bibinfo {author} {\bibfnamefont {S.}~\bibnamefont
  {Wengerowsky}}, \bibinfo {author} {\bibfnamefont {J.}~\bibnamefont
  {Handsteiner}}, \bibinfo {author} {\bibfnamefont {A.}~\bibnamefont
  {Hochrainer}}, \bibinfo {author} {\bibfnamefont {K.}~\bibnamefont {Phelan}},
  \bibinfo {author} {\bibfnamefont {F.}~\bibnamefont {Steinlechner}}, \bibinfo
  {author} {\bibfnamefont {J.}~\bibnamefont {Kofler}}, \bibinfo {author}
  {\bibfnamefont {J.-A.}\ \bibnamefont {Larsson}}, \bibinfo {author}
  {\bibfnamefont {C.}~\bibnamefont {Abell\'an}}, \bibinfo {author}
  {\bibfnamefont {W.}~\bibnamefont {Amaya}}, \bibinfo {author} {\bibfnamefont
  {V.}~\bibnamefont {Pruneri}}, \bibinfo {author} {\bibfnamefont {M.~W.}\
  \bibnamefont {Mitchell}}, \bibinfo {author} {\bibfnamefont {J.}~\bibnamefont
  {Beyer}}, \bibinfo {author} {\bibfnamefont {T.}~\bibnamefont {Gerrits}},
  \bibinfo {author} {\bibfnamefont {A.~E.}\ \bibnamefont {Lita}}, \bibinfo
  {author} {\bibfnamefont {L.~K.}\ \bibnamefont {Shalm}}, \bibinfo {author}
  {\bibfnamefont {S.~W.}\ \bibnamefont {Nam}}, \bibinfo {author} {\bibfnamefont
  {T.}~\bibnamefont {Scheidl}}, \bibinfo {author} {\bibfnamefont
  {R.}~\bibnamefont {Ursin}}, \bibinfo {author} {\bibfnamefont
  {B.}~\bibnamefont {Wittmann}},\ and\ \bibinfo {author} {\bibfnamefont
  {A.}~\bibnamefont {Zeilinger}},\ }\href
  {https://doi.org/10.1103/PhysRevLett.115.250401} {\bibfield  {journal}
  {\bibinfo  {journal} {Phys. Rev. Lett.}\ }\textbf {\bibinfo {volume} {115}},\
  \bibinfo {pages} {250401} (\bibinfo {year} {2015})}\BibitemShut {NoStop}%
\bibitem [{\citenamefont {Hensen}\ \emph {et~al.}(2015)\citenamefont {Hensen},
  \citenamefont {Bernien}, \citenamefont {Dr{\'{e}}au}, \citenamefont
  {Reiserer}, \citenamefont {Kalb}, \citenamefont {Blok}, \citenamefont
  {Ruitenberg}, \citenamefont {Vermeulen}, \citenamefont {Schouten},
  \citenamefont {Abell{\'{a}}n}, \citenamefont {Amaya}, \citenamefont
  {Pruneri}, \citenamefont {Mitchell}, \citenamefont {Markham}, \citenamefont
  {Twitchen}, \citenamefont {Elkouss}, \citenamefont {Wehner}, \citenamefont
  {Taminiau},\ and\ \citenamefont {Hanson}}]{Hensen2015loop_hole_free}%
  \BibitemOpen
  \bibfield  {author} {\bibinfo {author} {\bibfnamefont {B.}~\bibnamefont
  {Hensen}}, \bibinfo {author} {\bibfnamefont {H.}~\bibnamefont {Bernien}},
  \bibinfo {author} {\bibfnamefont {A.~E.}\ \bibnamefont {Dr{\'{e}}au}},
  \bibinfo {author} {\bibfnamefont {A.}~\bibnamefont {Reiserer}}, \bibinfo
  {author} {\bibfnamefont {N.}~\bibnamefont {Kalb}}, \bibinfo {author}
  {\bibfnamefont {M.~S.}\ \bibnamefont {Blok}}, \bibinfo {author}
  {\bibfnamefont {J.}~\bibnamefont {Ruitenberg}}, \bibinfo {author}
  {\bibfnamefont {R.~F.~L.}\ \bibnamefont {Vermeulen}}, \bibinfo {author}
  {\bibfnamefont {R.~N.}\ \bibnamefont {Schouten}}, \bibinfo {author}
  {\bibfnamefont {C.}~\bibnamefont {Abell{\'{a}}n}}, \bibinfo {author}
  {\bibfnamefont {W.}~\bibnamefont {Amaya}}, \bibinfo {author} {\bibfnamefont
  {V.}~\bibnamefont {Pruneri}}, \bibinfo {author} {\bibfnamefont {M.~W.}\
  \bibnamefont {Mitchell}}, \bibinfo {author} {\bibfnamefont {M.}~\bibnamefont
  {Markham}}, \bibinfo {author} {\bibfnamefont {D.~J.}\ \bibnamefont
  {Twitchen}}, \bibinfo {author} {\bibfnamefont {D.}~\bibnamefont {Elkouss}},
  \bibinfo {author} {\bibfnamefont {S.}~\bibnamefont {Wehner}}, \bibinfo
  {author} {\bibfnamefont {T.~H.}\ \bibnamefont {Taminiau}},\ and\ \bibinfo
  {author} {\bibfnamefont {R.}~\bibnamefont {Hanson}},\ }\href
  {https://doi.org/10.1038/nature15759} {\bibfield  {journal} {\bibinfo
  {journal} {Nature}\ }\textbf {\bibinfo {volume} {526}},\ \bibinfo {pages}
  {682} (\bibinfo {year} {2015})}\BibitemShut {NoStop}%
\bibitem [{\citenamefont {Shalm}\ \emph {et~al.}(2015)\citenamefont {Shalm},
  \citenamefont {Meyer-Scott}, \citenamefont {Christensen}, \citenamefont
  {Bierhorst}, \citenamefont {Wayne}, \citenamefont {Stevens}, \citenamefont
  {Gerrits}, \citenamefont {Glancy}, \citenamefont {Hamel}, \citenamefont
  {Allman}, \citenamefont {Coakley}, \citenamefont {Dyer}, \citenamefont
  {Hodge}, \citenamefont {Lita}, \citenamefont {Verma}, \citenamefont
  {Lambrocco}, \citenamefont {Tortorici}, \citenamefont {Migdall},
  \citenamefont {Zhang}, \citenamefont {Kumor}, \citenamefont {Farr},
  \citenamefont {Marsili}, \citenamefont {Shaw}, \citenamefont {Stern},
  \citenamefont {Abell\'an}, \citenamefont {Amaya}, \citenamefont {Pruneri},
  \citenamefont {Jennewein}, \citenamefont {Mitchell}, \citenamefont {Kwiat},
  \citenamefont {Bienfang}, \citenamefont {Mirin}, \citenamefont {Knill},\ and\
  \citenamefont {Nam}}]{Shalm2015_loophole_free}%
  \BibitemOpen
  \bibfield  {author} {\bibinfo {author} {\bibfnamefont {L.~K.}\ \bibnamefont
  {Shalm}}, \bibinfo {author} {\bibfnamefont {E.}~\bibnamefont {Meyer-Scott}},
  \bibinfo {author} {\bibfnamefont {B.~G.}\ \bibnamefont {Christensen}},
  \bibinfo {author} {\bibfnamefont {P.}~\bibnamefont {Bierhorst}}, \bibinfo
  {author} {\bibfnamefont {M.~A.}\ \bibnamefont {Wayne}}, \bibinfo {author}
  {\bibfnamefont {M.~J.}\ \bibnamefont {Stevens}}, \bibinfo {author}
  {\bibfnamefont {T.}~\bibnamefont {Gerrits}}, \bibinfo {author} {\bibfnamefont
  {S.}~\bibnamefont {Glancy}}, \bibinfo {author} {\bibfnamefont {D.~R.}\
  \bibnamefont {Hamel}}, \bibinfo {author} {\bibfnamefont {M.~S.}\ \bibnamefont
  {Allman}}, \bibinfo {author} {\bibfnamefont {K.~J.}\ \bibnamefont {Coakley}},
  \bibinfo {author} {\bibfnamefont {S.~D.}\ \bibnamefont {Dyer}}, \bibinfo
  {author} {\bibfnamefont {C.}~\bibnamefont {Hodge}}, \bibinfo {author}
  {\bibfnamefont {A.~E.}\ \bibnamefont {Lita}}, \bibinfo {author}
  {\bibfnamefont {V.~B.}\ \bibnamefont {Verma}}, \bibinfo {author}
  {\bibfnamefont {C.}~\bibnamefont {Lambrocco}}, \bibinfo {author}
  {\bibfnamefont {E.}~\bibnamefont {Tortorici}}, \bibinfo {author}
  {\bibfnamefont {A.~L.}\ \bibnamefont {Migdall}}, \bibinfo {author}
  {\bibfnamefont {Y.}~\bibnamefont {Zhang}}, \bibinfo {author} {\bibfnamefont
  {D.~R.}\ \bibnamefont {Kumor}}, \bibinfo {author} {\bibfnamefont {W.~H.}\
  \bibnamefont {Farr}}, \bibinfo {author} {\bibfnamefont {F.}~\bibnamefont
  {Marsili}}, \bibinfo {author} {\bibfnamefont {M.~D.}\ \bibnamefont {Shaw}},
  \bibinfo {author} {\bibfnamefont {J.~A.}\ \bibnamefont {Stern}}, \bibinfo
  {author} {\bibfnamefont {C.}~\bibnamefont {Abell\'an}}, \bibinfo {author}
  {\bibfnamefont {W.}~\bibnamefont {Amaya}}, \bibinfo {author} {\bibfnamefont
  {V.}~\bibnamefont {Pruneri}}, \bibinfo {author} {\bibfnamefont
  {T.}~\bibnamefont {Jennewein}}, \bibinfo {author} {\bibfnamefont {M.~W.}\
  \bibnamefont {Mitchell}}, \bibinfo {author} {\bibfnamefont {P.~G.}\
  \bibnamefont {Kwiat}}, \bibinfo {author} {\bibfnamefont {J.~C.}\ \bibnamefont
  {Bienfang}}, \bibinfo {author} {\bibfnamefont {R.~P.}\ \bibnamefont {Mirin}},
  \bibinfo {author} {\bibfnamefont {E.}~\bibnamefont {Knill}},\ and\ \bibinfo
  {author} {\bibfnamefont {S.~W.}\ \bibnamefont {Nam}},\ }\href
  {https://doi.org/10.1103/PhysRevLett.115.250402} {\bibfield  {journal}
  {\bibinfo  {journal} {Phys. Rev. Lett.}\ }\textbf {\bibinfo {volume} {115}},\
  \bibinfo {pages} {250402} (\bibinfo {year} {2015})}\BibitemShut {NoStop}%
\bibitem [{\citenamefont {Pironio}\ \emph {et~al.}(2010)\citenamefont
  {Pironio}, \citenamefont {Ac{\'\i}n}, \citenamefont {Massar}, \citenamefont
  {de~La~Giroday}, \citenamefont {Matsukevich}, \citenamefont {Maunz},
  \citenamefont {Olmschenk}, \citenamefont {Hayes}, \citenamefont {Luo},
  \citenamefont {Manning} \emph {et~al.}}]{pironio2010random}%
  \BibitemOpen
  \bibfield  {author} {\bibinfo {author} {\bibfnamefont {S.}~\bibnamefont
  {Pironio}}, \bibinfo {author} {\bibfnamefont {A.}~\bibnamefont {Ac{\'\i}n}},
  \bibinfo {author} {\bibfnamefont {S.}~\bibnamefont {Massar}}, \bibinfo
  {author} {\bibfnamefont {A.~B.}\ \bibnamefont {de~La~Giroday}}, \bibinfo
  {author} {\bibfnamefont {D.~N.}\ \bibnamefont {Matsukevich}}, \bibinfo
  {author} {\bibfnamefont {P.}~\bibnamefont {Maunz}}, \bibinfo {author}
  {\bibfnamefont {S.}~\bibnamefont {Olmschenk}}, \bibinfo {author}
  {\bibfnamefont {D.}~\bibnamefont {Hayes}}, \bibinfo {author} {\bibfnamefont
  {L.}~\bibnamefont {Luo}}, \bibinfo {author} {\bibfnamefont {T.~A.}\
  \bibnamefont {Manning}}, \emph {et~al.},\ }\href@noop {} {\bibfield
  {journal} {\bibinfo  {journal} {Nature}\ }\textbf {\bibinfo {volume} {464}},\
  \bibinfo {pages} {1021} (\bibinfo {year} {2010})}\BibitemShut {NoStop}%
\bibitem [{\citenamefont {Gallego}\ \emph {et~al.}(2010)\citenamefont
  {Gallego}, \citenamefont {Brunner}, \citenamefont {Hadley},\ and\
  \citenamefont {Ac\'{\i}n}}]{gallego2010_dim_witnessing}%
  \BibitemOpen
  \bibfield  {author} {\bibinfo {author} {\bibfnamefont {R.}~\bibnamefont
  {Gallego}}, \bibinfo {author} {\bibfnamefont {N.}~\bibnamefont {Brunner}},
  \bibinfo {author} {\bibfnamefont {C.}~\bibnamefont {Hadley}},\ and\ \bibinfo
  {author} {\bibfnamefont {A.}~\bibnamefont {Ac\'{\i}n}},\ }\href
  {https://doi.org/10.1103/PhysRevLett.105.230501} {\bibfield  {journal}
  {\bibinfo  {journal} {Phys. Rev. Lett.}\ }\textbf {\bibinfo {volume} {105}},\
  \bibinfo {pages} {230501} (\bibinfo {year} {2010})}\BibitemShut {NoStop}%
\bibitem [{\citenamefont {Renou}\ \emph {et~al.}(2018)\citenamefont {Renou},
  \citenamefont {Kaniewski},\ and\ \citenamefont
  {Brunner}}]{Renou2018_self-test_network}%
  \BibitemOpen
  \bibfield  {author} {\bibinfo {author} {\bibfnamefont {M.~O.}\ \bibnamefont
  {Renou}}, \bibinfo {author} {\bibfnamefont {J.~m.~k.}\ \bibnamefont
  {Kaniewski}},\ and\ \bibinfo {author} {\bibfnamefont {N.}~\bibnamefont
  {Brunner}},\ }\href {https://doi.org/10.1103/PhysRevLett.121.250507}
  {\bibfield  {journal} {\bibinfo  {journal} {Phys. Rev. Lett.}\ }\textbf
  {\bibinfo {volume} {121}},\ \bibinfo {pages} {250507} (\bibinfo {year}
  {2018})}\BibitemShut {NoStop}%
\bibitem [{\citenamefont {Bancal}\ \emph {et~al.}(2018)\citenamefont {Bancal},
  \citenamefont {Sangouard},\ and\ \citenamefont
  {Sekatski}}]{bancal2018_self-test_network}%
  \BibitemOpen
  \bibfield  {author} {\bibinfo {author} {\bibfnamefont {J.-D.}\ \bibnamefont
  {Bancal}}, \bibinfo {author} {\bibfnamefont {N.}~\bibnamefont {Sangouard}},\
  and\ \bibinfo {author} {\bibfnamefont {P.}~\bibnamefont {Sekatski}},\ }\href
  {https://doi.org/10.1103/PhysRevLett.121.250506} {\bibfield  {journal}
  {\bibinfo  {journal} {Phys. Rev. Lett.}\ }\textbf {\bibinfo {volume} {121}},\
  \bibinfo {pages} {250506} (\bibinfo {year} {2018})}\BibitemShut {NoStop}%
\bibitem [{\citenamefont {{\v{S}}upi{\'{c}}}\ and\ \citenamefont
  {Bowles}(2020)}]{supic2020}%
  \BibitemOpen
  \bibfield  {author} {\bibinfo {author} {\bibfnamefont {I.}~\bibnamefont
  {{\v{S}}upi{\'{c}}}}\ and\ \bibinfo {author} {\bibfnamefont {J.}~\bibnamefont
  {Bowles}},\ }\href {https://doi.org/10.22331/q-2020-09-30-337} {\bibfield
  {journal} {\bibinfo  {journal} {Quantum}\ }\textbf {\bibinfo {volume} {4}},\
  \bibinfo {pages} {337} (\bibinfo {year} {2020})}\BibitemShut {NoStop}%
\bibitem [{\citenamefont {\ifmmode \check{S}\else
  \v{S}\fi{}upi\ifmmode~\acute{c}\else \'{c}\fi{}}\ \emph
  {et~al.}(2022)\citenamefont {\ifmmode \check{S}\else
  \v{S}\fi{}upi\ifmmode~\acute{c}\else \'{c}\fi{}}, \citenamefont {Bancal},
  \citenamefont {Cai},\ and\ \citenamefont
  {Brunner}}]{Supic2022_network_self-testing}%
  \BibitemOpen
  \bibfield  {author} {\bibinfo {author} {\bibfnamefont {I.}~\bibnamefont
  {\ifmmode \check{S}\else \v{S}\fi{}upi\ifmmode~\acute{c}\else \'{c}\fi{}}},
  \bibinfo {author} {\bibfnamefont {J.-D.}\ \bibnamefont {Bancal}}, \bibinfo
  {author} {\bibfnamefont {Y.}~\bibnamefont {Cai}},\ and\ \bibinfo {author}
  {\bibfnamefont {N.}~\bibnamefont {Brunner}},\ }\href
  {https://doi.org/10.1103/PhysRevA.105.022206} {\bibfield  {journal} {\bibinfo
   {journal} {Phys. Rev. A}\ }\textbf {\bibinfo {volume} {105}},\ \bibinfo
  {pages} {022206} (\bibinfo {year} {2022})}\BibitemShut {NoStop}%
\bibitem [{\citenamefont {Ekert}(1991)}]{ekert1991}%
  \BibitemOpen
  \bibfield  {author} {\bibinfo {author} {\bibfnamefont {A.~K.}\ \bibnamefont
  {Ekert}},\ }\href {https://doi.org/10.1103/PhysRevLett.67.661} {\bibfield
  {journal} {\bibinfo  {journal} {Phys. Rev. Lett.}\ }\textbf {\bibinfo
  {volume} {67}},\ \bibinfo {pages} {661} (\bibinfo {year} {1991})}\BibitemShut
  {NoStop}%
\bibitem [{\citenamefont {Barrett}\ \emph {et~al.}(2005)\citenamefont
  {Barrett}, \citenamefont {Hardy},\ and\ \citenamefont {Kent}}]{Barrett2005}%
  \BibitemOpen
  \bibfield  {author} {\bibinfo {author} {\bibfnamefont {J.}~\bibnamefont
  {Barrett}}, \bibinfo {author} {\bibfnamefont {L.}~\bibnamefont {Hardy}},\
  and\ \bibinfo {author} {\bibfnamefont {A.}~\bibnamefont {Kent}},\ }\href
  {https://doi.org/10.1103/PhysRevLett.95.010503} {\bibfield  {journal}
  {\bibinfo  {journal} {Phys. Rev. Lett.}\ }\textbf {\bibinfo {volume} {95}},\
  \bibinfo {pages} {010503} (\bibinfo {year} {2005})}\BibitemShut {NoStop}%
\bibitem [{\citenamefont {Ac\'{\i}n}\ \emph {et~al.}(2006)\citenamefont
  {Ac\'{\i}n}, \citenamefont {Gisin},\ and\ \citenamefont
  {Masanes}}]{acin2006_di}%
  \BibitemOpen
  \bibfield  {author} {\bibinfo {author} {\bibfnamefont {A.}~\bibnamefont
  {Ac\'{\i}n}}, \bibinfo {author} {\bibfnamefont {N.}~\bibnamefont {Gisin}},\
  and\ \bibinfo {author} {\bibfnamefont {L.}~\bibnamefont {Masanes}},\ }\href
  {https://doi.org/10.1103/PhysRevLett.97.120405} {\bibfield  {journal}
  {\bibinfo  {journal} {Phys. Rev. Lett.}\ }\textbf {\bibinfo {volume} {97}},\
  \bibinfo {pages} {120405} (\bibinfo {year} {2006})}\BibitemShut {NoStop}%
\bibitem [{\citenamefont {Ac\'{\i}n}\ \emph {et~al.}(2007)\citenamefont
  {Ac\'{\i}n}, \citenamefont {Brunner}, \citenamefont {Gisin}, \citenamefont
  {Massar}, \citenamefont {Pironio},\ and\ \citenamefont
  {Scarani}}]{acin2007_di}%
  \BibitemOpen
  \bibfield  {author} {\bibinfo {author} {\bibfnamefont {A.}~\bibnamefont
  {Ac\'{\i}n}}, \bibinfo {author} {\bibfnamefont {N.}~\bibnamefont {Brunner}},
  \bibinfo {author} {\bibfnamefont {N.}~\bibnamefont {Gisin}}, \bibinfo
  {author} {\bibfnamefont {S.}~\bibnamefont {Massar}}, \bibinfo {author}
  {\bibfnamefont {S.}~\bibnamefont {Pironio}},\ and\ \bibinfo {author}
  {\bibfnamefont {V.}~\bibnamefont {Scarani}},\ }\href
  {https://doi.org/10.1103/PhysRevLett.98.230501} {\bibfield  {journal}
  {\bibinfo  {journal} {Phys. Rev. Lett.}\ }\textbf {\bibinfo {volume} {98}},\
  \bibinfo {pages} {230501} (\bibinfo {year} {2007})}\BibitemShut {NoStop}%
\bibitem [{\citenamefont {Liu}\ \emph {et~al.}(2018)\citenamefont {Liu},
  \citenamefont {Yuan}, \citenamefont {Li}, \citenamefont {Zhang},
  \citenamefont {Zhao}, \citenamefont {Zhong}, \citenamefont {Cao},
  \citenamefont {Li}, \citenamefont {Chen}, \citenamefont {Li}, \citenamefont
  {Peng}, \citenamefont {Chen}, \citenamefont {Peng}, \citenamefont {Shi},
  \citenamefont {Wang}, \citenamefont {You}, \citenamefont {Ma}, \citenamefont
  {Fan}, \citenamefont {Zhang},\ and\ \citenamefont {Pan}}]{Liu2018}%
  \BibitemOpen
  \bibfield  {author} {\bibinfo {author} {\bibfnamefont {Y.}~\bibnamefont
  {Liu}}, \bibinfo {author} {\bibfnamefont {X.}~\bibnamefont {Yuan}}, \bibinfo
  {author} {\bibfnamefont {M.-H.}\ \bibnamefont {Li}}, \bibinfo {author}
  {\bibfnamefont {W.}~\bibnamefont {Zhang}}, \bibinfo {author} {\bibfnamefont
  {Q.}~\bibnamefont {Zhao}}, \bibinfo {author} {\bibfnamefont {J.}~\bibnamefont
  {Zhong}}, \bibinfo {author} {\bibfnamefont {Y.}~\bibnamefont {Cao}}, \bibinfo
  {author} {\bibfnamefont {Y.-H.}\ \bibnamefont {Li}}, \bibinfo {author}
  {\bibfnamefont {L.-K.}\ \bibnamefont {Chen}}, \bibinfo {author}
  {\bibfnamefont {H.}~\bibnamefont {Li}}, \bibinfo {author} {\bibfnamefont
  {T.}~\bibnamefont {Peng}}, \bibinfo {author} {\bibfnamefont {Y.-A.}\
  \bibnamefont {Chen}}, \bibinfo {author} {\bibfnamefont {C.-Z.}\ \bibnamefont
  {Peng}}, \bibinfo {author} {\bibfnamefont {S.-C.}\ \bibnamefont {Shi}},
  \bibinfo {author} {\bibfnamefont {Z.}~\bibnamefont {Wang}}, \bibinfo {author}
  {\bibfnamefont {L.}~\bibnamefont {You}}, \bibinfo {author} {\bibfnamefont
  {X.}~\bibnamefont {Ma}}, \bibinfo {author} {\bibfnamefont {J.}~\bibnamefont
  {Fan}}, \bibinfo {author} {\bibfnamefont {Q.}~\bibnamefont {Zhang}},\ and\
  \bibinfo {author} {\bibfnamefont {J.-W.}\ \bibnamefont {Pan}},\ }\href
  {https://doi.org/10.1103/PhysRevLett.120.010503} {\bibfield  {journal}
  {\bibinfo  {journal} {Phys. Rev. Lett.}\ }\textbf {\bibinfo {volume} {120}},\
  \bibinfo {pages} {010503} (\bibinfo {year} {2018})}\BibitemShut {NoStop}%
\bibitem [{\citenamefont {Lee}\ and\ \citenamefont
  {Hoban}(2018{\natexlab{b}})}]{Lee2018}%
  \BibitemOpen
  \bibfield  {author} {\bibinfo {author} {\bibfnamefont {C.~M.}\ \bibnamefont
  {Lee}}\ and\ \bibinfo {author} {\bibfnamefont {M.~J.}\ \bibnamefont
  {Hoban}},\ }\href {https://doi.org/10.1103/PhysRevLett.120.020504} {\bibfield
   {journal} {\bibinfo  {journal} {Phys. Rev. Lett.}\ }\textbf {\bibinfo
  {volume} {120}},\ \bibinfo {pages} {020504} (\bibinfo {year}
  {2018}{\natexlab{b}})}\BibitemShut {NoStop}%
\bibitem [{\citenamefont {Vazirani}\ and\ \citenamefont
  {Vidick}(2019)}]{vazirani2019diqkd}%
  \BibitemOpen
  \bibfield  {author} {\bibinfo {author} {\bibfnamefont {U.}~\bibnamefont
  {Vazirani}}\ and\ \bibinfo {author} {\bibfnamefont {T.}~\bibnamefont
  {Vidick}},\ }\href {https://doi.org/https://doi.org/10.1145/3310974}
  {\bibfield  {journal} {\bibinfo  {journal} {Communications of the ACM}\
  }\textbf {\bibinfo {volume} {62}},\ \bibinfo {pages} {133} (\bibinfo {year}
  {2019})}\BibitemShut {NoStop}%
\bibitem [{\citenamefont {Deng}(2018)}]{deng2018machine}%
  \BibitemOpen
  \bibfield  {author} {\bibinfo {author} {\bibfnamefont {D.-L.}\ \bibnamefont
  {Deng}},\ }\href {https://doi.org/10.1103/PhysRevLett.120.240402} {\bibfield
  {journal} {\bibinfo  {journal} {Phys. Rev. Lett.}\ }\textbf {\bibinfo
  {volume} {120}},\ \bibinfo {pages} {240402} (\bibinfo {year}
  {2018})}\BibitemShut {NoStop}%
\bibitem [{\citenamefont {Canabarro}\ \emph {et~al.}(2019)\citenamefont
  {Canabarro}, \citenamefont {Brito},\ and\ \citenamefont
  {Chaves}}]{canabarro2019ml_nonlocality}%
  \BibitemOpen
  \bibfield  {author} {\bibinfo {author} {\bibfnamefont {A.}~\bibnamefont
  {Canabarro}}, \bibinfo {author} {\bibfnamefont {S.}~\bibnamefont {Brito}},\
  and\ \bibinfo {author} {\bibfnamefont {R.}~\bibnamefont {Chaves}},\ }\href
  {https://doi.org/10.1103/PhysRevLett.122.200401} {\bibfield  {journal}
  {\bibinfo  {journal} {Phys. Rev. Lett.}\ }\textbf {\bibinfo {volume} {122}},\
  \bibinfo {pages} {200401} (\bibinfo {year} {2019})}\BibitemShut {NoStop}%
\bibitem [{\citenamefont {Bharti}\ \emph {et~al.}(2019)\citenamefont {Bharti},
  \citenamefont {Haug}, \citenamefont {Vedral},\ and\ \citenamefont
  {Kwek}}]{bharti2019teach}%
  \BibitemOpen
  \bibfield  {author} {\bibinfo {author} {\bibfnamefont {K.}~\bibnamefont
  {Bharti}}, \bibinfo {author} {\bibfnamefont {T.}~\bibnamefont {Haug}},
  \bibinfo {author} {\bibfnamefont {V.}~\bibnamefont {Vedral}},\ and\ \bibinfo
  {author} {\bibfnamefont {L.-C.}\ \bibnamefont {Kwek}},\ }\href@noop {}
  {\bibfield  {journal} {\bibinfo  {journal} {arXiv preprint arXiv:1912.10783}\
  } (\bibinfo {year} {2019})}\BibitemShut {NoStop}%
\bibitem [{\citenamefont {Suprano}\ \emph {et~al.}(2021)\citenamefont
  {Suprano}, \citenamefont {Zia}, \citenamefont {Polino}, \citenamefont
  {Giordani}, \citenamefont {Innocenti}, \citenamefont {Ferraro}, \citenamefont
  {Paternostro}, \citenamefont {Spagnolo},\ and\ \citenamefont
  {Sciarrino}}]{Suprano2021}%
  \BibitemOpen
  \bibfield  {author} {\bibinfo {author} {\bibfnamefont {A.}~\bibnamefont
  {Suprano}}, \bibinfo {author} {\bibfnamefont {D.}~\bibnamefont {Zia}},
  \bibinfo {author} {\bibfnamefont {E.}~\bibnamefont {Polino}}, \bibinfo
  {author} {\bibfnamefont {T.}~\bibnamefont {Giordani}}, \bibinfo {author}
  {\bibfnamefont {L.}~\bibnamefont {Innocenti}}, \bibinfo {author}
  {\bibfnamefont {A.}~\bibnamefont {Ferraro}}, \bibinfo {author} {\bibfnamefont
  {M.}~\bibnamefont {Paternostro}}, \bibinfo {author} {\bibfnamefont
  {N.}~\bibnamefont {Spagnolo}},\ and\ \bibinfo {author} {\bibfnamefont
  {F.}~\bibnamefont {Sciarrino}},\ }\bibfield  {journal} {\bibinfo  {journal}
  {Advanced Photonics}\ }\textbf {\bibinfo {volume} {3}},\ \href
  {https://doi.org/10.1117/1.ap.3.6.066002} {10.1117/1.ap.3.6.066002} (\bibinfo
  {year} {2021})\BibitemShut {NoStop}%
\bibitem [{\citenamefont {Poderini}\ \emph {et~al.}(2022)\citenamefont
  {Poderini}, \citenamefont {Polino}, \citenamefont {Rodari}, \citenamefont
  {Suprano}, \citenamefont {Chaves},\ and\ \citenamefont
  {Sciarrino}}]{Poderini2022_black-box}%
  \BibitemOpen
  \bibfield  {author} {\bibinfo {author} {\bibfnamefont {D.}~\bibnamefont
  {Poderini}}, \bibinfo {author} {\bibfnamefont {E.}~\bibnamefont {Polino}},
  \bibinfo {author} {\bibfnamefont {G.}~\bibnamefont {Rodari}}, \bibinfo
  {author} {\bibfnamefont {A.}~\bibnamefont {Suprano}}, \bibinfo {author}
  {\bibfnamefont {R.}~\bibnamefont {Chaves}},\ and\ \bibinfo {author}
  {\bibfnamefont {F.}~\bibnamefont {Sciarrino}},\ }\href
  {https://doi.org/10.1103/PhysRevResearch.4.013159} {\bibfield  {journal}
  {\bibinfo  {journal} {Phys. Rev. Research}\ }\textbf {\bibinfo {volume}
  {4}},\ \bibinfo {pages} {013159} (\bibinfo {year} {2022})}\BibitemShut
  {NoStop}%
\bibitem [{\citenamefont {Branciard}\ \emph {et~al.}(2010)\citenamefont
  {Branciard}, \citenamefont {Gisin},\ and\ \citenamefont
  {Pironio}}]{Branciard_2010_bilocal_correlations}%
  \BibitemOpen
  \bibfield  {author} {\bibinfo {author} {\bibfnamefont {C.}~\bibnamefont
  {Branciard}}, \bibinfo {author} {\bibfnamefont {N.}~\bibnamefont {Gisin}},\
  and\ \bibinfo {author} {\bibfnamefont {S.}~\bibnamefont {Pironio}},\ }\href
  {https://doi.org/10.1103/PhysRevLett.104.170401} {\bibfield  {journal}
  {\bibinfo  {journal} {Phys. Rev. Lett.}\ }\textbf {\bibinfo {volume} {104}},\
  \bibinfo {pages} {170401} (\bibinfo {year} {2010})}\BibitemShut {NoStop}%
\bibitem [{\citenamefont {Branciard}\ \emph {et~al.}(2012)\citenamefont
  {Branciard}, \citenamefont {Rosset}, \citenamefont {Gisin},\ and\
  \citenamefont {Pironio}}]{Branciard_2012_bilocal_v_nonbilocal}%
  \BibitemOpen
  \bibfield  {author} {\bibinfo {author} {\bibfnamefont {C.}~\bibnamefont
  {Branciard}}, \bibinfo {author} {\bibfnamefont {D.}~\bibnamefont {Rosset}},
  \bibinfo {author} {\bibfnamefont {N.}~\bibnamefont {Gisin}},\ and\ \bibinfo
  {author} {\bibfnamefont {S.}~\bibnamefont {Pironio}},\ }\href
  {https://doi.org/10.1103/PhysRevA.85.032119} {\bibfield  {journal} {\bibinfo
  {journal} {Phys. Rev. A}\ }\textbf {\bibinfo {volume} {85}},\ \bibinfo
  {pages} {032119} (\bibinfo {year} {2012})}\BibitemShut {NoStop}%
\bibitem [{\citenamefont {Tavakoli}\ \emph {et~al.}(2014)\citenamefont
  {Tavakoli}, \citenamefont {Skrzypczyk}, \citenamefont {Cavalcanti},\ and\
  \citenamefont {Ac\'{\i}n}}]{Tavakoli_2014_star}%
  \BibitemOpen
  \bibfield  {author} {\bibinfo {author} {\bibfnamefont {A.}~\bibnamefont
  {Tavakoli}}, \bibinfo {author} {\bibfnamefont {P.}~\bibnamefont
  {Skrzypczyk}}, \bibinfo {author} {\bibfnamefont {D.}~\bibnamefont
  {Cavalcanti}},\ and\ \bibinfo {author} {\bibfnamefont {A.}~\bibnamefont
  {Ac\'{\i}n}},\ }\href {https://doi.org/10.1103/PhysRevA.90.062109} {\bibfield
   {journal} {\bibinfo  {journal} {Phys. Rev. A}\ }\textbf {\bibinfo {volume}
  {90}},\ \bibinfo {pages} {062109} (\bibinfo {year} {2014})}\BibitemShut
  {NoStop}%
\bibitem [{\citenamefont {Mukherjee}\ \emph {et~al.}(2015)\citenamefont
  {Mukherjee}, \citenamefont {Paul},\ and\ \citenamefont
  {Sarkar}}]{Mukherjee2015chain}%
  \BibitemOpen
  \bibfield  {author} {\bibinfo {author} {\bibfnamefont {K.}~\bibnamefont
  {Mukherjee}}, \bibinfo {author} {\bibfnamefont {B.}~\bibnamefont {Paul}},\
  and\ \bibinfo {author} {\bibfnamefont {D.}~\bibnamefont {Sarkar}},\ }\href
  {https://doi.org/10.1007/s11128-015-0971-7} {\bibfield  {journal} {\bibinfo
  {journal} {Quantum Information Processing}\ }\textbf {\bibinfo {volume}
  {14}},\ \bibinfo {pages} {2025} (\bibinfo {year} {2015})}\BibitemShut
  {NoStop}%
\bibitem [{\citenamefont {Rosset}\ \emph {et~al.}(2016)\citenamefont {Rosset},
  \citenamefont {Branciard}, \citenamefont {Barnea}, \citenamefont {P\"utz},
  \citenamefont {Brunner},\ and\ \citenamefont
  {Gisin}}]{Rosset_2016_nonlinear_bell_inequalities}%
  \BibitemOpen
  \bibfield  {author} {\bibinfo {author} {\bibfnamefont {D.}~\bibnamefont
  {Rosset}}, \bibinfo {author} {\bibfnamefont {C.}~\bibnamefont {Branciard}},
  \bibinfo {author} {\bibfnamefont {T.~J.}\ \bibnamefont {Barnea}}, \bibinfo
  {author} {\bibfnamefont {G.}~\bibnamefont {P\"utz}}, \bibinfo {author}
  {\bibfnamefont {N.}~\bibnamefont {Brunner}},\ and\ \bibinfo {author}
  {\bibfnamefont {N.}~\bibnamefont {Gisin}},\ }\href
  {https://doi.org/10.1103/PhysRevLett.116.010403} {\bibfield  {journal}
  {\bibinfo  {journal} {Phys. Rev. Lett.}\ }\textbf {\bibinfo {volume} {116}},\
  \bibinfo {pages} {010403} (\bibinfo {year} {2016})}\BibitemShut {NoStop}%
\bibitem [{\citenamefont {Tavakoli}(2016)}]{Tavakoli2016tree}%
  \BibitemOpen
  \bibfield  {author} {\bibinfo {author} {\bibfnamefont {A.}~\bibnamefont
  {Tavakoli}},\ }\href {https://doi.org/10.1103/PhysRevA.93.030101} {\bibfield
  {journal} {\bibinfo  {journal} {Phys. Rev. A}\ }\textbf {\bibinfo {volume}
  {93}},\ \bibinfo {pages} {030101} (\bibinfo {year} {2016})}\BibitemShut
  {NoStop}%
\bibitem [{\citenamefont {Yang}\ \emph {et~al.}(2021)\citenamefont {Yang},
  \citenamefont {Qi},\ and\ \citenamefont {Hou}}]{Yang2021_tree_network}%
  \BibitemOpen
  \bibfield  {author} {\bibinfo {author} {\bibfnamefont {L.}~\bibnamefont
  {Yang}}, \bibinfo {author} {\bibfnamefont {X.}~\bibnamefont {Qi}},\ and\
  \bibinfo {author} {\bibfnamefont {J.}~\bibnamefont {Hou}},\ }\href
  {https://doi.org/10.1103/PhysRevA.104.042405} {\bibfield  {journal} {\bibinfo
   {journal} {Phys. Rev. A}\ }\textbf {\bibinfo {volume} {104}},\ \bibinfo
  {pages} {042405} (\bibinfo {year} {2021})}\BibitemShut {NoStop}%
\bibitem [{\citenamefont {Tavakoli}\ \emph {et~al.}(2021)\citenamefont
  {Tavakoli}, \citenamefont {Pozas-Kerstjens}, \citenamefont {Luo},\ and\
  \citenamefont {Renou}}]{tavakoli2021network_nonlocality}%
  \BibitemOpen
  \bibfield  {author} {\bibinfo {author} {\bibfnamefont {A.}~\bibnamefont
  {Tavakoli}}, \bibinfo {author} {\bibfnamefont {A.}~\bibnamefont
  {Pozas-Kerstjens}}, \bibinfo {author} {\bibfnamefont {M.-X.}\ \bibnamefont
  {Luo}},\ and\ \bibinfo {author} {\bibfnamefont {M.-O.}\ \bibnamefont
  {Renou}},\ }\href@noop {} {\bibfield  {journal} {\bibinfo  {journal} {arXiv
  preprint arXiv:2104.10700}\ } (\bibinfo {year} {2021})}\BibitemShut {NoStop}%
\bibitem [{\citenamefont {Carvacho}\ \emph {et~al.}(2017)\citenamefont
  {Carvacho}, \citenamefont {Andreoli}, \citenamefont {Santodonato},
  \citenamefont {Bentivegna}, \citenamefont {Chaves},\ and\ \citenamefont
  {Sciarrino}}]{carvacho2017experimental_bilocal}%
  \BibitemOpen
  \bibfield  {author} {\bibinfo {author} {\bibfnamefont {G.}~\bibnamefont
  {Carvacho}}, \bibinfo {author} {\bibfnamefont {F.}~\bibnamefont {Andreoli}},
  \bibinfo {author} {\bibfnamefont {L.}~\bibnamefont {Santodonato}}, \bibinfo
  {author} {\bibfnamefont {M.}~\bibnamefont {Bentivegna}}, \bibinfo {author}
  {\bibfnamefont {R.}~\bibnamefont {Chaves}},\ and\ \bibinfo {author}
  {\bibfnamefont {F.}~\bibnamefont {Sciarrino}},\ }\href
  {https://doi.org/https://doi.org/10.1038/ncomms14775} {\bibfield  {journal}
  {\bibinfo  {journal} {Nature communications}\ }\textbf {\bibinfo {volume}
  {8}},\ \bibinfo {pages} {1} (\bibinfo {year} {2017})}\BibitemShut {NoStop}%
\bibitem [{\citenamefont {Saunders}\ \emph {et~al.}(2017)\citenamefont
  {Saunders}, \citenamefont {Bennet}, \citenamefont {Branciard},\ and\
  \citenamefont {Pryde}}]{saunders2017_bilocal_expt}%
  \BibitemOpen
  \bibfield  {author} {\bibinfo {author} {\bibfnamefont {D.~J.}\ \bibnamefont
  {Saunders}}, \bibinfo {author} {\bibfnamefont {A.~J.}\ \bibnamefont
  {Bennet}}, \bibinfo {author} {\bibfnamefont {C.}~\bibnamefont {Branciard}},\
  and\ \bibinfo {author} {\bibfnamefont {G.~J.}\ \bibnamefont {Pryde}},\ }\href
  {https://doi.org/10.1126/sciadv.1602743} {\bibfield  {journal} {\bibinfo
  {journal} {Science Advances}\ }\textbf {\bibinfo {volume} {3}},\ \bibinfo
  {pages} {e1602743} (\bibinfo {year} {2017})}\BibitemShut {NoStop}%
\bibitem [{\citenamefont {Andreoli}\ \emph
  {et~al.}(2017{\natexlab{a}})\citenamefont {Andreoli}, \citenamefont
  {Carvacho}, \citenamefont {Santodonato}, \citenamefont {Bentivegna},
  \citenamefont {Chaves},\ and\ \citenamefont
  {Sciarrino}}]{andreoli2017_bilocal_expt}%
  \BibitemOpen
  \bibfield  {author} {\bibinfo {author} {\bibfnamefont {F.}~\bibnamefont
  {Andreoli}}, \bibinfo {author} {\bibfnamefont {G.}~\bibnamefont {Carvacho}},
  \bibinfo {author} {\bibfnamefont {L.}~\bibnamefont {Santodonato}}, \bibinfo
  {author} {\bibfnamefont {M.}~\bibnamefont {Bentivegna}}, \bibinfo {author}
  {\bibfnamefont {R.}~\bibnamefont {Chaves}},\ and\ \bibinfo {author}
  {\bibfnamefont {F.}~\bibnamefont {Sciarrino}},\ }\href
  {https://doi.org/10.1103/PhysRevA.95.062315} {\bibfield  {journal} {\bibinfo
  {journal} {Phys. Rev. A}\ }\textbf {\bibinfo {volume} {95}},\ \bibinfo
  {pages} {062315} (\bibinfo {year} {2017}{\natexlab{a}})}\BibitemShut
  {NoStop}%
\bibitem [{\citenamefont {Sun}\ \emph {et~al.}(2019)\citenamefont {Sun},
  \citenamefont {Jiang}, \citenamefont {Bai}, \citenamefont {Zhang},
  \citenamefont {Li}, \citenamefont {Jiang}, \citenamefont {Zhang},
  \citenamefont {You}, \citenamefont {Chen}, \citenamefont {Wang} \emph
  {et~al.}}]{sun2019experimental_bilocal}%
  \BibitemOpen
  \bibfield  {author} {\bibinfo {author} {\bibfnamefont {Q.-C.}\ \bibnamefont
  {Sun}}, \bibinfo {author} {\bibfnamefont {Y.-F.}\ \bibnamefont {Jiang}},
  \bibinfo {author} {\bibfnamefont {B.}~\bibnamefont {Bai}}, \bibinfo {author}
  {\bibfnamefont {W.}~\bibnamefont {Zhang}}, \bibinfo {author} {\bibfnamefont
  {H.}~\bibnamefont {Li}}, \bibinfo {author} {\bibfnamefont {X.}~\bibnamefont
  {Jiang}}, \bibinfo {author} {\bibfnamefont {J.}~\bibnamefont {Zhang}},
  \bibinfo {author} {\bibfnamefont {L.}~\bibnamefont {You}}, \bibinfo {author}
  {\bibfnamefont {X.}~\bibnamefont {Chen}}, \bibinfo {author} {\bibfnamefont
  {Z.}~\bibnamefont {Wang}}, \emph {et~al.},\ }\href
  {https://doi.org/https://doi.org/10.1038/s41566-019-0502-7} {\bibfield
  {journal} {\bibinfo  {journal} {Nature Photonics}\ }\textbf {\bibinfo
  {volume} {13}},\ \bibinfo {pages} {687} (\bibinfo {year} {2019})}\BibitemShut
  {NoStop}%
\bibitem [{\citenamefont {Poderini}\ \emph {et~al.}(2020)\citenamefont
  {Poderini}, \citenamefont {Agresti}, \citenamefont {Marchese}, \citenamefont
  {Polino}, \citenamefont {Giordani}, \citenamefont {Suprano}, \citenamefont
  {Valeri}, \citenamefont {Milani}, \citenamefont {Spagnolo}, \citenamefont
  {Carvacho} \emph {et~al.}}]{poderini2020experimental}%
  \BibitemOpen
  \bibfield  {author} {\bibinfo {author} {\bibfnamefont {D.}~\bibnamefont
  {Poderini}}, \bibinfo {author} {\bibfnamefont {I.}~\bibnamefont {Agresti}},
  \bibinfo {author} {\bibfnamefont {G.}~\bibnamefont {Marchese}}, \bibinfo
  {author} {\bibfnamefont {E.}~\bibnamefont {Polino}}, \bibinfo {author}
  {\bibfnamefont {T.}~\bibnamefont {Giordani}}, \bibinfo {author}
  {\bibfnamefont {A.}~\bibnamefont {Suprano}}, \bibinfo {author} {\bibfnamefont
  {M.}~\bibnamefont {Valeri}}, \bibinfo {author} {\bibfnamefont
  {G.}~\bibnamefont {Milani}}, \bibinfo {author} {\bibfnamefont
  {N.}~\bibnamefont {Spagnolo}}, \bibinfo {author} {\bibfnamefont
  {G.}~\bibnamefont {Carvacho}}, \emph {et~al.},\ }\href
  {https://doi.org/https://doi.org/10.1038/s41467-020-16189-6} {\bibfield
  {journal} {\bibinfo  {journal} {Nature communications}\ }\textbf {\bibinfo
  {volume} {11}},\ \bibinfo {pages} {1} (\bibinfo {year} {2020})}\BibitemShut
  {NoStop}%
\bibitem [{\citenamefont {Cabello}\ \emph {et~al.}(2005)\citenamefont
  {Cabello}, \citenamefont {Feito},\ and\ \citenamefont
  {Lamas-Linares}}]{cabello2005_colored_noise}%
  \BibitemOpen
  \bibfield  {author} {\bibinfo {author} {\bibfnamefont {A.}~\bibnamefont
  {Cabello}}, \bibinfo {author} {\bibfnamefont {A.}~\bibnamefont {Feito}},\
  and\ \bibinfo {author} {\bibfnamefont {A.}~\bibnamefont {Lamas-Linares}},\
  }\href {https://doi.org/10.1103/PhysRevA.72.052112} {\bibfield  {journal}
  {\bibinfo  {journal} {Phys. Rev. A}\ }\textbf {\bibinfo {volume} {72}},\
  \bibinfo {pages} {052112} (\bibinfo {year} {2005})}\BibitemShut {NoStop}%
\bibitem [{\citenamefont {Pal}\ and\ \citenamefont {Ghosh}(2015)}]{Pal2015}%
  \BibitemOpen
  \bibfield  {author} {\bibinfo {author} {\bibfnamefont {R.}~\bibnamefont
  {Pal}}\ and\ \bibinfo {author} {\bibfnamefont {S.}~\bibnamefont {Ghosh}},\
  }\href {https://doi.org/10.1088/1751-8113/48/15/155302} {\bibfield  {journal}
  {\bibinfo  {journal} {Journal of Physics A: Mathematical and Theoretical}\
  }\textbf {\bibinfo {volume} {48}},\ \bibinfo {pages} {155302} (\bibinfo
  {year} {2015})}\BibitemShut {NoStop}%
\bibitem [{\citenamefont {Zhang}\ \emph {et~al.}(2020)\citenamefont {Zhang},
  \citenamefont {Bravo}, \citenamefont {Lorenz},\ and\ \citenamefont
  {Chitambar}}]{Zhang2020}%
  \BibitemOpen
  \bibfield  {author} {\bibinfo {author} {\bibfnamefont {Y.}~\bibnamefont
  {Zhang}}, \bibinfo {author} {\bibfnamefont {R.~A.}\ \bibnamefont {Bravo}},
  \bibinfo {author} {\bibfnamefont {V.~O.}\ \bibnamefont {Lorenz}},\ and\
  \bibinfo {author} {\bibfnamefont {E.}~\bibnamefont {Chitambar}},\ }\href
  {https://doi.org/10.1088/1367-2630/ab7bef} {\bibfield  {journal} {\bibinfo
  {journal} {New Journal of Physics}\ }\textbf {\bibinfo {volume} {22}},\
  \bibinfo {pages} {043003} (\bibinfo {year} {2020})}\BibitemShut {NoStop}%
\bibitem [{\citenamefont {Doolittle}\ and\ \citenamefont
  {Bromley}(2022)}]{qNetVO}%
  \BibitemOpen
  \bibfield  {author} {\bibinfo {author} {\bibfnamefont {B.}~\bibnamefont
  {Doolittle}}\ and\ \bibinfo {author} {\bibfnamefont {T.}~\bibnamefont
  {Bromley}},\ }\href {https://doi.org/10.5281/zenodo.6345834} {\bibinfo
  {title} {{qNetVO: the Quantum Network Variational Optimizer, v0.1.3}}},\
  \bibinfo {howpublished} {\url{https://github.com/ChitambarLab/qNetVO}}
  (\bibinfo {year} {2022})\BibitemShut {NoStop}%
\bibitem [{\citenamefont {Bergholm}\ \emph {et~al.}(2018)\citenamefont
  {Bergholm}, \citenamefont {Izaac}, \citenamefont {Schuld}, \citenamefont
  {Gogolin}, \citenamefont {Alam}, \citenamefont {Ahmed}, \citenamefont
  {Arrazola}, \citenamefont {Blank}, \citenamefont {Delgado}, \citenamefont
  {Jahangiri} \emph {et~al.}}]{pennylane2018}%
  \BibitemOpen
  \bibfield  {author} {\bibinfo {author} {\bibfnamefont {V.}~\bibnamefont
  {Bergholm}}, \bibinfo {author} {\bibfnamefont {J.}~\bibnamefont {Izaac}},
  \bibinfo {author} {\bibfnamefont {M.}~\bibnamefont {Schuld}}, \bibinfo
  {author} {\bibfnamefont {C.}~\bibnamefont {Gogolin}}, \bibinfo {author}
  {\bibfnamefont {M.~S.}\ \bibnamefont {Alam}}, \bibinfo {author}
  {\bibfnamefont {S.}~\bibnamefont {Ahmed}}, \bibinfo {author} {\bibfnamefont
  {J.~M.}\ \bibnamefont {Arrazola}}, \bibinfo {author} {\bibfnamefont
  {C.}~\bibnamefont {Blank}}, \bibinfo {author} {\bibfnamefont
  {A.}~\bibnamefont {Delgado}}, \bibinfo {author} {\bibfnamefont
  {S.}~\bibnamefont {Jahangiri}}, \emph {et~al.},\ }\href@noop {} {\bibfield
  {journal} {\bibinfo  {journal} {arXiv preprint arXiv:1811.04968}\ } (\bibinfo
  {year} {2018})}\BibitemShut {NoStop}%
\bibitem [{\citenamefont {Doolittle}(2022)}]{supp_codebase}%
  \BibitemOpen
  \bibfield  {author} {\bibinfo {author} {\bibfnamefont {B.}~\bibnamefont
  {Doolittle}},\ }\href {https://doi.org/10.5281/zenodo.6519147} {\bibinfo
  {title} {{Supplemental Code: Variational Quantum Optimization of Nonlocality
  in Noisy Quantum Networks, v0.1.0}}},\ \bibinfo {howpublished}
  {\url{https://github.com/ChitambarLab/vqo-nonlocality-noisy-quantum-networks}}
  (\bibinfo {year} {2022})\BibitemShut {NoStop}%
\bibitem [{\citenamefont {Nielsen}\ and\ \citenamefont
  {Chuang}(2009)}]{Nielsen2009}%
  \BibitemOpen
  \bibfield  {author} {\bibinfo {author} {\bibfnamefont {M.~A.}\ \bibnamefont
  {Nielsen}}\ and\ \bibinfo {author} {\bibfnamefont {I.~L.}\ \bibnamefont
  {Chuang}},\ }\href {https://doi.org/10.1017/cbo9780511976667} {\emph
  {\bibinfo {title} {Quantum Computation and Quantum Information}}}\ (\bibinfo
  {publisher} {Cambridge University Press},\ \bibinfo {year}
  {2009})\BibitemShut {NoStop}%
\bibitem [{\citenamefont {Kraus}(1983)}]{Kraus1983}%
  \BibitemOpen
  \bibfield  {author} {\bibinfo {author} {\bibfnamefont {K.}~\bibnamefont
  {Kraus}},\ }\href@noop {} {\emph {\bibinfo {title} {States, Effects, and
  Operations: Fundamental Notions of Quantum Theory (Lecture Notes in Physics,
  190)}}},\ edited by\ \bibinfo {editor} {\bibfnamefont {B.}~\bibnamefont
  {A.}}, \bibinfo {editor} {\bibfnamefont {J.~D.}\ \bibnamefont {Dollard}},\
  and\ \bibinfo {editor} {\bibfnamefont {W.~H.}\ \bibnamefont {Wootters}}\
  (\bibinfo  {publisher} {Springer},\ \bibinfo {year} {1983})\ p.\ \bibinfo
  {pages} {151}\BibitemShut {NoStop}%
\bibitem [{\citenamefont {Stinespring}(1955)}]{stinespring1955}%
  \BibitemOpen
  \bibfield  {author} {\bibinfo {author} {\bibfnamefont {W.~F.}\ \bibnamefont
  {Stinespring}},\ }\href {https://doi.org/https://doi.org/10.2307/2032342}
  {\bibfield  {journal} {\bibinfo  {journal} {Proceedings of the American
  Mathematical Society}\ }\textbf {\bibinfo {volume} {6}},\ \bibinfo {pages}
  {211} (\bibinfo {year} {1955})}\BibitemShut {NoStop}%
\bibitem [{\citenamefont {Ruder}(2016)}]{ruder2016gradient_descent}%
  \BibitemOpen
  \bibfield  {author} {\bibinfo {author} {\bibfnamefont {S.}~\bibnamefont
  {Ruder}},\ }\href@noop {} {\bibfield  {journal} {\bibinfo  {journal} {arXiv
  preprint arXiv:1609.04747}\ } (\bibinfo {year} {2016})}\BibitemShut {NoStop}%
\bibitem [{\citenamefont {Bottou}\ \emph {et~al.}(2018)\citenamefont {Bottou},
  \citenamefont {Curtis},\ and\ \citenamefont
  {Nocedal}}]{Bottou2018gradient_descent}%
  \BibitemOpen
  \bibfield  {author} {\bibinfo {author} {\bibfnamefont {L.}~\bibnamefont
  {Bottou}}, \bibinfo {author} {\bibfnamefont {F.~E.}\ \bibnamefont {Curtis}},\
  and\ \bibinfo {author} {\bibfnamefont {J.}~\bibnamefont {Nocedal}},\ }\href
  {https://doi.org/10.1137/16M1080173} {\bibfield  {journal} {\bibinfo
  {journal} {SIAM Review}\ }\textbf {\bibinfo {volume} {60}},\ \bibinfo {pages}
  {223} (\bibinfo {year} {2018})}\BibitemShut {NoStop}%
\bibitem [{\citenamefont {Griewank}\ \emph {et~al.}(1989)\citenamefont
  {Griewank} \emph {et~al.}}]{griewank1989autodiff}%
  \BibitemOpen
  \bibfield  {author} {\bibinfo {author} {\bibfnamefont {A.}~\bibnamefont
  {Griewank}} \emph {et~al.},\ }\href@noop {} {\bibfield  {journal} {\bibinfo
  {journal} {Mathematical Programming: recent developments and applications}\
  }\textbf {\bibinfo {volume} {6}},\ \bibinfo {pages} {83} (\bibinfo {year}
  {1989})}\BibitemShut {NoStop}%
\bibitem [{\citenamefont {Baydin}\ \emph {et~al.}(2018)\citenamefont {Baydin},
  \citenamefont {Pearlmutter}, \citenamefont {Radul},\ and\ \citenamefont
  {Siskind}}]{Baydin2018autodiff}%
  \BibitemOpen
  \bibfield  {author} {\bibinfo {author} {\bibfnamefont {A.~G.}\ \bibnamefont
  {Baydin}}, \bibinfo {author} {\bibfnamefont {B.~A.}\ \bibnamefont
  {Pearlmutter}}, \bibinfo {author} {\bibfnamefont {A.~A.}\ \bibnamefont
  {Radul}},\ and\ \bibinfo {author} {\bibfnamefont {J.~M.}\ \bibnamefont
  {Siskind}},\ }\href {http://jmlr.org/papers/v18/17-468.html} {\bibfield
  {journal} {\bibinfo  {journal} {Journal of Machine Learning Research}\
  }\textbf {\bibinfo {volume} {18}},\ \bibinfo {pages} {1} (\bibinfo {year}
  {2018})}\BibitemShut {NoStop}%
\bibitem [{\citenamefont {Schuld}\ \emph {et~al.}(2019)\citenamefont {Schuld},
  \citenamefont {Bergholm}, \citenamefont {Gogolin}, \citenamefont {Izaac},\
  and\ \citenamefont {Killoran}}]{Schuld2019_parameter_shift}%
  \BibitemOpen
  \bibfield  {author} {\bibinfo {author} {\bibfnamefont {M.}~\bibnamefont
  {Schuld}}, \bibinfo {author} {\bibfnamefont {V.}~\bibnamefont {Bergholm}},
  \bibinfo {author} {\bibfnamefont {C.}~\bibnamefont {Gogolin}}, \bibinfo
  {author} {\bibfnamefont {J.}~\bibnamefont {Izaac}},\ and\ \bibinfo {author}
  {\bibfnamefont {N.}~\bibnamefont {Killoran}},\ }\href
  {https://doi.org/10.1103/PhysRevA.99.032331} {\bibfield  {journal} {\bibinfo
  {journal} {Phys. Rev. A}\ }\textbf {\bibinfo {volume} {99}},\ \bibinfo
  {pages} {032331} (\bibinfo {year} {2019})}\BibitemShut {NoStop}%
\bibitem [{\citenamefont {Fritz}(2012)}]{Fritz_2012}%
  \BibitemOpen
  \bibfield  {author} {\bibinfo {author} {\bibfnamefont {T.}~\bibnamefont
  {Fritz}},\ }\href {https://doi.org/10.1088/1367-2630/14/10/103001} {\bibfield
   {journal} {\bibinfo  {journal} {New Journal of Physics}\ }\textbf {\bibinfo
  {volume} {14}},\ \bibinfo {pages} {103001} (\bibinfo {year}
  {2012})}\BibitemShut {NoStop}%
\bibitem [{\citenamefont {Chaves}(2016)}]{chaves_2016_polynomial}%
  \BibitemOpen
  \bibfield  {author} {\bibinfo {author} {\bibfnamefont {R.}~\bibnamefont
  {Chaves}},\ }\href {https://doi.org/10.1103/PhysRevLett.116.010402}
  {\bibfield  {journal} {\bibinfo  {journal} {Phys. Rev. Lett.}\ }\textbf
  {\bibinfo {volume} {116}},\ \bibinfo {pages} {010402} (\bibinfo {year}
  {2016})}\BibitemShut {NoStop}%
\bibitem [{\citenamefont {Clauser}\ \emph {et~al.}(1969)\citenamefont
  {Clauser}, \citenamefont {Horne}, \citenamefont {Shimony},\ and\
  \citenamefont {Holt}}]{chsh-inequality1969}%
  \BibitemOpen
  \bibfield  {author} {\bibinfo {author} {\bibfnamefont {J.~F.}\ \bibnamefont
  {Clauser}}, \bibinfo {author} {\bibfnamefont {M.~A.}\ \bibnamefont {Horne}},
  \bibinfo {author} {\bibfnamefont {A.}~\bibnamefont {Shimony}},\ and\ \bibinfo
  {author} {\bibfnamefont {R.~A.}\ \bibnamefont {Holt}},\ }\href
  {https://doi.org/10.1103/PhysRevLett.23.880} {\bibfield  {journal} {\bibinfo
  {journal} {Phys. Rev. Lett.}\ }\textbf {\bibinfo {volume} {23}},\ \bibinfo
  {pages} {880} (\bibinfo {year} {1969})}\BibitemShut {NoStop}%
\bibitem [{\citenamefont {Cirel{\textquotesingle}son}(1980)}]{Cirelson1980}%
  \BibitemOpen
  \bibfield  {author} {\bibinfo {author} {\bibfnamefont {B.~S.}\ \bibnamefont
  {Cirel{\textquotesingle}son}},\ }\href {https://doi.org/10.1007/bf00417500}
  {\bibfield  {journal} {\bibinfo  {journal} {Letters in Mathematical Physics}\
  }\textbf {\bibinfo {volume} {4}},\ \bibinfo {pages} {93} (\bibinfo {year}
  {1980})}\BibitemShut {NoStop}%
\bibitem [{\citenamefont {Dasgupta}\ and\ \citenamefont
  {Humble}(2021)}]{dasgupta2021stability}%
  \BibitemOpen
  \bibfield  {author} {\bibinfo {author} {\bibfnamefont {S.}~\bibnamefont
  {Dasgupta}}\ and\ \bibinfo {author} {\bibfnamefont {T.~S.}\ \bibnamefont
  {Humble}},\ }\href@noop {} {\bibfield  {journal} {\bibinfo  {journal} {arXiv
  preprint arXiv:2105.09472}\ } (\bibinfo {year} {2021})}\BibitemShut {NoStop}%
\bibitem [{\citenamefont {Gerritsma}\ \emph {et~al.}(2010)\citenamefont
  {Gerritsma}, \citenamefont {Kirchmair}, \citenamefont {Z\"{a}hringer},
  \citenamefont {Solano}, \citenamefont {Blatt},\ and\ \citenamefont
  {Roos}}]{Gerritsma2010_simulation}%
  \BibitemOpen
  \bibfield  {author} {\bibinfo {author} {\bibfnamefont {R.}~\bibnamefont
  {Gerritsma}}, \bibinfo {author} {\bibfnamefont {G.}~\bibnamefont
  {Kirchmair}}, \bibinfo {author} {\bibfnamefont {F.}~\bibnamefont
  {Z\"{a}hringer}}, \bibinfo {author} {\bibfnamefont {E.}~\bibnamefont
  {Solano}}, \bibinfo {author} {\bibfnamefont {R.}~\bibnamefont {Blatt}},\ and\
  \bibinfo {author} {\bibfnamefont {C.~F.}\ \bibnamefont {Roos}},\ }\href
  {https://doi.org/10.1038/nature08688} {\bibfield  {journal} {\bibinfo
  {journal} {Nature}\ }\textbf {\bibinfo {volume} {463}},\ \bibinfo {pages}
  {68} (\bibinfo {year} {2010})}\BibitemShut {NoStop}%
\bibitem [{\citenamefont {Lanyon}\ \emph {et~al.}(2011)\citenamefont {Lanyon},
  \citenamefont {Hempel}, \citenamefont {Nigg}, \citenamefont {M\"{u}ller},
  \citenamefont {Gerritsma}, \citenamefont {Z\"{a}hringer}, \citenamefont
  {Schindler}, \citenamefont {Barreiro}, \citenamefont {Rambach}, \citenamefont
  {Kirchmair}, \citenamefont {Hennrich}, \citenamefont {Zoller}, \citenamefont
  {Blatt},\ and\ \citenamefont {Roos}}]{Lanyon2011_simulation}%
  \BibitemOpen
  \bibfield  {author} {\bibinfo {author} {\bibfnamefont {B.~P.}\ \bibnamefont
  {Lanyon}}, \bibinfo {author} {\bibfnamefont {C.}~\bibnamefont {Hempel}},
  \bibinfo {author} {\bibfnamefont {D.}~\bibnamefont {Nigg}}, \bibinfo {author}
  {\bibfnamefont {M.}~\bibnamefont {M\"{u}ller}}, \bibinfo {author}
  {\bibfnamefont {R.}~\bibnamefont {Gerritsma}}, \bibinfo {author}
  {\bibfnamefont {F.}~\bibnamefont {Z\"{a}hringer}}, \bibinfo {author}
  {\bibfnamefont {P.}~\bibnamefont {Schindler}}, \bibinfo {author}
  {\bibfnamefont {J.~T.}\ \bibnamefont {Barreiro}}, \bibinfo {author}
  {\bibfnamefont {M.}~\bibnamefont {Rambach}}, \bibinfo {author} {\bibfnamefont
  {G.}~\bibnamefont {Kirchmair}}, \bibinfo {author} {\bibfnamefont
  {M.}~\bibnamefont {Hennrich}}, \bibinfo {author} {\bibfnamefont
  {P.}~\bibnamefont {Zoller}}, \bibinfo {author} {\bibfnamefont
  {R.}~\bibnamefont {Blatt}},\ and\ \bibinfo {author} {\bibfnamefont {C.~F.}\
  \bibnamefont {Roos}},\ }\href {https://doi.org/10.1126/science.1208001}
  {\bibfield  {journal} {\bibinfo  {journal} {Science}\ }\textbf {\bibinfo
  {volume} {334}},\ \bibinfo {pages} {57} (\bibinfo {year} {2011})}\BibitemShut
  {NoStop}%
\bibitem [{\citenamefont {Barends}\ \emph {et~al.}(2015)\citenamefont
  {Barends}, \citenamefont {Lamata}, \citenamefont {Kelly}, \citenamefont
  {Garc{\'{\i}}a-{\'{A}}lvarez}, \citenamefont {Fowler}, \citenamefont
  {Megrant}, \citenamefont {Jeffrey}, \citenamefont {White}, \citenamefont
  {Sank}, \citenamefont {Mutus}, \citenamefont {Campbell}, \citenamefont
  {Chen}, \citenamefont {Chen}, \citenamefont {Chiaro}, \citenamefont
  {Dunsworth}, \citenamefont {Hoi}, \citenamefont {Neill}, \citenamefont
  {O'Malley}, \citenamefont {Quintana}, \citenamefont {Roushan}, \citenamefont
  {Vainsencher}, \citenamefont {Wenner}, \citenamefont {Solano},\ and\
  \citenamefont {Martinis}}]{Barends2015}%
  \BibitemOpen
  \bibfield  {author} {\bibinfo {author} {\bibfnamefont {R.}~\bibnamefont
  {Barends}}, \bibinfo {author} {\bibfnamefont {L.}~\bibnamefont {Lamata}},
  \bibinfo {author} {\bibfnamefont {J.}~\bibnamefont {Kelly}}, \bibinfo
  {author} {\bibfnamefont {L.}~\bibnamefont {Garc{\'{\i}}a-{\'{A}}lvarez}},
  \bibinfo {author} {\bibfnamefont {A.~G.}\ \bibnamefont {Fowler}}, \bibinfo
  {author} {\bibfnamefont {A.}~\bibnamefont {Megrant}}, \bibinfo {author}
  {\bibfnamefont {E.}~\bibnamefont {Jeffrey}}, \bibinfo {author} {\bibfnamefont
  {T.~C.}\ \bibnamefont {White}}, \bibinfo {author} {\bibfnamefont
  {D.}~\bibnamefont {Sank}}, \bibinfo {author} {\bibfnamefont {J.~Y.}\
  \bibnamefont {Mutus}}, \bibinfo {author} {\bibfnamefont {B.}~\bibnamefont
  {Campbell}}, \bibinfo {author} {\bibfnamefont {Y.}~\bibnamefont {Chen}},
  \bibinfo {author} {\bibfnamefont {Z.}~\bibnamefont {Chen}}, \bibinfo {author}
  {\bibfnamefont {B.}~\bibnamefont {Chiaro}}, \bibinfo {author} {\bibfnamefont
  {A.}~\bibnamefont {Dunsworth}}, \bibinfo {author} {\bibfnamefont {I.-C.}\
  \bibnamefont {Hoi}}, \bibinfo {author} {\bibfnamefont {C.}~\bibnamefont
  {Neill}}, \bibinfo {author} {\bibfnamefont {P.~J.~J.}\ \bibnamefont
  {O'Malley}}, \bibinfo {author} {\bibfnamefont {C.}~\bibnamefont {Quintana}},
  \bibinfo {author} {\bibfnamefont {P.}~\bibnamefont {Roushan}}, \bibinfo
  {author} {\bibfnamefont {A.}~\bibnamefont {Vainsencher}}, \bibinfo {author}
  {\bibfnamefont {J.}~\bibnamefont {Wenner}}, \bibinfo {author} {\bibfnamefont
  {E.}~\bibnamefont {Solano}},\ and\ \bibinfo {author} {\bibfnamefont {J.~M.}\
  \bibnamefont {Martinis}},\ }\bibfield  {journal} {\bibinfo  {journal} {Nature
  Communications}\ }\textbf {\bibinfo {volume} {6}},\ \href
  {https://doi.org/10.1038/ncomms8654} {10.1038/ncomms8654} (\bibinfo {year}
  {2015})\BibitemShut {NoStop}%
\bibitem [{\citenamefont {Martinez}\ \emph {et~al.}(2016)\citenamefont
  {Martinez}, \citenamefont {Muschik}, \citenamefont {Schindler}, \citenamefont
  {Nigg}, \citenamefont {Erhard}, \citenamefont {Heyl}, \citenamefont {Hauke},
  \citenamefont {Dalmonte}, \citenamefont {Monz}, \citenamefont {Zoller},\ and\
  \citenamefont {Blatt}}]{Martinez2016_schwinger_simulation}%
  \BibitemOpen
  \bibfield  {author} {\bibinfo {author} {\bibfnamefont {E.~A.}\ \bibnamefont
  {Martinez}}, \bibinfo {author} {\bibfnamefont {C.~A.}\ \bibnamefont
  {Muschik}}, \bibinfo {author} {\bibfnamefont {P.}~\bibnamefont {Schindler}},
  \bibinfo {author} {\bibfnamefont {D.}~\bibnamefont {Nigg}}, \bibinfo {author}
  {\bibfnamefont {A.}~\bibnamefont {Erhard}}, \bibinfo {author} {\bibfnamefont
  {M.}~\bibnamefont {Heyl}}, \bibinfo {author} {\bibfnamefont {P.}~\bibnamefont
  {Hauke}}, \bibinfo {author} {\bibfnamefont {M.}~\bibnamefont {Dalmonte}},
  \bibinfo {author} {\bibfnamefont {T.}~\bibnamefont {Monz}}, \bibinfo {author}
  {\bibfnamefont {P.}~\bibnamefont {Zoller}},\ and\ \bibinfo {author}
  {\bibfnamefont {R.}~\bibnamefont {Blatt}},\ }\href
  {https://doi.org/10.1038/nature18318} {\bibfield  {journal} {\bibinfo
  {journal} {Nature}\ }\textbf {\bibinfo {volume} {534}},\ \bibinfo {pages}
  {516} (\bibinfo {year} {2016})}\BibitemShut {NoStop}%
\bibitem [{\citenamefont {Rost}\ \emph {et~al.}(2020)\citenamefont {Rost},
  \citenamefont {Jones}, \citenamefont {Vyushkova}, \citenamefont {Ali},
  \citenamefont {Cullip}, \citenamefont {Vyushkov},\ and\ \citenamefont
  {Nabrzyski}}]{rost2020simulation}%
  \BibitemOpen
  \bibfield  {author} {\bibinfo {author} {\bibfnamefont {B.}~\bibnamefont
  {Rost}}, \bibinfo {author} {\bibfnamefont {B.}~\bibnamefont {Jones}},
  \bibinfo {author} {\bibfnamefont {M.}~\bibnamefont {Vyushkova}}, \bibinfo
  {author} {\bibfnamefont {A.}~\bibnamefont {Ali}}, \bibinfo {author}
  {\bibfnamefont {C.}~\bibnamefont {Cullip}}, \bibinfo {author} {\bibfnamefont
  {A.}~\bibnamefont {Vyushkov}},\ and\ \bibinfo {author} {\bibfnamefont
  {J.}~\bibnamefont {Nabrzyski}},\ }\href@noop {} {\bibfield  {journal}
  {\bibinfo  {journal} {arXiv preprint arXiv:2001.00794}\ } (\bibinfo {year}
  {2020})}\BibitemShut {NoStop}%
\bibitem [{\citenamefont {Rost}\ \emph {et~al.}(2021)\citenamefont {Rost},
  \citenamefont {Del~Re}, \citenamefont {Earnest}, \citenamefont {Kemper},
  \citenamefont {Jones},\ and\ \citenamefont
  {Freericks}}]{rost2021demonstrating}%
  \BibitemOpen
  \bibfield  {author} {\bibinfo {author} {\bibfnamefont {B.}~\bibnamefont
  {Rost}}, \bibinfo {author} {\bibfnamefont {L.}~\bibnamefont {Del~Re}},
  \bibinfo {author} {\bibfnamefont {N.}~\bibnamefont {Earnest}}, \bibinfo
  {author} {\bibfnamefont {A.~F.}\ \bibnamefont {Kemper}}, \bibinfo {author}
  {\bibfnamefont {B.}~\bibnamefont {Jones}},\ and\ \bibinfo {author}
  {\bibfnamefont {J.~K.}\ \bibnamefont {Freericks}},\ }\href@noop {} {\bibfield
   {journal} {\bibinfo  {journal} {arXiv preprint arXiv:2108.01183}\ }
  (\bibinfo {year} {2021})}\BibitemShut {NoStop}%
\bibitem [{\citenamefont {Childs}\ \emph {et~al.}(2018)\citenamefont {Childs},
  \citenamefont {Maslov}, \citenamefont {Nam}, \citenamefont {Ross},\ and\
  \citenamefont {Su}}]{Childs2018}%
  \BibitemOpen
  \bibfield  {author} {\bibinfo {author} {\bibfnamefont {A.~M.}\ \bibnamefont
  {Childs}}, \bibinfo {author} {\bibfnamefont {D.}~\bibnamefont {Maslov}},
  \bibinfo {author} {\bibfnamefont {Y.}~\bibnamefont {Nam}}, \bibinfo {author}
  {\bibfnamefont {N.~J.}\ \bibnamefont {Ross}},\ and\ \bibinfo {author}
  {\bibfnamefont {Y.}~\bibnamefont {Su}},\ }\href
  {https://doi.org/10.1073/pnas.1801723115} {\bibfield  {journal} {\bibinfo
  {journal} {Proceedings of the National Academy of Sciences}\ }\textbf
  {\bibinfo {volume} {115}},\ \bibinfo {pages} {9456} (\bibinfo {year}
  {2018})}\BibitemShut {NoStop}%
\bibitem [{\citenamefont {Peng}\ \emph {et~al.}(2020)\citenamefont {Peng},
  \citenamefont {Harrow}, \citenamefont {Ozols},\ and\ \citenamefont
  {Wu}}]{Peng2020_simulating_large_circuits}%
  \BibitemOpen
  \bibfield  {author} {\bibinfo {author} {\bibfnamefont {T.}~\bibnamefont
  {Peng}}, \bibinfo {author} {\bibfnamefont {A.~W.}\ \bibnamefont {Harrow}},
  \bibinfo {author} {\bibfnamefont {M.}~\bibnamefont {Ozols}},\ and\ \bibinfo
  {author} {\bibfnamefont {X.}~\bibnamefont {Wu}},\ }\href
  {https://doi.org/10.1103/PhysRevLett.125.150504} {\bibfield  {journal}
  {\bibinfo  {journal} {Phys. Rev. Lett.}\ }\textbf {\bibinfo {volume} {125}},\
  \bibinfo {pages} {150504} (\bibinfo {year} {2020})}\BibitemShut {NoStop}%
\bibitem [{\citenamefont {Barratt}\ \emph {et~al.}(2021)\citenamefont
  {Barratt}, \citenamefont {Dborin}, \citenamefont {Bal}, \citenamefont
  {Stojevic}, \citenamefont {Pollmann},\ and\ \citenamefont
  {Green}}]{Barratt2021_parallel_nisq_simulation}%
  \BibitemOpen
  \bibfield  {author} {\bibinfo {author} {\bibfnamefont {F.}~\bibnamefont
  {Barratt}}, \bibinfo {author} {\bibfnamefont {J.}~\bibnamefont {Dborin}},
  \bibinfo {author} {\bibfnamefont {M.}~\bibnamefont {Bal}}, \bibinfo {author}
  {\bibfnamefont {V.}~\bibnamefont {Stojevic}}, \bibinfo {author}
  {\bibfnamefont {F.}~\bibnamefont {Pollmann}},\ and\ \bibinfo {author}
  {\bibfnamefont {A.~G.}\ \bibnamefont {Green}},\ }\bibfield  {journal}
  {\bibinfo  {journal} {npj Quantum Information}\ }\textbf {\bibinfo {volume}
  {7}},\ \href {https://doi.org/10.1038/s41534-021-00420-3}
  {10.1038/s41534-021-00420-3} (\bibinfo {year} {2021})\BibitemShut {NoStop}%
\bibitem [{\citenamefont {Horodecki}\ \emph {et~al.}(1995)\citenamefont
  {Horodecki}, \citenamefont {Horodecki},\ and\ \citenamefont
  {Horodecki}}]{Horodecki1995}%
  \BibitemOpen
  \bibfield  {author} {\bibinfo {author} {\bibfnamefont {R.}~\bibnamefont
  {Horodecki}}, \bibinfo {author} {\bibfnamefont {P.}~\bibnamefont
  {Horodecki}},\ and\ \bibinfo {author} {\bibfnamefont {M.}~\bibnamefont
  {Horodecki}},\ }\href
  {https://doi.org/https://doi.org/10.1016/0375-9601(95)00214-N} {\bibfield
  {journal} {\bibinfo  {journal} {Physics Letters A}\ }\textbf {\bibinfo
  {volume} {200}},\ \bibinfo {pages} {340} (\bibinfo {year}
  {1995})}\BibitemShut {NoStop}%
\bibitem [{\citenamefont {Gisin}\ \emph {et~al.}(2017)\citenamefont {Gisin},
  \citenamefont {Mei}, \citenamefont {Tavakoli}, \citenamefont {Renou},\ and\
  \citenamefont {Brunner}}]{gisin2017_bilocal_criterion}%
  \BibitemOpen
  \bibfield  {author} {\bibinfo {author} {\bibfnamefont {N.}~\bibnamefont
  {Gisin}}, \bibinfo {author} {\bibfnamefont {Q.}~\bibnamefont {Mei}}, \bibinfo
  {author} {\bibfnamefont {A.}~\bibnamefont {Tavakoli}}, \bibinfo {author}
  {\bibfnamefont {M.~O.}\ \bibnamefont {Renou}},\ and\ \bibinfo {author}
  {\bibfnamefont {N.}~\bibnamefont {Brunner}},\ }\href
  {https://doi.org/10.1103/PhysRevA.96.020304} {\bibfield  {journal} {\bibinfo
  {journal} {Phys. Rev. A}\ }\textbf {\bibinfo {volume} {96}},\ \bibinfo
  {pages} {020304} (\bibinfo {year} {2017})}\BibitemShut {NoStop}%
\bibitem [{\citenamefont {Andreoli}\ \emph
  {et~al.}(2017{\natexlab{b}})\citenamefont {Andreoli}, \citenamefont
  {Carvacho}, \citenamefont {Santodonato}, \citenamefont {Chaves},\ and\
  \citenamefont {Sciarrino}}]{andreoli2017maximal_star_violation}%
  \BibitemOpen
  \bibfield  {author} {\bibinfo {author} {\bibfnamefont {F.}~\bibnamefont
  {Andreoli}}, \bibinfo {author} {\bibfnamefont {G.}~\bibnamefont {Carvacho}},
  \bibinfo {author} {\bibfnamefont {L.}~\bibnamefont {Santodonato}}, \bibinfo
  {author} {\bibfnamefont {R.}~\bibnamefont {Chaves}},\ and\ \bibinfo {author}
  {\bibfnamefont {F.}~\bibnamefont {Sciarrino}},\ }\href
  {https://doi.org/https://doi.org/10.1088/1367-2630/aa8b9b} {\bibfield
  {journal} {\bibinfo  {journal} {New Journal of Physics}\ }\textbf {\bibinfo
  {volume} {19}},\ \bibinfo {pages} {113020} (\bibinfo {year}
  {2017}{\natexlab{b}})}\BibitemShut {NoStop}%
\bibitem [{\citenamefont {Kundu}\ \emph {et~al.}(2020)\citenamefont {Kundu},
  \citenamefont {Molla}, \citenamefont {Chattopadhyay},\ and\ \citenamefont
  {Sarkar}}]{kundu2020_nlocal_max_qubit_violations}%
  \BibitemOpen
  \bibfield  {author} {\bibinfo {author} {\bibfnamefont {A.}~\bibnamefont
  {Kundu}}, \bibinfo {author} {\bibfnamefont {M.~K.}\ \bibnamefont {Molla}},
  \bibinfo {author} {\bibfnamefont {I.}~\bibnamefont {Chattopadhyay}},\ and\
  \bibinfo {author} {\bibfnamefont {D.}~\bibnamefont {Sarkar}},\ }\href
  {https://doi.org/10.1103/PhysRevA.102.052222} {\bibfield  {journal} {\bibinfo
   {journal} {Phys. Rev. A}\ }\textbf {\bibinfo {volume} {102}},\ \bibinfo
  {pages} {052222} (\bibinfo {year} {2020})}\BibitemShut {NoStop}%
\bibitem [{\citenamefont {Feynman}\ \emph {et~al.}(1982)\citenamefont {Feynman}
  \emph {et~al.}}]{feynman1982simulating}%
  \BibitemOpen
  \bibfield  {author} {\bibinfo {author} {\bibfnamefont {R.~P.}\ \bibnamefont
  {Feynman}} \emph {et~al.},\ }\href@noop {} {\bibfield  {journal} {\bibinfo
  {journal} {Int. j. Theor. phys}\ }\textbf {\bibinfo {volume} {21}} (\bibinfo
  {year} {1982})}\BibitemShut {NoStop}%
\bibitem [{\citenamefont {Lloyd}(1996)}]{Lloyd1996_simulation}%
  \BibitemOpen
  \bibfield  {author} {\bibinfo {author} {\bibfnamefont {S.}~\bibnamefont
  {Lloyd}},\ }\href {https://doi.org/10.1126/science.273.5278.1073} {\bibfield
  {journal} {\bibinfo  {journal} {Science}\ }\textbf {\bibinfo {volume}
  {273}},\ \bibinfo {pages} {1073} (\bibinfo {year} {1996})}\BibitemShut
  {NoStop}%
\end{thebibliography}%

\onecolumngrid

\newpage

\appendix

\section{Quantum Circuits for Preparation and Measurement Ansatzes}\label{appendix:ansatz_table}

\begin{table}[h]
    \centering
    \begin{tabular}{|C{0.15\linewidth}|C{0.3\linewidth}|C{0.13\linewidth}|C{0.4\linewidth}|}
        \hline
        \textbf{Ansatz Name} & \textbf{Ansatz Circuit} & \textbf{Number of Parameters} & \textbf{Ansatz Description} \\
        \hline
        Arbitrary $M$-Qubit Pure State Preparation & \begin{tikzcd}
            \ket{0}^{\otimes M}  & [4mm]\gate{U\left(\phiv\right)} \qwbundle{M} & \qwbundle{M}
        \end{tikzcd} & $\big|\phiv\big| = 2^{M+1} - 2$ & Parameterizes all $M$-qubit pure state preparations. See the \texttt{ArbitraryStatePreparation} method in PennyLane for implementation \cite{pennylane2018}.\\
        \hline
        $\ket{\Phi^+}$ State Preparation & \begin{tikzcd}
            \ket{0} & \gate{H} & \ctrl{1} & \qw \\
            \ket{0} & \qw & \targ{} & \qw
        \end{tikzcd} & None & Prepares the state $\ket{\Phi^+} = (\ket{00}+\ket{11})/\sqrt{2}$.\\
        \hline
        $\ket{\Psi^+}$ State Preparation & \begin{tikzcd}
            \ket{0} & \gate{H} & \ctrl{1} & \qw \\
            \ket{0} & \gate{\sigma_x} & \targ{} & \qw
        \end{tikzcd} & None & Prepares the state $\ket{\Psi^+} = (\ket{01}+\ket{10})/\sqrt{2}$.\\
        \hline
        Maximally Entangled State Preparation & \begin{tikzcd}
            \ket{0} & \gate{H} & \ctrl{1} & \gate{Rot\left(\phiv\right)} & \qw \\
            \ket{0} & \qw & \targ{} & \qw & \qw
        \end{tikzcd} & $\big|\phiv\big| = 3$ & Parameterizes all 2-qubit maximally entangled state preparations using an arbitrary qubit rotation $Rot(\phiv) = R_z(\phi_1)R_y(\phi_2)R_z(\phi_3)$. \\
        \hline
        Nonmaximally Entangled State Preparation & \begin{tikzcd}
            \ket{0} & \gate{R_y(\phi_1)} & \gate{R_z(\phi_2)} & \ctrl{1} & \qw \\
            \ket{0} & \qw & \qw \qw & \targ{} & \qw 
        \end{tikzcd} & $\big|\phiv\big| = 2$ & Parameterizes a family of separable through maximally entangled states as $\ket{\psi} = \cos(\phi_1/2)\ket{00} + \sin(\phi_1/2)e^{i\phi_2}\ket{11}$.  \\
        \hline
        Arbitrary $M$-Qubit Projective Measurement & \begin{tikzcd}
            \qwbundle{n} & [4mm]\gate{U\left(\thetav\right)}\qwbundle{M}  & [2mm]\meter{$\{\Pi_+,\Pi_-\}^{\otimes M}$} \qwbundle{M} & \cw
        \end{tikzcd} & $\big|\thetav\big| = 2^{2M} - 1$ & Parameterizes all $M$-qubit projective measurements. See the \texttt{ArbitraryUnitary} method in PennyLane for implementation \cite{pennylane2018}.  \\
        \hline
        $M$-Qubit Local $R_y$ Measurement & \begin{tikzcd}
            \ket{\psi^{q_1}} & \gate{R_y(\theta_1)} & \meter{} & \cw \\
            [-0.5cm]\vdots& \vdots& \vdots  &  \\
            [-0.3cm]\ket{\psi^{q_M}} & \gate{R_y(\theta_M)} & \meter{} & \cw 
        \end{tikzcd} & $\big|\thetav\big| = M$ & Parameterizes all $M$-qubit projective measurements that decompose as $\Pi^A_{\zv|\xv} =\bigotimes_{i=1}^M R_y(\phi_i)\op{z^{q_i}}{z^{q_i}}$.  \\
        \hline
        Arbitrary Local Qubit Measurement & \begin{tikzcd}
            \ket{\psi^{q_1}} & \gate{Rot(\thetav_1)} & \meter{} & \cw \\
            [-0.5cm]\vdots& \vdots& \vdots  &  \\
            [-0.3cm]\ket{\psi^{q_M}} & \gate{Rot(\thetav_M)} & \meter{} & \cw 
        \end{tikzcd} & $\big|\thetav\big| = 3M$ & Parameterizes all $M$-qubit projective measurements that decompose as $\Pi^A_{\zv|\xv} =\bigotimes_{i=1}^M Rot(\phiv_i)\op{z^{q_i}}{z^{q_i}}$ where $Rot(\phiv) = R_z(\phi_1)R_y(\phi_2)R_z(\phi_3)$. \\
        \hline
    \end{tabular}
     \caption{\linespread{1}\selectfont{
        \small This table shows the various preparation and measurement ansatzes used throughout this work.
        For each ansatz, we provide the circuit diagram, the number of tunable parameters, and a short description of the parameterization.
    }}
    \label{Table:prepare_measure_ansatzes}
\end{table}

\section{Maximally Entangled States are Optimal for CHSH Violation in the Presence of Two-Sided Qubit Unital Noise} \label{appendix:unital_channels}

In this section, we show that maximally entangled states optimally violate the CHSH inequality described by Eq. \eqref{eq:chsh-inequality} in the presence of two-sided unital qubit noise $\mc{U}_1\otimes\mc{U}_2$.
We first present a theorem that proves the previous statement.
Then we apply our theorem to derive the maximal violation of the CHSH inequality when two-sided dephasing noise is present.

\begin{theorem}\label{thm:two-sided_unital_chsh}
    Consider the two-sided unital channel acting upon an arbitrary state preparation $\mc{U}_1\otimes\mc{U}_2(\rho^{AB})$ where  $\mc{U}_1$ and $\mc{U}_2$ are unital qubit channels. 
    For any two unital channels $\mc{U}_1$, and $\mc{U}_2$, the maximal CHSH score $S^{\star}_{\text{CHSH}}$ is achieved by the maximally entangled state $\rho^{AB}=\op{\Phi^+}{\Phi^+}$.
    
    \begin{proof}
        The CHSH score of the noisy state $\mc{U}_1\otimes\mc{U}_2(\rho^{AB})$ is defined as
        \begin{equation}
            S_{\text{CHSH}} = \tr{O^{AB}_{\text{CHSH}}\mc{U}_1\otimes\mc{U}_2(\rho^{AB})},
        \end{equation}
        with the CHSH operator $O^{AB}_{\text{CHSH}}$ is a two-body correlator
        \begin{equation}
            O^{AB}_{\text{CHSH}} = \av_0 \cdot \sigmav\otimes (\bv_0 + \bv_1)\cdot\sigmav + \av_1 \cdot \sigmav \otimes (\bv_0 - \bv_1) \cdot \sigmav,
        \end{equation}
        where $\sigmav = (\sigma_x, \sigma_y, \sigma_z)$ and $\av,\bv\in\mbb{R}^3$ satisfy $|\av|,|\bv|\leq1$.
        The operator $O_{\text{CHSH}}^{AB}$ admits the decomposition
        \begin{equation}
           O^{AB}_{\text{CHSH}} = \sum_{i,j}t_{i,j}\sigma^A_i\otimes\sigma^B_j,
        \end{equation}
        where $t_{i,j} =  \tr{\sigma^A_i\otimes\sigma^B_jO^{AB}_{\text{CHSH}}}$ and the values $t_{i,j}$ form the $3 \times 3$ correlation matrix $T_{\text{CHSH}}$.
        In Proposition 1 of \cite{Zhang2020} it was proven that for any correlation matrix $T$, if $\text{Rank}(T)\leq 2$, then its corresponding two-body correlator $\tr{O^{AB} \rho^{AB}}$ is maximized by a maximally entangled state.
        The authors then prove that the unital channel $\mc{U}_1\otimes \mbb{I}^B$ preserves the optimality of the Bell state $\op{\Phi^+}{\Phi^+}$.
        We adapt their proof to the two-sided case to find that
        \begin{align}
            \text{Tr}\big[O^{AB}_{\text{CHSH}}\mc{U}_1\otimes\mc{U}_2(\rho^{AB})\big] = \tr{\mc{U}_1^{\dagger}\otimes\mc{U}_2^{\dagger}(O_{\text{CHSH}}^{AB})\rho^{AB}},
        \end{align}
        where the adjoint channel $\mc{U}_i^\dagger$ is also unital.
        Noting that the CHSH operator can be diagonalized as
        \begin{equation}
            O^{AB}_{\text{CHSH}} = \alpha_x \sigma^A_x\otimes \sigma^B_x + \alpha_z \sigma^A_z \otimes\sigma^B_z,
        \end{equation}
        where $\alpha_x$ and $\alpha_z$ are eigenvalues of $O^{AB}_{\text{CHSH}}$.
        It follows that,
        \begin{align}
            \mc{U}^\dagger_1\otimes\mc{U}^\dagger_2(O^{AB}_{\text{CHSH}}) &= \mc{U}_1^\dagger\otimes\mc{U}_2^\dagger(\alpha_x \sigma^A_x\otimes \sigma^B_x + \alpha_z \sigma^A_z \otimes\sigma^B_z) \\
            &= \alpha_x \mc{U}_1^\dagger(\sigma^A_x)\otimes\mc{U}^\dagger_2(\sigma^B_x) + \alpha_z \mc{U}^\dagger_1(\sigma^A_z)\otimes\mc{U}^\dagger_2(\sigma^B_z) \\
            &= \alpha_x \widetilde{\sigma}^A_{x,1}\otimes \widetilde\sigma^B_{x,2} + \alpha_z \widetilde{\sigma}^A_{z,1}\otimes \widetilde\sigma^B_{z,2} \\
            &=\widetilde{O}_{\text{CHSH}}^{AB}, 
        \end{align}
        where $\widetilde{\sigma}_{i,j} = \mc{U}^\dagger_j(\sigma_i)= \vec{s}\cdot\sigmav$ for some $\vec{s}\in\mbb{R}^3$ satisfying $|\vec{s}|\leq1$ and $\widetilde{O}_{\text{CHSH}}^{AB}$ is a correlation operator with correlation matrix $\widetilde{T}_{\text{CHSH}}$ satisfying $\text{Rank}(\widetilde{T}_{\text{CHSH}})\leq2$.
        By Proposition 1 of \cite{Zhang2020}, the rank constraint proves that a maximally entangled state maximizes the CHSH inequality.
        Furthermore, the Bell state $\op{\Phi^+}{\Phi^+}$ can always achieve the optimal CHSH score because
        the CHSH operator can always be transformed by local qubit unitaries  $\widetilde{O}_{\text{CHSH}}^{AB'} = U\otimes V \widetilde{O}_{\text{CHSH}}^{AB} U^\dagger\otimes V^\dagger$  such that the largest eigenvalue of $\widetilde{O}_{\text{CHSH}}^{AB'}$ corresponds to the eigenvector $\ket{\phi^+}$.
    \end{proof}
\end{theorem}

\begin{proposition} \label{prop:two-sided_dephasing}
    Given two-sided dephasing noise, $\mc{P}_{\gamma_1}\otimes\mc{P}_{\gamma_2}$ the maximal violation of the CHSH inequality is 
    \begin{equation}\label{eq:two-sided_dephasing_bell_state_boundary}
        S_{\text{CHSH}}^\star = 2\sqrt{ 1 + (1-\gamma_1)(1-\gamma_2)}.
    \end{equation}
    
    \begin{proof}
        Since the phase damping channel is unital, Theorem \ref{thm:two-sided_unital_chsh} states that the maximal CHSH violation can be achieved using the Bell state preparation $\ket{\Phi^+} = (\ket{00}+\ket{11})/\sqrt{2}$.
        By direct application of the dephasing Kraus operators in Eq. \eqref{eq:dephasing_kraus_ops} the channel's output is found to be
        \begin{align}
            \rho_{\N} &= \mc{P}_{\gamma_1}\otimes\mc{P}_{\gamma_2}(\op{\Phi^+}{\Phi^+}) \\
            &= \frac{1}{2}\begin{pmatrix}
                1 & 0 & 0 & \sqrt{1-\gamma_1}\sqrt{1-\gamma_2} \\
                0 & 0 & 0 & 0 \\
                0 & 0 & 0 & 0 \\
                \sqrt{1-\gamma_1}\sqrt{1-\gamma_2} & 0 & 0 & 1
            \end{pmatrix}.
        \end{align}
        To decide whether $\rho_{\N}$ can violate the CHSH inequality we apply the necessary and sufficient conditions introduced by Horodecki \textit{et al.} \cite{Horodecki1995}.
        First, we construct the real-valued $3\times 3$ correlation matrix $T_{\rho_{\mc{N}}}$  using Eq. \eqref{eq:correlation_matrix}
        \begin{equation}
            T_{\rho_{\N}} = \begin{pmatrix}
                \sqrt{1-\gamma_1}\sqrt{1-\gamma_2} & 0 & 0 \\
                0 & -\sqrt{1-\gamma_1}\sqrt{1-\gamma_2} & 0 \\
                0 & 0 & 1\\
            \end{pmatrix}.
        \end{equation}
        Next, we find
        \begin{align}
            R_{\rho_{\N}} &= T_{\rho\N}^TT_{\rho\N} = \begin{pmatrix}
                (1-\gamma_1)(1-\gamma_2) & 0 & 0 \\
                0 & (1-\gamma_1)(1-\gamma_2) & 0 \\
                0 & 0 & 1
            \end{pmatrix}.
        \end{align}
        Since $\gamma_1,\gamma_2\in[0,1]$, the two maximal eigenvalues are $\mu_1(R_{\rho\N})=1$ and $\mu_2(R_{\rho\N}) = (1-\gamma_1)(1-\gamma_2)$.
        It follows that the maximal CHSH violation is 
        \begin{equation}
            S^{\star}_{\text{CHSH}} = 2\sqrt{1 + (1-\gamma_1)(1-\gamma_2)}.
        \end{equation}
    \end{proof}
\end{proposition}

\section{The Unitality of Detector White Noise and Maximal Bell Violations}\label{appendix:unitality of detector white noise}

\begin{proposition}\label{prop:detector_white_noise_unitality}
    The white noise detector error $\mbf{W}_\gamma$ is unital and equivalent to the $M$-qubit depolarizing channel $\mc{D}_\gamma(X) = (1-\gamma)X + \frac{\gamma}{2^M}\mbb{I}_{2^M} \tr{X}$ acting upon the detector's local state as $\mc{D}_{\gamma}(\rho^{A_j})$.
    
    \begin{proof}
        Let $\Pi_{+|x}$ and $\Pi_{-|x}$ constitute an $M$-qubit PVM where $\Pi_{+|x}$ and $\Pi_{-|x}$ measure even and odd parity respectively.
        It follows that $\tr{\Pi_{+|x}}=\tr{\Pi_{-|x}}=(2^M)/2$.
        Furthermore, a detector that outputs white noise is expressed by the POVM elements $\Pi_{+|x}=\Pi_{-|x}=\mbb{I}_M/2$.
        Thus, the white noise detector error in Eq. \eqref{eq:white_noise_detector_error} is equivalent to the dichotomic POVM $\Pi'_{+|x}=(1-\gamma)\Pi_{+|x} + \frac{\gamma}{2}\mbb{I}_M$ and $\Pi'_{-|x}=(1-\gamma)\Pi_{-|x} + \frac{\gamma}{2}\mbb{I}_M$.
        Let the $M$-qubit depolarizing channel be defined as $\mc{D}_{\gamma}(X) = (1-\gamma) X + \frac{\gamma}{2^M}\mbb{I}_M\tr{X}$ where $\mc{D}(X)=\mc{D}^{\dagger}(X)$ because the Kraus operators of the depolarizing channel are all Hermitian $K_i=K_i^\dagger$.
        Then,
        \begin{align}
            \Pi'_{\pm|x} &= \mc{D}^{\dagger}_{\gamma}(\Pi_{\pm|x}) \\
            &= (1-\gamma) \Pi_{\pm|x} + \frac{\gamma}{2^M}\mbb{I}_M\tr{\Pi_{\pm|x}} \\
            &= (1-\gamma) \Pi_{\pm|x} + \frac{\gamma}{2}\mbb{I}_M.
        \end{align}
        It follows that,
        \begin{equation}
            \tr{\mc{D}_{\gamma}^{\dagger}(\Pi^{A_j}_{\pm|x})\rho^{A_j}} = \tr{\Pi^{A_j}_{\pm|x}\mc{D}_{\gamma}(\rho^{A_j})},
        \end{equation}
        where $\rho^{A_j}$ is the $M$-qubit state local to the detector.
        Thus, we conclude that detector white noise is equivalent to depolarizing noise applied to $\rho^{A_j}$.
        Hence, detector white noise is unital.
    \end{proof}
\end{proposition}

\begin{proposition}\label{prop:theoretical_white_noise_detector_error}
    Consider an $n$-local network with a detector noise model $\NNet_{\gammav} = \bigotimes_{j=1}^m \mbf{W}_{\gamma_j}^{A_j}$.
    The maximal Bell score for the star and chain inequalities are
    \begin{align}
        &S^\star_{n\text{-Star}} = \sqrt{2} \left(\prod_{j=1}^{n+1} (1-\gamma_j)\right)^{1/n}, \\
        &S^\star_{n\text{-Chain}} = \sqrt{2}\sqrt{\prod_{j=1}^{n+1}(1-\gamma_j)}.
    \end{align}
    
    \begin{proof}
        Within this work the $m$-partite correlators in Eq. \eqref{eq:I_correlator_combo} are parity observables that decompose as Eq. \eqref{eq:m-partite_correlator}.
        If a measurement device $A_j$ outputs uniform white noise such that $P(1|x)=P(-1|x)$, then $\langle O^{A_j}_x\rangle=0$ and, as a result, $\langle O^{A_1}_{x_1}\dots O^{A_m}_{x_m}\rangle=0$.
        Thus, if
        \begin{equation}
            \mbf{E}^{\Net}_{\gammav} = \bigotimes_{j=1}^m \mbf{W}_{\gamma_j}^{A_j}=\bigotimes_{j=1}^m\left( (1-\gamma_j)\mbb{I}_2 + \frac{\gamma_j}{2}\begin{pmatrix}1 & 1 \\ 1 & 1 \end{pmatrix}\right),
        \end{equation}
        then all terms in the tensor product sum that contain uniform white noise on at least one detector vanish.
        The only term that is nonzero is the identity, $\left(\prod_{j=1}^m(1-\gamma_j)\right)\mbb{I}_{2^m}$.
        It follows that
        \begin{equation}
            |\langle O^{A_1}_{x_1}\dots O^{A_m}_{x_m}\rangle| \leq \prod_{j=1}^m(1-\gamma_j).
        \end{equation}
        If the optimal noiseless strategy for violation of the star inequality is used \cite{Tavakoli_2014_star}, then
        \begin{equation}
            |I^\star_{n,y}| = \frac{1}{\sqrt{2^n}}\prod_{j=1}^{n+1}(1-\gamma_j),
        \end{equation}
        and the max violation of the star Bell inequality in Eq. \eqref{eq:n-local_star_inequality} is found to be
        \begin{align}
            S^\star_{n\text{-Star}} = \sqrt{2}\left( \prod_{j=1}^{n+1}(1-\gamma_j)\right)^{1/n}.
        \end{align}
        Likewise, if the optimal strategy for the chain inequality is used \cite{Mukherjee2015chain}, then
        \begin{equation}
            |I^\star_{2,y}| = \frac{1}{2}\prod_{j=1}^{n+1}(1-\gamma_j),
        \end{equation}
        and the max violation of the chain Bell inequality in  Eq. \eqref{eq:n-local-chain-bell-inequality} is found to be
        \begin{align}
            S^\star_{n\text{-Chain}} = \sqrt{2}\sqrt{ \prod_{j=1}^{n+1}(1-\gamma_j)}.
        \end{align}
    \end{proof}
\end{proposition}

\section{Nonmaximally Entangled States Optimally Violate the CHSH Inequality in the Presence of Two-Sided Qubit Amplitude Damping Noise}\label{appendix:amplitude_damping_channel}

\begin{figure}[h] 
    \centering
    \includegraphics[width=.8\textwidth]{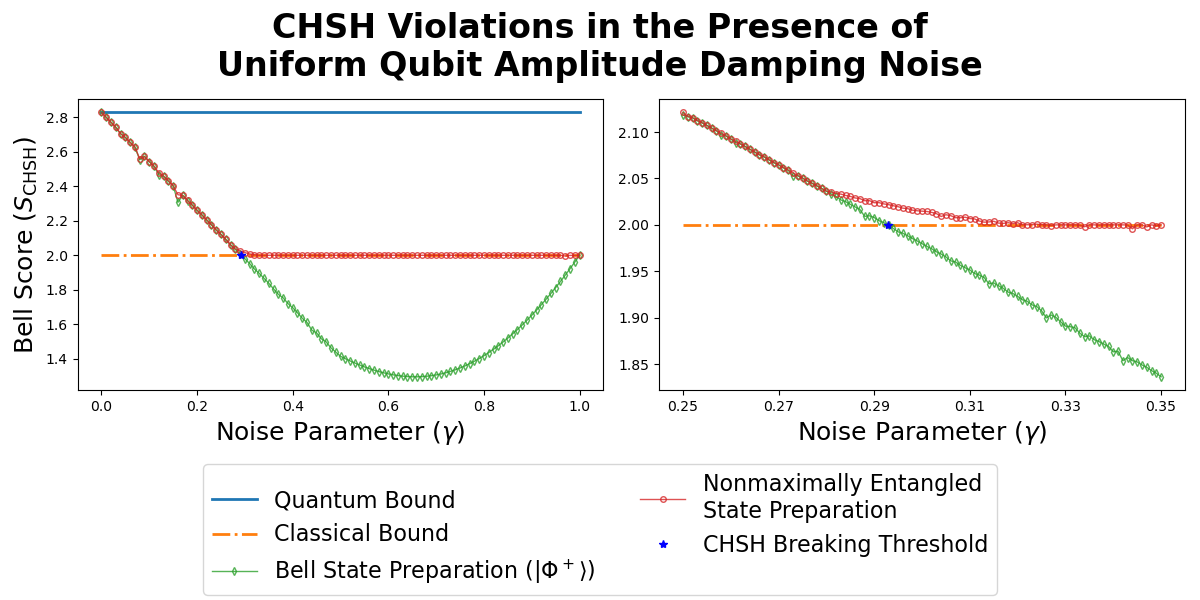}
    \caption{\linespread{1}\selectfont{\small
        \textbf{Uniform qubit amplitude damping in the CHSH scenario.}
        VQO is used on a classical simulator to find the maximal violation of the CHSH inequality when uniform amplitude damping is applied to the two source qubits.
        (Left) The noise parameter $\gamma\in[0,1]$ is scanned over an interval of $0.01$.
        (Right) The noise parameter $\gamma\in[0.25,0.35]$ is scanned over an interval of $0.001$.
        The green diamonds show the maximal violation optimized for $\ket{\Phi^+}$ state preparation.
        The red circles show the maximal violation optimized for nonmaximally entangled state preparations as described by the preparation ansatz in Table \ref{Table:prepare_measure_ansatzes}.
        In each case, arbitrary local qubit measurements are considered.
        The blue star denotes the value $\gamma = (1- 1/\sqrt{2})$ where nonlocality with respect to the CHSH inequality is broken for maximally entangled states.
    }
}
\label{fig:chsh_uniform_amplitude_damping_high_res}
\end{figure}

In this section, we show that nonmaximally entangled state preparations can achieve higher CHSH scores than maximally entangled states in the presence of two-sided qubit amplitude damping noise $\mc{A}_{\gamma_1}\otimes \mc{A}_{\gamma_2}$.
In Fig. \ref{fig:chsh_uniform_amplitude_damping_high_res}, we show that nonmaximally entangled state preparations parameterized as $\ket{\psi} = \cos(\phi_1/2)\ket{00} + \sin(\phi_1/2)e^{i\phi_2} \ket{11}$ can score better than the Bell state preparation $\ket{\Phi^+} = (\ket{00} + \ket{11})/\sqrt{2}$.
We then prove the maximally entangled states cannot perform better than Bell state preparation.
Thus, we confirm that the maximal violation of the CHSH inequality is not achieved by maximally entangled states when two-sided amplitude damping noise is present.
However, as pointed out in reference \cite{Pal2015}, maximally entangled states are optimal when amplitude damping is applied to one side $\mc{A}_{\gamma}\otimes \text{id}(\rho^{AB})$.

\begin{proposition}\label{prop:chsh-breaking_amplitude_damping}
    For maximally entangled state preparations, the two-sided amplitude damping channel $\mc{A}_{\gamma_1}\otimes\mc{A}_{\gamma_2}$ breaks nonlocality with respect to the CHSH inequality if and only if
    \begin{equation}\label{eq:two-sided_amplitude_damping_bell_state_boundary}
        \frac{1}{2} \geq (1-\gamma_1)(1-\gamma_2).
    \end{equation}
    
    \begin{proof}
        We first derive the critical boundary at which the channel $\mc{A}_{\gamma_1}\otimes\mc{A}_{\gamma_2}$ breaks the nonlocality of the bell state $\ket{\Phi^+} = \frac{1}{\sqrt{2}}(\ket{00} + \ket{11})$.
        By direct application of the amplitude damping Kraus operators in Eq. \eqref{eq:amplitude_damping_kraus_ops} the channel's output is found to be
        \begin{align}
            \rho_{\N} &= \mc{A}_{\gamma_1}\otimes\mc{A}_{\gamma_2}(\op{\Phi^+}{\Phi^+}) \\
            &= \frac{1}{2}\begin{pmatrix}
                1 + \gamma_1\gamma_2 & 0 & 0 & \sqrt{1-\gamma_1}\sqrt{1-\gamma_2} \\
                0 & \gamma_1(1 - \gamma_2) & 0 & 0 \\
                0 & 0 & (1-\gamma_1)\gamma_2 & 0 \\
                \sqrt{1-\gamma_1}\sqrt{1-\gamma_2} & 0 & 0 & (1-\gamma_1)(1-\gamma_2)
            \end{pmatrix}.
        \end{align}
        To decide whether $\rho_{\N}$ can violate the CHSH inequality we apply the necessary and sufficient conditions introduced by Horodecki \textit{et al.} \cite{Horodecki1995}.
        First, we construct the real-valued $3\times 3$ correlation matrix $T_{\rho_{\mc{N}}}$  using Eq. \eqref{eq:correlation_matrix}
        \begin{equation}
            T_{\rho_{\N}} = \begin{pmatrix}
                \sqrt{1-\gamma_1}\sqrt{1-\gamma_2} & 0 & 0 \\
                0 & -\sqrt{1-\gamma_1}\sqrt{1-\gamma_2} & 0 \\
                0 & 0 & 1 - \gamma_1 - \gamma_2 + 2\gamma_1\gamma_2\\
            \end{pmatrix}.
        \end{equation}
        Next, we find
        \begin{align}
            R_{\rho_{\N}} &= T_{\rho\N}^T T_{\rho\N} = \begin{pmatrix}
                (1-\gamma_1)(1-\gamma_2) & 0 & 0 \\
                0 & (1-\gamma_1)(1-\gamma_2) & 0 \\
                0 & 0 & (1 - \gamma_1 - \gamma_2 + 2\gamma_1\gamma_2)^2
            \end{pmatrix}.
        \end{align}
        Finally, we calculate the quantity maximal CHSH violation as
        \begin{equation}
            S^{\star}_{\text{CHSH}} = 2\sqrt{\mu_1(R_{\rho\N}) + \mu_2(R_{\rho\N})},
        \end{equation}
        where $\mu_1(R_{\rho\N})$ and $\mu_1(R_{\rho\N})$ denote the two largest eigenvalues of $R_{\rho\N}$.
        If the sum of eigenvalues satisfies $\mu_1(R_{\rho\N}) + \mu_2(R_{\rho\N})>1$, then the CHSH inequality is violated.
        For all $\gamma_1,\gamma_2 \in [0,0.5]$ the two largest eigenvalues of $R_{\rho\N}$ are $\mu_1(R_{\rho\N}) = \mu_2(R_{\rho\N})= (1-\gamma_1)(1-\gamma_2$.
        It follows that the CHSH inequality cannot be violated when
        \begin{equation}
            1 \geq 2(1-\gamma_1)(1-\gamma_2) \to \frac{1}{2}\geq (1-\gamma_1)(1-\gamma_2).
        \end{equation}
        Thus, we have proven the critical boundary at which the nonlocality of the Bell state $\ket{\Phi^+}$ is broken by the two-sided amplitude damping channel.
        
        The question still remains as to whether there exists a maximally entangled state which performs better than the Bell state $\ket{\Phi^+}$.
         To this end we consider the family of maximally entangled states as $\ket{\psi}=U\otimes\mbb{I}\ket{\Phi^+}$ where the arbitrary unitary $U$ is expressed as
        \begin{equation}
            U = \begin{pmatrix}
                e^{-i(\phi+\omega)/2}\cos(\theta/2) & -e^{i(\phi-\omega)/2}\sin(\theta/2)\\
                e^{-i(\phi-\omega)/2}\sin(\theta/2) & e^{i(\phi+\omega)/2}\cos(\theta/2)
            \end{pmatrix}.
        \end{equation}
        Using Mathematica, we calculate the eigenvalues of $R_{\rho\N}$ and find that they depend only on the parameters $\gamma_1$, $\gamma_2$, and $\theta$.
        We use Mathematica to maximize these eigenvalues over the parameters where nonlocality is broken for the $\ket{\Phi^+}$ state, \textit{i.e.}, $\theta\in[0,2\pi]$, $\gamma_1,\gamma_2\in[0,1]$, and $\frac{1}{2}\geq(1-\gamma_1)(1-\gamma_2)$.
        In this optimization we find no example violation of the CHSH inequality.
        Hence, the nonlocality of arbitrary maximally entangled states is also broken in the region specified by Eq. \eqref{eq:two-sided_amplitude_damping_bell_state_boundary}.
        
        Finally, within the region where the nonlocality of the Bell state is not broken, we maximize
        \begin{equation}
            \max{\gamma_1,\gamma_2,\theta}2(1-\gamma_1)(1-\gamma_2) - (\mu_1(R_{\rho\N}) + \mu_2(R_{\rho\N}))
        \end{equation}
        subject to the constraints $\gamma_1,\gamma_2\in[0,0.5]$, $\theta\in[0,2\pi]$ and $\frac{1}{2}<(1-\gamma_1)(1-\gamma_2)$ where $\rho\N$ is taken to be an arbitrary maximally entangled states.
        Using Mathematica to solve the maximization problem, we find no example where maximally entangled states achieves a larger violation than the Bell state.
        However, we do find examples where, once the nonlocality is broken, a larger Bell score can be obtained using a maximally entangled state that is not the Bell state.
        Hence we confirm that the nonlocality of all maximally entangled states is broken with respect to the CHSH inequality if Eq. \eqref{eq:two-sided_amplitude_damping_bell_state_boundary} is satisfied.
        The interested reader can find our Mathematica code in our supplementary codebase \cite{supp_codebase}.
    \end{proof}
\end{proposition}

\section{The Nonunitality of Biased Detector Noise}\label{appendix:nonunitality_biased_detector_noise}

\begin{proposition}\label{prop:equivalence_between_replacer_and_biased_detector_errors}
    The biased detector error $\mbf{R}_\gamma$ is nonunital and equivalent to a partial replacer channel \begin{equation}
        \mc{R}_{\gamma,x}(X) = (1-\gamma)X + \gamma \rho'_{x}\tr{X}
    \end{equation}
    applied to the quantum state $\rho^{A_j}_{x_j}$ local to node $A_j$ where the replacer state $\rho'_x$ is density operator contained by the projective subspace of $\Pi^{A_j}_{+|x}$.

    \begin{proof}
        For an $M$-qubit partial replacer channel with a pure replacer state $\rho_x = \op{\psi'_x}{\psi'_x}$, the Kraus operators are
        \begin{equation}
            K_i = \sqrt{\gamma}\op{\psi'_x}{i} \; \forall \; i\in[0,2^M), \; K_{2^M} = \sqrt{1-\gamma} \mbb{I},
        \end{equation}
        where $\{\ket{i}\}_{i=0}^{2^M-1}$ form an orthonormal basis.
        Then, for a measurement operator $\Pi_{a|x}$ having $a\in\{\pm\}$, we find
        \begin{align}
            \tr{\Pi_{a|x}\mc{R}_{\gamma}(\rho)} &= \sum_i \tr{\Pi_{a|x} K_i\rho K_i^\dagger}\\
            &= \sum_i \tr{K_i^\dagger \Pi_{a|x} K_i \rho} \\
            &= \tr{\mc{R}^\dagger_\gamma(\Pi_{a|x})\rho}.
        \end{align}
        If the replacer state $\ket{\psi_x}$ lies within the projective subspace of $\Pi_{+|x}$ we find that
        \begin{align}
            \mc{R}^{\dagger}_{\gamma,x}(\Pi_{+|x}) &= \sum_i K_i^\dagger \Pi_{+|x} K_i \\
            &= (1-\gamma)\Pi_{+|x} + \notag \\  
            &\quad\qquad+\gamma\sum_{i=0}^{2^M-1} \op{i}{\psi'_x}\Pi_{+|x}\op{\psi'_x}{i} \\
            &= (1-\gamma)\Pi_{+|x} + \gamma\sum_{i=0}^{2^M-1} \op{i}{i}\\
            &=(1-\gamma)\Pi_{+|x} + \gamma\mbb{I}.
        \end{align}
        However, repeating the procedure for $\Pi_{-|x}$ yields $\mc{R}_{\gamma,x}(\Pi_{-|x}) = (1-\gamma)\Pi_{-|x}$ because $\bra{\psi'_x}\Pi_{-|x}\ket{\psi'_x} = 0$.
        Thus, we recover the POVM in Eq. \eqref{eq:replacer_povm} for the biased detector error.
    \end{proof}
\end{proposition}

\end{document}